\documentclass{article}

\usepackage{graphicx}
\usepackage{natbib}
\usepackage{verbatim}
\usepackage{tikz}
\usepackage{pgfplots}
\usepackage{pgfplotstable}
\usepackage{amsmath}

\newcommand\be{\begin{equation}}
\newcommand\ee{\end{equation}}
\newcommand\inviscid{{\rm inviscid}}
\newcommand\base{{\rm base}}
\newcommand\summer{{\rm Sumer}}
\newcommand\pse{{\rm PSE}}
\newcommand\ose{{\rm OSE}}
\newcommand\ts{{\rm TS}}
\newcommand\bl{{\rm BL}}
\newcommand\rf{{\rm ref}}

\makeatletter
\newcommand{\pushright}[1]{\ifmeasuring@#1\else\omit\hfill$\displaystyle#1$\fi\ignorespaces}
\newcommand{\pushleft}[1]{\ifmeasuring@#1\else\omit$\displaystyle#1$\hfill\fi\ignorespaces}
\makeatother

\newcommand\Real{\mbox{Re}} 
\newcommand\Imag{\mbox{Im}} 
\newcommand\Rey{\mbox{\textit{Re}}}  

\newsavebox{\astrutbox}
\sbox{\astrutbox}{\rule[-5pt]{0pt}{20pt}}

\begin{document}

\title{Linear stability of the boundary layer under a solitary wave}

\author
{Joris\, C.\, G.\, Verschaeve$^1$\thanks{Email address for correspondence: joris@math.uio.no}
\, \& \,
Geir\, K.\, Pedersen$^2$}

\date{$^1$Norwegian Geotechnical Institute, Po.Box 3930, Ullev\r{a}l Stadion, 0806 Oslo, Norway\\[2ex]
$^2$University of Oslo, Po.Box 1072 Blindern, 0316 Oslo, Norway\\[5ex]\today}


\maketitle

\begin{abstract}
A theoretical and numerical analysis of 
the linear stability of the boundary layer flow under a solitary wave is
presented. In the present
work, the nonlinear boundary layer equations are
solved. The result is compared to the linear
boundary layer solution in \cite{LiuParkCowen2007} revealing
that both profiles are disagreeing more than has been found 
before. A change of frame of reference
has been used to allow for a classical linear stability
analysis without the need to redefine the notion of stability
for this otherwise unsteady flow. 
For the linear stability
the Orr-Sommerfeld
equation and the parabolic stability equation were used. The results
are compared to key results of inviscid stability theory and validated 
by means of a direct numerical simulation using a 
Legendre-Galerkin spectral element Navier-Stokes solver. 
Special care has been taken
to ensure that the numerical results are valid. 
Linear stability predicts that the boundary layer
flow is unstable for the entire parameter range considered,
confirming qualitatively the results
by \cite{BlondeauxPralitsVittori2012}. As a result 
of this analysis the
stability of this flow cannot be described by 
a critical Reynolds number
unlike what is atempted in previous publications. 
This apparent contradiction can be resolved by looking at the 
amplification factor responsible for the amplification of 
the perturbation. For lower
Reynolds numbers, the boundary layer flow becomes
unstable in the deceleration region of the flow. For higher
Reynolds numbers, instability arises also in the acceleration
region of the flow, confirming, albeit only qualitatively, an
observation in the experiments by
\cite{SumerJensenSorensenFredsoeLiuCarstensen2010}. 

\end{abstract}

\section{Introduction}

Solitary waves are frequently encountered in
experimental and theoretical fluid mechanics. This is not only
due to the existence of approximate solutions for 
solitary waves, see for instance \cite{Grimshaw1971} or \cite{Fenton1972}, but
also a consequence of the fact that solitary waves are relatively simple
to generate under laboratory conditions with good reproducibility. 
Solitary waves display a series of remarkable properties, cf \cite{Miles1980}. 
The key feature herein is the preservation of their shape
during propagation. This is, however, only true in the limit of
vanishing frictional effects by the air and the bottom. In reality, due
to nonzero viscosity a thin boundary layer will develop between 
water and air and between water and bottom. These boundary layers will lead to
a drain of energy finally dissipating the solitary wave \cite{Shuto1976,Miles1980}. Although 
very small for solitary waves on relatively large depths, such frictional
effects become more important for cases where the layer of water becomes
thin. This can be the case for solitary waves on small depths or
when the solitary wave is running up a beach which leads to a large
discrepancy between theoretical and experimental run up heights
\cite{PedersenLindstromBertelsenJensenSaelevik2013}.\\
The  bottom boundary layer has been considered 
to be the more relevant, cf. 
\cite{LiuOrfila2004}, and research has focused on it.
Investigation of the bottom boundary layer under a solitary wave has been 
initiated by \cite{LiuParkCowen2007} when they published theoretical
and experimental results concerning the shape of the boundary layer profile. 
This work has led to subsequent publications by 
\cite{SumerJensenSorensenFredsoeLiuCarstensen2010},
\cite{VittoriBlondeaux2008,VittoriBlondeaux2011} 
and \cite{BlondeauxPralitsVittori2012} investigating the 
transitions in the boundary layer. 
\cite{SumerJensenSorensenFredsoeLiuCarstensen2010} investigated
experimentally the stability of the
boundary layer flow under a solitary wave,
\cite{VittoriBlondeaux2008,VittoriBlondeaux2011} performed
direct numerical simulations to this end and
\cite{BlondeauxPralitsVittori2012} were the first to perform 
a linear stability analysis on this type of flow. A result of the works by 
\cite{SumerJensenSorensenFredsoeLiuCarstensen2010} and 
\cite{VittoriBlondeaux2011} is that the regimes of 
the boundary layer flow can be categorized into three to four regimes
\cite{SumerJensenSorensenFredsoeLiuCarstensen2010}:
laminar, laminar with regular vortex tubes, transitional and fully turbulent.
The transition between the first and the second regime is predicted by
\cite{VittoriBlondeaux2008} to happen at a Reynolds number $ \Rey_\summer $
somewhat below $ \Rey_\summer = 5 \times 10^5 $, 
whereas \cite{SumerJensenSorensenFredsoeLiuCarstensen2010}
measured it to be lower, namely at $ \Rey_\summer = 2 \times 10^5 $. Here
$ \Rey_\summer $ is a Reynolds number defined by \cite{SumerJensenSorensenFredsoeLiuCarstensen2010} which is based on particle displacement and maximum velocity in the outer flow as length and velocity scale, respectively. 
\cite{VittoriBlondeaux2011} proposed that 
circumstantial
laboratory conditions, such as wall roughness or vibrations, perturbed
the system and led to a lowering of the critical Reynolds number. As 
\cite{BlondeauxPralitsVittori2012} concluded later, the flow under a solitary
wave is always unstable in the sense of linear stability and 
they suggested to use the growth of the kinetic energy attached
to the perturbations as a measure for the appearance of transitions in the
flow. However, a linear stability analysis as presented in 
\cite{BlondeauxPralitsVittori2012} cannot predict whether the flow
after transition is turbulent or not, contrary to what the
title of \cite{BlondeauxPralitsVittori2012} suggests.
In a recent study   \cite{PedersenLindstromBertelsenJensenSaelevik2013} 
measured boundary layers during runup of solitary waves on a beach. They 
observed instabilities (undulations and vortices) for Reynolds numbers
which were higher than those of \cite{SumerJensenSorensenFredsoeLiuCarstensen2010}, when defined in a corresponding manner. However, the runup flow is
different from that under a solitary wave on constant depth, due to the
 moving shoreline and the
relatively longer retardation phase as compared to the acceleration phase,
which makes direct comparison difficult.    \\
In spite of the progress made in the aforementioned
references, a number of issues remain and need to
be addressed. The outer velocity field in 
\cite{SumerJensenSorensenFredsoeLiuCarstensen2010},
\cite{VittoriBlondeaux2008,VittoriBlondeaux2011} 
and \cite{BlondeauxPralitsVittori2012}
was either given by 
the simple secant hyperbolic formula \cite{Miles1980}
or the third order approximate 
formula by \cite{Grimshaw1971}. Both velocity fields deviate markedly
from the true velocity field. In addition, 
for the experiments in \cite{SumerJensenSorensenFredsoeLiuCarstensen2010},
and the numerical simulations in 
\cite{VittoriBlondeaux2008,VittoriBlondeaux2011} 
and \cite{BlondeauxPralitsVittori2012}, the outer
velocity field was made 'spatially uniform'. 
A result of the process of uniformization is that nonlinear
terms of the boundary layer equations are neglected and the
wall normal velocity component is put to zero. This results in 
a different boundary layer flow, thereby excluding
nonlinear and nonparallel effects. 
A justification for this uniformization was founded on the
conclusion by \cite{LiuParkCowen2007}, that the linear
boundary layer flow approximates the nonlinear one very closely. 
This conclusion is, however, based on an erroneous formula
given in \cite{LiuOrfila2004}
and therefore not properly justified. In addition, a
common difficulty encountered in the works by
\cite{SumerJensenSorensenFredsoeLiuCarstensen2010},
\cite{VittoriBlondeaux2008,VittoriBlondeaux2011} 
and \cite{BlondeauxPralitsVittori2012} is that the flow 
of a solitary wave is time dependent
and therefore the notion of hydrodynamic stability needed to be redefined. 
However,
the risk is then that the resulting definition is
of descriptive nature, rather  than being mathematically
concise, as for example in \cite{SumerJensenSorensenFredsoeLiuCarstensen2010}
and \cite{VittoriBlondeaux2008,VittoriBlondeaux2011}, where
instability simply meant that something unexpected became visible.
The relation between local instabilities, either temporal or spatial, to
a  global instability of a non-uniform flow may in general 
be complex \cite[see, for instance][]{HuerreMonkewitz1990} and the application
of approximate stability analysis, such as one involving a uniform flow
assumption, must be carefully checked. 
\cite{BlondeauxPralitsVittori2012} justify their application of the 
Orr-Sommerfeld
equation for the transient flow under a solitary wave by assuming that the
growth of instabilities takes  place on a
time scale much faster than the time scale of the basic flow. However, this 
assumption is incorrect, at least close to neutral stability. On the other 
hand, as will be shown subsequently, through comparison  with more general
 theories, their version of the Orr-Sommerfeld
equation still performs reasonably well. Another issue, which is not sufficiently elaborated in the references,
 is the seeding, or triggering, of the perturbation in the flow. 
\cite{VittoriBlondeaux2008,VittoriBlondeaux2011} applied white
noise with an amplitude of $ 10^{-4} $ as a seeding for the 
perturbation before the arrival of the solitary wave. 
\cite{SumerJensenSorensenFredsoeLiuCarstensen2010} did not
introduce any perturbation in their experiments at all, but relied instead 
on a natural seeding by the experimental environment.
 In general neither the frequency
nor the amplitude of the perturbation have been controlled in
\cite{VittoriBlondeaux2008,VittoriBlondeaux2011} and 
\cite{SumerJensenSorensenFredsoeLiuCarstensen2010}.\\
In the present treatise, the incorrect formulae in 
\cite{LiuParkCowen2007} are discussed and the corrected nonlinear
boundary layer solution is presented. To avoid 
dealing with a transient boundary value problem,
a simple change of frame of reference is made. 
In the frame of reference of the solitary wave,
the boundary layer flow is stationary and the 
entire range of classical theory of hydrodynamic stability
\cite{DrazinReid1981} can be applied. 
The stability properties of the viscous boundary layer under
a solitary wave are thus obtained using classical methods of
linear stability theory. In particular the Orr-Sommerfeld equation \cite{DrazinReid1981} and the parabolic stability equation \cite{BertolottiHerbertSpalart1992} were used to find the unstable regions of the boundary layer flow. 
For this type of flow, 
the criterion of 
amplification \cite{Jordinson1970,BertolottiHerbertSpalart1992}  
can be used to characterize the flow. 
The results by means of
this amplification criterion indicate that
the transition in the boundary layer flow under a solitary wave might
neither be characterized by a critical Reynolds number $ \Rey_\summer $
\cite{SumerJensenSorensenFredsoeLiuCarstensen2010,
VittoriBlondeaux2011} nor by a 
critical set of parameters $ (\delta_c , \epsilon_c ) $
\cite{VittoriBlondeaux2008,BlondeauxPralitsVittori2012}. 
Instead, the mere appearance of vortex tubes, say,
will depend in a large amount on the initial amplitude of the perturbation.
In addition, as was observed by \cite{SumerJensenSorensenFredsoeLiuCarstensen2010} for higher Reynolds numbers, the transition is characterized by 
a growth of perturbations in the acceleration region of the boundary layer, 
i.e. where the pressure gradient favors stability. 
This can be explained by the development 
of a 'viscous' instability, akin to that for the Blasius profile,
as shown by linear stability in the present treatise.
A direct numerical simulation by means of a Navier-Stokes solver was performed
to verify the results obtained by linear stability analysis.\\
The present discussion is organized as follows. 
The physical problem 
and the equations treated in the present discussion are briefly 
explained in section \ref{sec:problem}.
The numerical schemes used
to solve the equations are presented in section \ref{sec:numerics}. 
Section \ref{sec:results} presents results and discussion 
of the present investigation.
A short summary and the final conclusions are given in
section \ref{sec:conclusions}.\\
Since the present work is based on a numerical analysis using different
techniques we decided to give a short description of the numerical
methods used in section \ref{sec:numerics}. Verification and validation
are important for all numerical applications, but should be particularly
indispensable for stability models. Still, in the literature these
aspects are often superficially treated and the trustworthiness
of the numerical results cannot be properly assessed.
Hence, the schemes are concisely outlined in section 
\ref{sec:numerics}, while benchmarking is left for the 
appendices \ref{sec:appendixBoundaryLayer}-
\ref{sec:appendixNS}. This should enable the reader to
make an independent judgment of the strengths and limitations
of the present numerical analysis. Readers mainly interested in the
physical content, on the other hand, might skip section
\ref{sec:numerics}, and go directly
to the results in section \ref{sec:results}. 

\section{Description of the problem} \label{sec:problem}

\begin{figure}


\centering
\begin{tikzpicture}[xscale=1,yscale=1]
\begin{axis}
  [
    xlabel={$x$},
    ylabel={$y$},
    xmax  = 1.25,
    xmin  = -1.25,
    ymin  = -0.25
  ]
  \node (begin) at (axis cs: -0.75 , 1){};
  \node (end) at (axis cs: -0.75 , 1.5){};
  \node[anchor=east] (mitte) at (axis cs: -0.75 , 1.25){$ \epsilon h_0 $};
  \draw[<->] (begin)--(end); 
  \node (begin2) at (axis cs: -0.75 , 1){};
  \node (end2) at (axis cs: -0.75 , 0.){};
  \node[anchor=east] (mitte2) at (axis cs: -0.75 , 0.5){$ h_0 $};
  \draw[<->] (begin2)--(end2); 
  \node (begin3) at (axis cs: -0.5 , 0.75){};
  \node (end3) at (axis cs: 0.5 , 0.75){};
  \node[anchor=south] (mitte3) at (axis cs: 0. , 0.75){$ c $};
  \draw[<-] (begin3)--(end3); 
  \node (begin4) at (axis cs: -1. , 0.){};
  \node (end4) at (axis cs: 1. , 0.){};
  \draw[thick] (begin4)--(end4);

  \addplot[mark=none] table[x index=0,y expr=\thisrowno{1}+1]{dataPicture1.dat};

\end{axis}

\end{tikzpicture}
\caption{A solitary wave with height $ \epsilon h_0 $ traveling from 
right to left on constant depth $ h_0 $ at speed $ c $. The axes are
scaled according to (\ref{eq:scaling1}).}
\label{fig:solitarywave}
\end{figure}
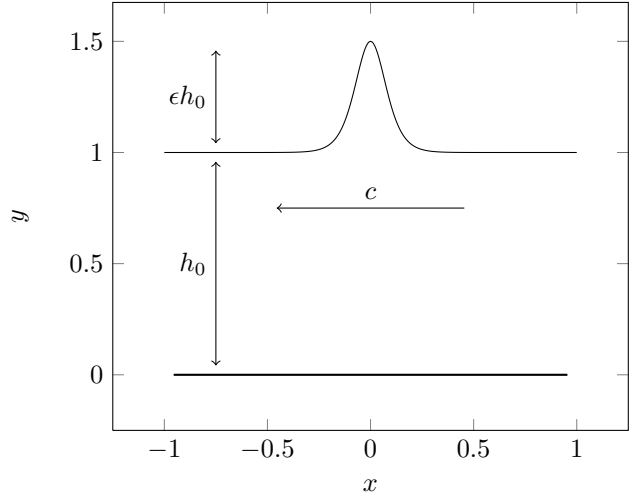

\begin{figure}
\centerline{
\includegraphics[width=\linewidth]{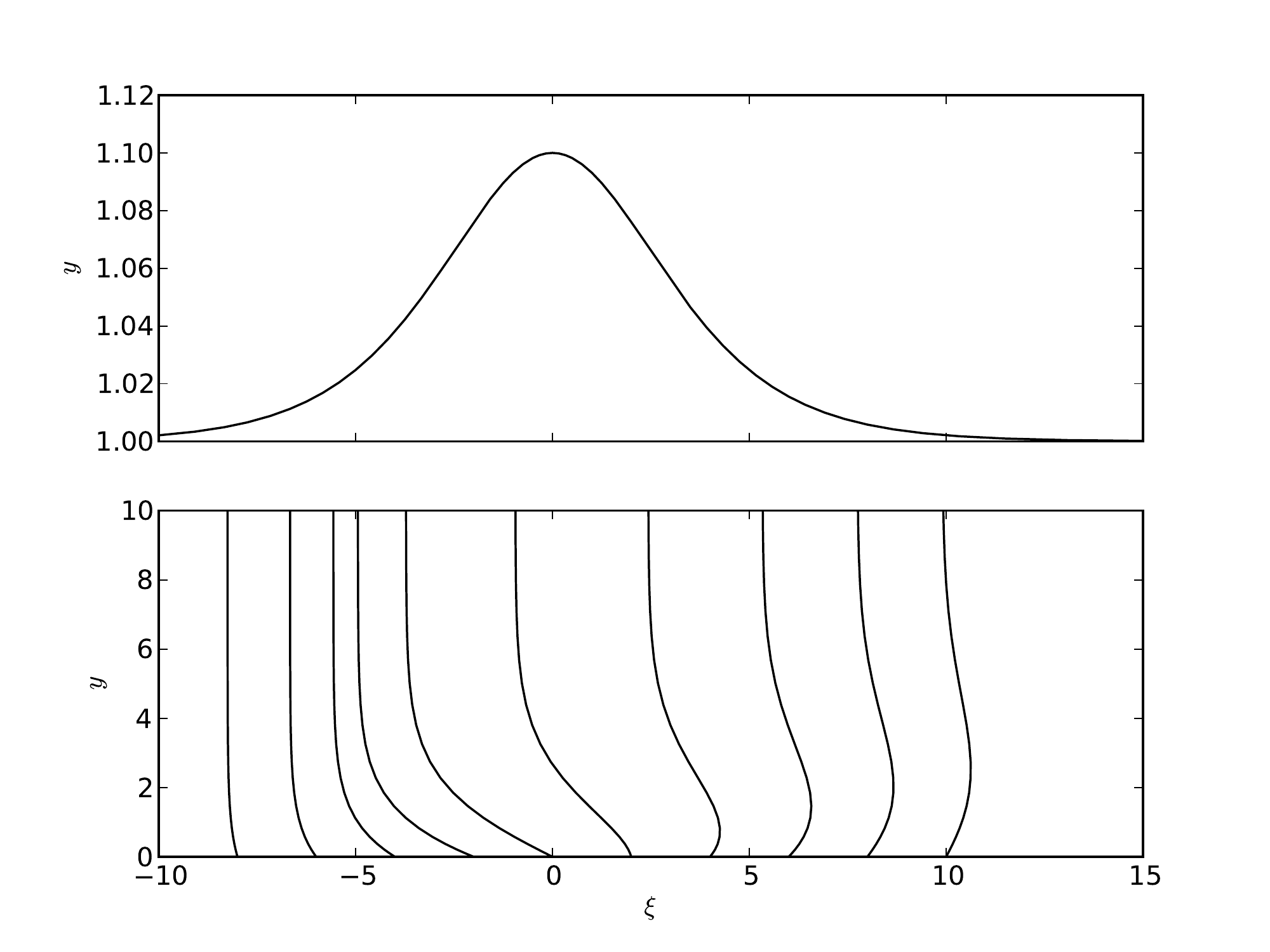}
}
\caption{Surface elevation $ \eta $ and profiles of the horizontal velocity component in the boundary layer under a solitary wave moving from right to left, $ \epsilon = 0.1 $, 
$ \delta = 8\times10^{-3} $. The profiles 
have been multiplied by 40. Upper panel is scaled according to (\ref{eq:scaling1}), while $ y $ in lower panel is scaled by (\ref{eq:scaling2}).}
\label{fig:profilesAndSolitaryWave}
\end{figure}

Herein, we investigate a solitary wave
with amplitude $ \epsilon h_0 $ propagating from
right to left on a flat bottom at depth $ h_0 $, cf. figure 
\ref{fig:solitarywave}. Neglecting friction, different
formulations for the inviscid solution to the problem exists.
These are briefly discussed in subsection \ref{sec:inviscid}. 
Common to all of these formulations is that the velocity field
$ (U_\inviscid,V_\inviscid) $ under the solitary wave is derived from
a potential $ \Phi $:
\be
U_\inviscid(x,y,t) = \frac{\partial \Phi}{\partial x}(x,y,t) \quad
V_\inviscid(x,y,t) = \frac{\partial \Phi}{\partial y}(x,y,t).
\label{eq:inviscidFlow}
\ee
The quantities
in (\ref{eq:inviscidFlow}) are scaled by the water depth 
$ h_0 $ and
the shallow water speed $ \sqrt{g h_0} $:
\be
x = \frac{x^*}{ h_0} \quad y = \frac{y^*}{h_0} 
\quad \Phi = \frac{\Phi^*}{\sqrt{gh_0} h_0} \quad t = \frac{t^*\sqrt{gh_0}}{h_0}
\label{eq:scaling1},
\ee
where the asterisk $ ^* $ designates dimensional quantities. 
In this idealized description, solitary waves
propagate without change in shape and velocity. 
In reality, however,
frictional effects slowly dissipate the energy of the solitary wave.
As mentioned in the introduction, most of the 
dissipation happens in the viscous boundary layer at the bottom. 
In this bottom boundary layer the velocity profile
decreases to zero in order to satisfy the no-slip boundary condition
at the wall. The flow in this thin layer is described by the
boundary layer equations which are presented in 
subsection \ref{sec:boundaryLayer}. 
These boundary layer equations are then solved
numerically, cf. subsection \ref{sec:boundaryLayerSolver},
in order to obtain an accurate velocity profile of the 
boundary layer under a solitary wave. 
As an illustration,
some profiles of the horizontal velocity component
in the boundary layer under a solitary wave are displayed
in figure \ref{fig:profilesAndSolitaryWave}. In 
addition to the boundary layer equations,
the incompressible Navier-Stokes equations 
used for the validation by direct numerical simulation
are presented in subsection \ref{sec:boundaryLayer}
for the present problem.  
The central question
in the present context is then to decide if the solution
of the boundary layer equations
is linearly stable and, if not, which are the parameters
governing its instability. 
Linear stability of flows, cf. subsection \ref{sec:linearStability},
is traditionally investigated by means
of the Orr-Sommerfeld equation (\ref{eq:OSE}) \cite{Jordinson1970,Orszag1971,VanStijnVanDeVooren1980} and/or the parabolic
stability equation (\ref{eq:PSE}) \cite{BertolottiHerbertSpalart1992,Herbert1997}. We 
present these two equations in subsections \ref{sec:OrrSommerfeld}
and \ref{sec:ParabolicStability}, respectively. 

\subsection{Inviscid solitary wave solutions} \label{sec:inviscid}

\begin{figure}
  \centering
\includegraphics[width=0.7\linewidth]{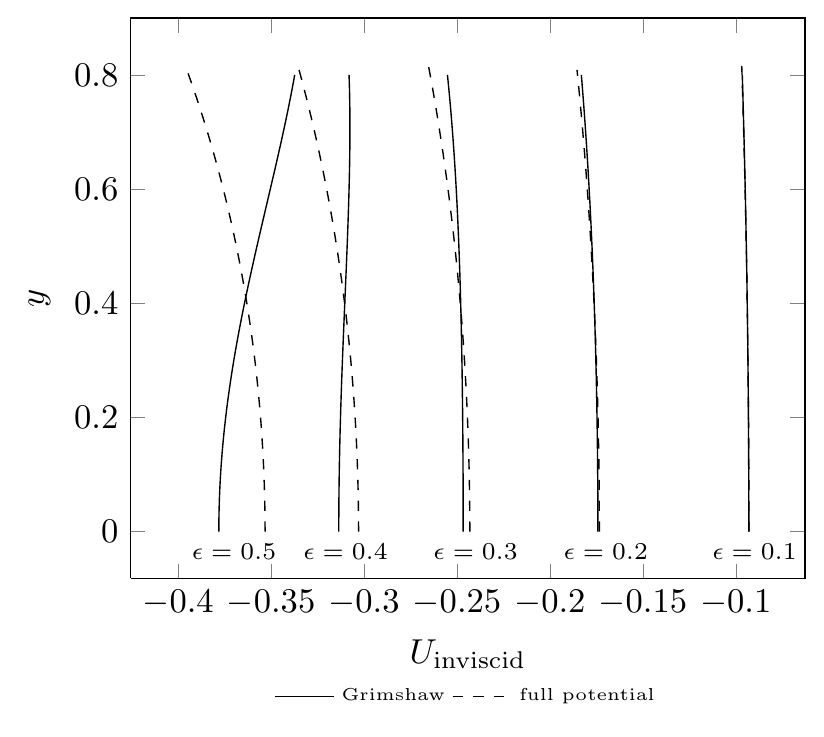}
\caption{Profiles of the inviscid horizontal velocity $ U_\inviscid $
under the crest of a solitary wave for different amplitudes $ \epsilon $. 
The profiles have been computed by means of the
third order approximate formula by
\cite{Grimshaw1971} and by means of the full
potential solution solved by the method of \cite{Tanaka1986}. The scaling
is given by (\ref{eq:scaling1}). }
\label{fig:ProfilesGrimshawVSPotential}
\end{figure}

\begin{figure}
\centering
\includegraphics[width=0.7\linewidth]{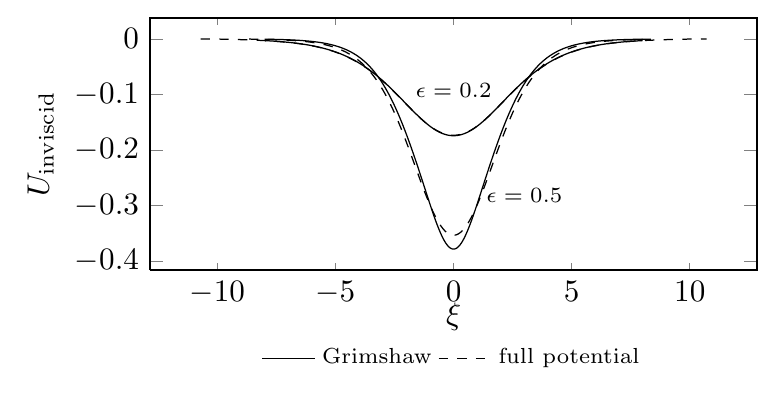}
\caption{The inviscid horizontal velocity $ U_\inviscid $
at the bottom wall for a solitary wave for different amplitudes $ \epsilon $. 
The horizontal velocity has been computed by means of the
third order approximate formula by
\cite{Grimshaw1971} and by means of the full
potential solution solved by the method of \cite{Tanaka1986}. The scaling
is given by (\ref{eq:scaling1}). }
\label{fig:BottomGrimshawVSPotential}
\end{figure}

The third order approximate solution by \cite{Grimshaw1971}
for the potential $ \Phi $ in \ref{eq:inviscidFlow} has been used
by \cite{LiuParkCowen2007} and
\cite{VittoriBlondeaux2008,VittoriBlondeaux2011} 
for their
computations of the boundary layer profile. 
The velocity $ (U_\inviscid,V_\inviscid) $
is thereby approximated by \cite{Grimshaw1971}:
\begin{eqnarray}
U_\inviscid &=& -\epsilon s^2 \\
           &+& 
\epsilon^2 \left[ - \frac{1}{4} s^2 + s^4 + y^2 \left( \frac{3}{2} s^2 - \frac{9}{4} s^4 \right) \right] \\
           &+&
\epsilon^3 \Big[ \frac{19}{40} s^2 + \frac{1}{5} s^4 - \frac{6}{5} s^6 + 
y^2 \left( -\frac{3}{2} s^2 - \frac{15}{4} s^4 + \frac{15}{2} s^6 \right) \\
& & \, \,\,\, \,\,\,\,\,+ \, y^4 \left( - \frac{3}{8} s^2 + \frac{45}{16} s^4 - \frac{45}{16} s^6 \right) \Big] \label{eq:grimshaw1} \\
V_\inviscid/\left(\sqrt{3\epsilon}yq\right) &=& - \epsilon s^2 \\
&+& \epsilon^2 \left[ \frac{3}{8} s^2 + 2 s^4 + y^2 \left(\frac{1}{2}s^2 - \frac{3}{2} s^4 \right) \right] \\
&+& \epsilon^3 \Big[ \frac{49}{640}s^2 - \frac{17}{20} s^4 
- \frac{18}{5} s^6 + y^2 \left( - \frac{13}{16} s^2 - \frac{25}{16} s^4 + \frac{15}{2} s^6 \right) \\
& & \, \,\,\, \,\,\,\,\,+ \, y^4 \left( - \frac{3}{40} s^2 + \frac{9}{8} s^4 - \frac{27}{16} s^6 \right) \Big],\label{eq:grimshaw2}
\end{eqnarray}
where
\begin{equation}
s = \mathrm{sech} \left( \alpha \left( x + c t \right) \right), \quad
q = \tanh \left( \alpha \left( x + c t \right) \right). 
\end{equation}
The wavenumber $ \alpha $ of the solitary wave is given by:
\be
 \alpha = \sqrt{\frac{3}{4} \epsilon } \left( 1 - \frac{5}{8} \epsilon + \frac{71}{128} \epsilon^2 \right), 
\ee
and its celerity $ c $ by:
\be
c^2 = { 1 + \epsilon - \frac{1}{20} \epsilon^2 - \frac{3}{70} \epsilon^3 }. 
\ee
Plots of the horizontal velocity profile under the crest of
a solitary wave using the solution of \cite{Grimshaw1971}
and the full potential solution described below
are given in figure \ref{fig:ProfilesGrimshawVSPotential}. 
As can be observed from figure \ref{fig:ProfilesGrimshawVSPotential},
the profile given by means of Grimshaw's solution deviates
markedly from the full potential one for $ \epsilon > 0.2 $. For
$ \epsilon > 0.3 $ Grimshaw's profile does even qualitatively
not follow the potential solution. Instead of increasing with the
distance from the bottom, Grimshaw's profile decreases.
The situation is, however, not as dramatic for the computation of the
boundary layer flow, since in this case only the value of $ U_\inviscid $ 
at the bottom is important. In figure \ref{fig:BottomGrimshawVSPotential},
$ U_\inviscid $ at the bottom wall is plotted for different values of
$ \epsilon $, the curves by Grimshaw's solution and the 
full potential solution are similar. However, 
for the $ \epsilon = 0.5 $ case, for example, Grimshaw's solution 
overpredicts the
magnitude of the bottom velocity under the crest 
of the solitary wave by approximately $  7\% $. 
Hence, 
we instead use a numerical solution for the full potential $ \Phi $
by a method derived by \cite{Tanaka1986} combined with a straightforward
application of Cauchy's formula \cite{PedersenLindstromBertelsenJensenSaelevik2013}. As mentioned above frictional
effects will give rise to a thin viscous boundary layer at the bottom.
The inviscid solution $ (U_{inviscid}, V_{inviscid})$ obtained by the method of \cite{Tanaka1986} is then used to compute the boundary layer flow,
equations (\ref{eq:BL1}-\ref{eq:BL3}),
as described in the next subsection. 

\subsection{Navier-Stokes and Boundary layer equations} \label{sec:boundaryLayer}

In the present work, we use two different scalings. 
The first one, (\ref{eq:scaling1}), is based on
the equilibrium water depth $ h_0 $ as a length scale and the shallow water
speed $ \sqrt{gh_0} $ as a velocity scale. The velocity scale
for the second scaling is the same as for the first one. However,
the length scale shall be given by $ \delta^* $, which
is a viscous length scale defined by 
\cite{VittoriBlondeaux2008,VittoriBlondeaux2011}, and which
characterizes the thickness of the boundary layer:
\be
\delta^* = \sqrt{ \frac{ 2 \nu h_0 }{ \sqrt{ g h_0 }}}.
\label{eq:deltaStar}
\ee
Spatial variables are scaled by $ \delta^* $:
\be
x = x^*/\delta^*, \quad y = y^*/\delta^*. \label{eq:scaling2}
\ee
We do not use different sub- or superscripts in order to label
how variables are scaled. 
In general, in the remaining figures presented in
this treatise the horizontal lengths are scaled by $ h_0 $,
whereas the vertical lengths are scaled by $ \delta^* $. 
As before velocities are scaled by $ \sqrt{gh_0} $, which leads to 
time scaled the following way:
\be
t = \frac{t^* \sqrt{gh_0}}{\delta^*}. \label{eq:scaling2time}
\ee
The Navier-Stokes equations appear then 
as usual:
\begin{eqnarray}
\frac{\partial u}{\partial x} + \frac{\partial v}{\partial y} & = & 0 \label{eq:NS1} \\
\frac{\partial u}{\partial t} + 
 u \frac{\partial u}{\partial x} + v \frac{\partial u}{\partial y} 
&=& - \frac{\partial p}{\partial x} 
+ \frac{1}{\rm Re} \nabla^2 u  \label{eq:NS2} \\
\frac{\partial v}{\partial t} +
u \frac{\partial v}{\partial x} + v \frac{\partial v} { \partial y} 
&=& - \frac{\partial p}{\partial y} 
+ \frac{1}{\rm Re} \nabla^2 v,  \label{eq:NS3} 
\end{eqnarray}
where the Reynolds number $ \rm Re $ is given by:
\be
{\rm Re} = \frac{\delta^* \sqrt{gh_0}}{\nu }.
\ee
Since $ \delta^* $ is a viscous length scale and
given by (\ref{eq:deltaStar}), the
Reynolds number can be further simplified. 
\be
{\rm Re} = \frac{2 h_0 }{\delta^* } = \frac{2}{\delta},
\ee
where $ \delta = \delta^*/h_0 $. Following \cite{VittoriBlondeaux2011}
we will use $\delta$ and $\epsilon$ to identify the investigated cases.
The list of employed $\delta$ values, together with the corresponding
value of $h_0$ for water is\vspace{2mm}\\
\begin{center}
\begin{tabular}{l|c|c|c|c|c|c}
$\delta$ &$1 \cdot 10^{-5}$ &$4\cdot 10^{-5}$&$8\cdot 10^{-5}$&1$\cdot 10^{-4}$&$4.75\cdot 10^{-4}$&$8\cdot 10^{-4}$\\
$h_0$ (m) & 344.2 & 54.2 & 21.5 & 16.0 &2.0 & 1.0
\end{tabular}\vspace{2mm}
\end{center}
The smaller $\delta$ corresponds to rather deep water, whereas the larger ones 
approach values that are relevant for wave tank experiments.\\
In order to derive the boundary layer equations 
an inhomogeneous scaling is used. 
The changes of the boundary layer flow 
in the horizontal direction happen
on a length scale of $ h_0 $, whereas 
all variations of the boundary layer flow
in the vertical direction
are produced on a scale of $ \delta^* $. 
The coordinates are therefore scaled the following way:
\be
x = \frac{x^*}{h_0}, \quad y = \frac{y^*}{\delta h_0}.
\ee
The velocity components are scaled accordingly:
\be
u =  \frac{u^*}{\sqrt{gh_0}}, \quad v = \frac{v^*}{\sqrt{gh_0}\delta}. 
\ee
When we insert this into the Navier-Stokes equations and retain only the 
leading order terms in $ \delta^2$ terms, we arrive at the 
following boundary layer equations:
\begin{eqnarray}
\frac{\partial u}{\partial x} + \frac{\partial v}{\partial y} & = & 0,
\label{eq:BL1absolute} \\
\frac{\partial u}{\partial t} + u \frac{\partial u}{\partial x} + v \frac{\partial u}{\partial y} & = &
- \frac{\partial p^{\rm ext}}{\partial x} + \frac{1}{2} 
\frac{ \partial^2 u}{ \partial y^2}, \label{eq:BL2absolute}  \\
\frac{\partial p^{\rm ext}}{\partial y} & = & 0, \label{eq:BL3absolute} 
\end{eqnarray}
where the exterior pressure gradient is given by the inviscid bulk flow:
\be
-\frac{\partial p^{\rm ext}}{\partial x} = \frac{\partial }{\partial t}U_\inviscid(x,0,t) + U_\inviscid(x,0,t)
\frac{\partial}{\partial x} U_\inviscid (x,0,t).
\ee
However, the problem with
equations (\ref{eq:BL1absolute}-\ref{eq:BL3absolute}) is 
that its solution is not steady and we would be constrained to
redefine the notion of hydrodynamic stability as attempted in
\cite{SumerJensenSorensenFredsoeLiuCarstensen2010},
\cite{VittoriBlondeaux2008,VittoriBlondeaux2011} 
and \cite{BlondeauxPralitsVittori2012}. 
To avoid this difficulty
and to be able to use classical hydrodynamic stability theory
\cite{DrazinReid1981}, we employ a simple standard trick. Equations
(\ref{eq:NS1}-\ref{eq:NS2}) are valid for an observer in 
the frame of reference defined by the equilibrium state. This
observer sees a wave traveling with velocity $ -c \, \mathbf{e}_x $. However, for the 
remainder of the present work the frame of reference will be
defined by the solitary wave itself. The observer will then see
a bottom moving at velocity $ c \, \mathbf{e}_x $ and water entering at velocity 
$ c \, \mathbf{e}_x $.
The letter $ \xi $ shall be used to denote the moving coordinate:
\be
\xi = x + ct.
\ee
Since the Navier-Stokes equations are Galilean invariant they
appear the same as in equations (\ref{eq:NS1}-\ref{eq:NS2}) with 
only $ x $ replaced by $ \xi $. The Navier-Stokes equations
in the moving frame of reference with both spatial 
variables scaled by $ \delta^* $ are then used for the 
Orr-Sommerfeld equation (\ref{eq:OSE}), the parabolic stability
equation (\ref{eq:PSE}) and the Navier-Stokes solver, cf.
subsection \ref{sec:NavierStokesSolver}. \\
Neglecting the slow change of the inviscid base flow
due to viscous dissipation, the boundary layer flow can be regarded as steady
in this frame of reference. The final boundary layer equations are thus
given by:
\begin{eqnarray}
\frac{\partial u}{\partial \xi} + \frac{\partial v}{\partial y} & = & 0,
\label{eq:BL1} \\
u \frac{\partial u}{\partial \xi} + v \frac{\partial u}{\partial y} & = &
- \frac{\partial p^{\rm ext}}{\partial \xi} + \frac{1}{2} 
\frac{ \partial^2 u}{ \partial y^2}, \label{eq:BL2}  \\
\frac{\partial p^{\rm ext}}{\partial y} & = & 0, \label{eq:BL3} 
\end{eqnarray}
where the exterior pressure gradient is given by the inviscid bulk flow:
\be
-\frac{\partial p^{\rm ext}}{\partial \xi} = U_\inviscid(\xi,0)
\frac{\partial U_\inviscid}{\partial \xi}(\xi,0).
\ee
The boundary conditions for (\ref{eq:BL1}-\ref{eq:BL3})
in the vertical direction are given by:
\begin{eqnarray}
u & = & c \quad \mbox{at} \quad y = 0, \\
u & = & U_\inviscid(\xi,0)  \quad \mbox{at} \quad y= y^{\rm ext}\\
v & = & 0 \quad \mbox{at} \quad y = 0,
\end{eqnarray}
where $ y^{\rm ext} $ is the 'edge' of the boundary layer \cite{Keller1978}. 
We solve equations (\ref{eq:BL1}-\ref{eq:BL3}) numerically by a Chebyshev collocation method. The details of this method are presented in subsection 
\ref{sec:boundaryLayerSolver}. Once a solution to (\ref{eq:BL1}-\ref{eq:BL3})
has been found, we use linear stability to determine whether this
boundary layer flow is stable or not. This boundary layer flow is
thus the steady base flow for the remainder of this study. 
The solution of (\ref{eq:BL1}-\ref{eq:BL3}), marked by lower
case letters $ u $ and $ v $, shall subsequently be denoted by
upper case letters
\be
U_\base \quad \mbox{and} \quad V_\base. \label{eq:base}
\ee
The reason for this is to distinguish the boundary layer flow from the 
perturbed flow, which is the sum of the base flow and the perturbation
as explained in the following subsection. 

\subsection{Linear stability of the boundary layer} \label{sec:linearStability}

An in-depth review on linear stability of boundary layers is
given in the book by \cite{DrazinReid1981}. The following
subsections give a brief overview of the equations needed herein.\\
Since our problem treated is two dimensional,
the base flow, equation (\ref{eq:base}), can be derived from a 
stream function $ \Psi $ in the following way:
\be
U_{base} = \frac{\partial \Psi}{\partial y} \quad V_{base} 
= - \frac{\partial \Psi}{\partial \xi}.
\label{eq:base2}
\ee
The pressure can then be eliminated from the Navier-Stokes 
equations (\ref{eq:NS1}-\ref{eq:NS3}) leading to a governing equation
for the stream function $ \Psi $ \cite{White2005}:
\be
\left( \frac{\partial}{\partial t} - \frac{1}{\Rey} \nabla^2 + \frac{\partial \Psi}{\partial y} \frac{\partial}{\partial \xi} - \frac{\partial \Psi}{\partial \xi}\frac{\partial}{\partial y} \right) \nabla^2 \Psi = 0. \label{eq:NSPSI}
\ee
A common ansatz in linear stability theory is to add  a perturbation $(u',v')$
to the base flow (\ref{eq:base2}). As before we write this perturbation 
in stream function form:
\be
u' = \frac{\partial \psi'}{\partial y}
\quad v' = -\frac{\partial \psi'}{\partial \xi}. \label{eq:perturbation}
\ee
Injecting now the perturbed flow $ \Psi + \psi' $ into (\ref{eq:NSPSI}),
we obtain a nonlinear governing equation for the perturbation $ \psi' $. 
Since, however, the perturbation is assumed to be small in amplitude, we can 
neglect the nonlinear terms and find a linear equation for 
the perturbation $ \psi' $ for a given base flow $ \Psi $:
\be
\left( \frac{\partial}{\partial t} - \frac{1}{\Rey} \nabla^2 + \frac{\partial \Psi}{\partial y} 
\frac{\partial}{\partial \xi} - \frac{\partial \Psi}{\partial \xi} 
\frac{\partial}{\partial y} \right) \nabla^2 \psi' + 
\frac{\partial \nabla^2 \Psi}{\partial \xi} 
\frac{\partial \psi'}{\partial y} 
- \frac{\partial \nabla^2 \Psi}{\partial y}  \frac{\partial \psi'}{\partial \xi} = 0
\ee
This equation is the foundation for the two methods of linear stability
which shall be used in the present discussion. The first one is the
famous Orr-Sommerfeld equation, cf. subsection \ref{sec:OrrSommerfeld},
whereas the second one is the parabolic stability equation presented
briefly in subsection \ref{sec:ParabolicStability}. 

\subsection{Orr-Sommerfeld equation} \label{sec:OrrSommerfeld}
The Orr-Sommerfeld equation, see \cite{DrazinReid1981} for a more
detailed review, is based on the assumption of parallel flow. This means
that 
the normal component of the base flow is assumed to be
negligible, $ V_{base} = - \partial \Psi /\partial x = 0$ 
and that non parallel effects are ignored. Hence,
the stability of each profile for a given $ \xi $ is
analyzed independently. The resulting governing
equation for $ \psi' $ is thus given by:
\be
\left( \frac{\partial}{\partial t} 
- \frac{1}{\Rey} \nabla^2 + \frac{\partial \Psi}{\partial y} \frac{\partial}{\partial \xi}  \right) \nabla^2 \psi'  
- \frac{\partial^3 \Psi}{\partial y^3} \frac{\partial \psi'}{\partial \xi} = 0 \label{eq:preOSE}
\ee
Looking at each profile independently means that we assume the perturbation
to have a specific form. It is modeled as a Tollmien-Schlichting wave traveling along
the horizontal direction:
\be
\psi' = \phi(y) \exp \left( a \xi - {\rm i} \omega t\right), 
\label{eq:PsiOSE}
\ee
where $ \phi $ is an unknown function controlling the shape of the
wave in normal direction. The given real number $ \omega $ 
is the angular velocity of the wave. The
complex part of $ a $ is the wave number and 
its real part the growth rate of the wave. For a given 
angular velocity $ \omega $ and a given profile at some $ \xi $,
equation (\ref{eq:preOSE}) gives rise to an algebraic eigenvalue problem
for the eigenvalue $ a $ and the eigenfunction $ \phi $, the
famous Orr-Sommerfeld equation \cite{DrazinReid1981}:
\be
\frac{1}{\Rey} \left( D^2 + a^2 \right)^2\phi + \left( {\rm i} \omega - U_\base a \right) \left( D^2 + a^2 \right) \phi + \frac{\partial^2 U_\base}{\partial y^2} a \phi = 0,
\label{eq:OSE}
\ee
where $ D = d/dy $. The boundary conditions for $ \phi $ are given by:
\be
\phi(0) = D\phi(0) = 0 \quad \phi(y \rightarrow \infty) \rightarrow 0. \label{eq:bcOSE}
\ee 
The discrete spectrum of (\ref{eq:OSE})
will determine the stability of the flow. If there exists an
eigenvalue $ a $ with a positive real part, then we say that the
base flow is (becoming) unstable. This happens usually at a certain
value of $ \xi $ after which the Orr-Sommerfeld equation
gives rise to eigenvalues with positive real part. The numerical
details on how equation (\ref{eq:OSE}) is solved are given
in subsection \ref{sec:OrrSommerfeldSolver}.
Unstable regions along the horizontal axis $ \xi $ for a given $ \omega $
are then defined by:
\be
\Real \, a(\xi) > 0 . 
\ee
As in \cite{Jordinson1970}, amplification
of the perturbation is measured by
\be
\ln \frac{A}{A_0} = \int \limits_{\xi_0}^{\xi} \Real \, a(x) \, dx.
\label{eq:ampliOSE}
\ee
As shall 
be seen later on, the nonparallel effects are, however, significant
for the present boundary layer. Therefore, 
an additional method of linear stability shall be used,
the parabolic stability equation, presented in the next
subsection.

\subsection{Parabolic stability equation} \label{sec:ParabolicStability}
The parabolic stability equation was derived by 
\cite{BertolottiHerbertSpalart1992}. An in-depth discussion of this
method can be found in their article and in \cite{Herbert1997}.
In the present subsection only
a brief summary of the main elements is given.
This method pursues two goals. First, it weakens
the parallel flow assumption and only assumes that the flow is slowly 
varying in $ \xi $. Second, it reformulates the governing
equation as an initial value problem and not as an eigenvalue problem. 
As we do not assume that the base flow is parallel, the perturbation,
equation (\ref{eq:perturbation}) needs, opposed to
the Orr-Sommerfeld equation, to account for a variation in $ \xi$. 
\cite{BertolottiHerbertSpalart1992} proposed the following ansatz for
the Tollmien-Schlichting wave:
\be
\psi' = \phi(\xi,y) \exp \left( \int \limits_{\xi_{0}}^\xi a(\hat\xi) \, d\hat\xi - {\rm i} \omega t \right). \label{eq:PsiPSE}
\ee
Now the shape function $ \phi $ and the wave number and growth rate defined by 
$ a $ are dependent on $ \xi $. Although the flow is not assumed to be 
parallel, it is assumed that all flow variables vary
slowly with respect to $ \xi $, such that higher than first order derivatives
of $ \phi $ and $ a $ 
with respect to $ \xi $ can be neglected. This 
leads to the following nonlinear initial value problem for $ a $ and $ \phi$,
cf. \cite{BertolottiHerbertSpalart1992}:
\be
\left( L_0 + L_1 \right)  \phi + L_2 \frac{\partial \phi}{\partial \xi} + L_3 \phi \frac{d a}{d \xi} = 0, \label{eq:PSE}
\ee
where the operators $ L_i $, $ i = 0,1,2,3 $ operate on $ y $ only and are
given by:
\begin{eqnarray}
L_0 & = & - \frac{1}{\Rey} \left( D^2 + a^2 \right)^2 + \left( {\rm i} \omega - U_{base} a \right) \left( D^2 + a^2 \right) - \frac{ \partial^2 U_\base}{\partial y^2} a, \\
L_1 & = & - \frac{\partial^2 V_\base}{\partial y^2} D + V_\base \left( D^2 + a^2 \right) D,\\
L_2 & = & - \frac{4a}{\Rey } \left( D^2 + a^2 \right) + U_\base \left( D^2 + 3 a^2 \right) - 2{\rm i} \omega a - \frac{\partial^2 U_\base}{\partial y^2}\\
L_3 & = &  - \frac{2}{\Rey } \left( D^2 + 3 a^2 \right) - {\rm i} \omega +
3 a U_\base
\end{eqnarray}
The form of $ \psi' $, equation (\ref{eq:PsiPSE}), is not unique and an 
additional condition is needed to determine $ \phi $ and $ a$. 
The main idea for finding an additional constraint is to restrict the growth to the parameter $ a $ and let
$ \phi $ only have variations in shape. As mentioned
by \cite{BertolottiHerbertSpalart1992}, several choices are possible. 
In the present discussion, we adopt one of their choices, namely to 
require orthogonality between the horizontal velocity component and its
derivative with respect to $ \xi $ \cite{BertolottiHerbertSpalart1992}:
\be
\int \limits_{0}^{\infty}  \frac{\partial^2 \phi}{ \partial \xi \partial y}  \overline{\frac{\partial \phi}{ \partial y} } \, dy = 0. \label{eq:psiConstraint} 
\ee
In order to be able to measure the growth of the perturbation
independently of the constraint chosen, \cite{BertolottiHerbertSpalart1992} 
defined the amplitude $ A $ of the perturbation the following way:
\be
A = \max_y | \frac{\partial \phi}{\partial y}  |
\exp \int \limits_{\xi_0}^{\xi} \Real \, a(x) \, dx. \label{eq:amplitude}
\ee 
The amplification is then the ratio between the amplitudes
at two different points:
\be
\frac{A}{A_0} = \frac{ \max_y | \frac{\partial \phi}{\partial y}  |
\exp \int \limits_{\xi_0}^{\xi} \Real \,a(x) \, dx }{ 
\max_y | \frac{\partial \phi_0}{\partial y}  |}. \label{eq:amplificationPSE}
\ee
The unstable region along the horizontal axis for a given $ \omega $
is then bounded by the points $ \xi $, where
\be
\max_y | \frac{\partial \phi}{\partial y}  |
\exp \int \limits_{\xi_0}^{\xi} a(x) \, dx  \label{eq:stabilityRegionPSE}
\ee
is minimum or maximum. The last term in equation (\ref{eq:PSE}) given by 
$ L_3 {d a}/{d \xi} $ is neglected in \cite{BertolottiHerbertSpalart1992}
as well as in the present work. A back-calculation of the term after solution 
of the equations does indeed confirm that it is small. \\[1ex]
\clearpage

\section{Numerical methods} \label{sec:numerics}

Since the present investigation depends crucially on the quality of the 
numerical analysis employed, we present a brief description of each numerical method
used in this section. 


\subsection{Boundary layer equations solver} \label{sec:boundaryLayerSolver}

In order to solve the boundary layer equations (\ref{eq:BL1}-\ref{eq:BL2}),
we use a Chebyshev collocation method, cf. \cite{Trefethen2000}, since 
Chebyshev polynomials are superior in accuracy compared to classical
finite difference formulations \cite{Orszag1971}. The problem 
(\ref{eq:BL1}-\ref{eq:BL2}) is then solved on a 
$ (N_\bl+1) \times (N_\bl+1) $ grid
at the Gau{\ss} Lobatto Chebyshev knots. 
The differential operators in (\ref{eq:BL1}-\ref{eq:BL2}) are
replaced by their discrete representations:
\begin{eqnarray}
\frac{\partial}{\partial \xi} & \leftrightarrow & D_\xi = \frac{1}{L_x} D_N \otimes I, \\
\frac{\partial}{\partial y} & \leftrightarrow & D_y =  I \otimes \frac{2}{y^{\rm ext}} D_N,  \\
\frac{\partial^2}{\partial y^2}& \leftrightarrow & D_y^2 = I \otimes \frac{4}{{y^{\rm ext}}^2} D_N^2, 
\end{eqnarray}
where $ I $ is the identity matrix and $ D_N $ the Chebyshev collocation
differentiation matrix. The lengths $ L_x $ and $ y^{\rm ext} $ control the 
size of the domain $ [-L_x,L_x ] \times [ 0, y^{\rm ext} ] $. Once we are given,
an initial guess $ u_N^0 $ for the discrete solution $ u_N $, we solve
first the continuity equation for $ v_N^0 $:
\be
A v_N^0 = b^0
\ee
where $ A $ is essentially $ D_y $ and $ b  $ is essentially 
$ - D_\xi u^0_N $. The rows of $ A $ corresponding to the boundary $ y = 0 $,
are modified in order to account for $ v_N^0 = 0 $ at the lower boundary. 
The corresponding rows in $ b^0 $ are modified accordingly. The construction
of the momentum operator $ C $ is similar. First we set
\be
C^0 = {\rm diag}(u^0_N) D_\xi + \rm{diag(D_\xi u^0_N)} + {\rm diag}(v^0_N) D_y - \frac{1}{2} D_y^2,
\ee
where $ {\rm diag (u^0_N) } $ designates a diagonal matrix with the elements of $
u^0_N $ on the diagonal. 
Some rows of $ C $ are modified in order to account for:
\begin{eqnarray}
u_N^0 = c & \mbox{at} &  \xi = 0 \label{eq:bound1}\\ 
u_N^0 = c & \mbox{at} &  y = 0 \\
u_N^0 = U_\inviscid & \mbox{at} &  y = y^{\rm ext} \label{eq:bound3}
\end{eqnarray}
The second member $ d $ is obtained similarly by posing:
\be
d^0 = U_\inviscid \frac{\partial U_\inviscid}{\partial \xi}  - {\rm diag}(u^0_N) D_\xi u^0_N - {\rm diag}(v^0_N) D_\xi u^0_N + \frac{1}{2} D_y^2 u^0_N.
\ee
As before, some elements of $ d $ need to be modified 
in order to account for the boundary conditions
(\ref{eq:bound1}-\ref{eq:bound3}). The increment $ u'_N $ updating the 
solution $ u_N $ from $ u_N^0 $ to $ u^1_N = u_N^0 + u'_N $ is then
computed by solving: 
\be
C^0 u'_N = d^0. 
\ee
This procedure is iterated until the Euclidean norm of $ u'_N $ falls
below $ 10^{-12} $. The result gives then the base flow $ (U_\base, V_\base) $
we need for the methods described below. 
The code was written in MATLAB and we verified the above algorithm
by a test problem as well as comparison with the boundary layer model of
\cite{PedersenLindstromBertelsenJensenSaelevik2013}, which again was compared
to measurements. In addition the present boundary layer equations solver 
was validated 
by means of the Blasius boundary layer. The results of this
benchmarking can be found in appendix \ref{sec:appendixBoundaryLayer}.
The boundary layer
solutions used in section \ref{sec:results} were all computed
using a resolution of $ N_\bl = 80 $, which gives sufficiently 
accurate results as can be verified in appendix \ref{sec:appendixBoundaryLayer}.
The edge of the boundary layer, given by $ y^{\rm ext} $, is of course
chosen such
that the boundary layer width is smaller than $ y^{\rm ext} $. 
As a matter of fact the boundary layer width is often less
than the width of the Tollmien-Schlichting waves. Therefore,
we chose a larger than necessary
value for $ y^{\rm ext} $, namely $ y^{\rm ext} = 60 $,
cf. appendix \ref{sec:appendixBoundaryLayer}, such that the bulk of the 
Tollmien-Schlichting waves fits into the domain. 
However, for very low frequencies the Tollmien-Schlichting wave
displays a large width. Therefore an extrapolation of 
$ (U_\base,V_\base) $ for values of $ y $ larger 
than $ y^{\rm ext} $ is necessary.
This is done by means of the inviscid solution
$ (U_\inviscid,V_\inviscid) $ in such a way that at $ y^{\rm ext}$ the 
extrapolant is continuous in the horizontal and vertical velocity. 

\subsection{Orr-Sommerfeld equation solver} \label{sec:OrrSommerfeldSolver}

The numerical solution of the Orr-Sommerfeld equation has led to
a vast number of works published in literature \cite{Osborne1967,Jordinson1970,Orszag1971,VanStijnVanDeVooren1980}. In the
following we use as in \cite{Orszag1971} a Chebyshev collocation method
in order to solve equation (\ref{eq:OSE}) in the domain $ [0,L_y] $.
We choose to cut the domain at $ L_y $ instead of using an
algebra mapping. Once we are given the 
differentiation matrix $ D_N $, the discrete version of equation
(\ref{eq:OSE}) can be written as:
\begin{eqnarray}
 &-&\frac{1}{\Rey}\left(\frac{16}{L^4_y} D_N^4 + \frac{8}{L_y^2} a^2 D_N^2+ a^4 I \right) \phi_N \nonumber \\
& +&\left( {\rm diag}(U_\base) a - {\rm i}\omega I \right) \left(\frac{4}{L^2_y} D_N^2 + a^2 I\right) \phi_N - a \,{\rm diag}\left(\frac{\partial^2 U_\base}{\partial y^2 } \right) \phi_N = 0, 
\label{eq:discreteOSE}
\end{eqnarray}
where $ I $ is the identity matrix and $ \phi_N $ are the values
of $ \phi $ at the $ N_\ose +1 $ Gau{\ss} Lobatto Chebyshev nodes. In order to
solve (\ref{eq:discreteOSE}) together with the boundary conditions
(\ref{eq:bcOSE}) we use an approach described in \cite{Trefethen2000}. 
The fourth order differentiation matrix $ D_N^4 $ in (\ref{eq:discreteOSE})
is thereby replaced by a different matrix $ \tilde{D}_N^4 $ 
which ensures the correct boundary conditions, see 
\cite{Trefethen2000} for details. 
The
code was written in MATLAB. We used the method by \cite{Osborne1967}
and the inbuilt eigenvalue solver
by MATLAB to solve the eigenvalue equation (\ref{eq:discreteOSE}). 
This was done in order to double-check the correctness of the solution. 
As verification, some eigenvalues and eigenfunctions
for the Blasius boundary value were computed using the present solver and 
the results were compared to values in literature, cf. appendix \ref{sec:appendixOSE}. The important numerical parameters for the 
present Orr-Sommerfeld solver are the number $ N_\ose $ of 
Chebyshev collocation points and the extension of the domain $ L_y $. 
The determination of the extension $ L_y $ for the Orr-Sommerfeld
equation solver is opposed to the extension in $ y $ for
the boundary layer equations solver in subsection 
\ref{sec:boundaryLayerSolver} not determined by the thickness of the
boundary layer but instead by the width of the Tollmien-Schlichting wave,
which can be several times larger than the boundary layer
thickness. As mentioned in \cite{VanStijnVanDeVooren1980}, the 
dominant asymptotic solution to the Orr-Sommerfeld equation for $ y \rightarrow \infty $ in case for the Blasius boundary layer is given by:
\be
\phi \approx \exp \left( - \Imag( a) y \right), \label{eq:asymptotic}
\ee
where $ \phi $ and $ a $ are defined in (\ref{eq:PsiOSE}). 
As we shall see
in section \ref{sec:results}, the phase speed
of the Tollmien-Schlichting wave in the boundary layer
under a solitary wave is close 
to the celerity $ c $ of the solitary wave. 
Assuming that the asymptotic solution (\ref{eq:asymptotic})
is also approximately valid for a 
the Tollmien-Schlichting wave in the boundary layer
under a solitary wave, it
allows us to 
give an apriori estimation of $ L_y $ by posing
\be
L_y = c_\ts / \omega, \label{eq:extensionPSE}
\ee
where $ c_\ts $ is a constant. By numerical inspection we found that a 
value of $ c_\ts = 10 $ allows enough tolerance to capture the 
Tollmien-Schlichting wave for a broad range of $ \epsilon $ and $ \delta$. 
This value has been used for all simulations throughout the present treatise.
The number $ N_\ose $ of Chebyshev collocation points has
been determined by convergence tests, cf. appendix \ref{sec:appendixOSE},
and is fixed to the value $ N_\ose = 130 $ for the remainder of the
present work.

\subsection{Parabolic stability equation solver} \label{sec:ParabolicStabilityEquationSolver}

For the parabolic stability equation solver, we use basically the same
approach as in \cite{BertolottiHerbertSpalart1992}. The main difference
is that instead of using an algebraic mapping in order to map 
the interval $ [-1,1] $ onto $ [0,\infty) $, we use truncation of the 
domain at $ L_y $. In \cite{GroschOrszag1977} it was shown that
although truncation of the domain is less efficient than an
algebraic mapping, it does nevertheless produce accurate results. 
As in \cite{BertolottiHerbertSpalart1992}, the operators 
$ L_0$ , $ L_1 $ and $ L_2 $ are replaced by their discrete 
correspondence $ M_0 $, $ M_1 $ and $ M_2 $, respectively, defined the following
way:
\begin{eqnarray}
M_0 & = & 
-\frac{1}{\Rey}\left( \frac{16}{L^4_y} \tilde{D}_N^4 + \frac{8}{L^2_y} a^2 D_N^2 + a^4 I \right)
\nonumber \\
& +&  \left( {\rm diag}(U_\base) a - {\rm i}\omega I \right) \left(\frac{4}{L^2_y} D_N^2 + a^2 I\right) - a \,{\rm diag}\left(\frac{\partial^2 U_\base}{\partial y^2 } \right) 
\\
M_1 & = & - {\rm diag}\left( \frac{\partial^2 V_\base}{\partial y^2} \right) \frac{2}{L_y} D_N + {\rm diag} \left( V_\base \right) \left( \frac{8}{L^3_y} \tilde{D}^3_N + a^2\frac{2}{L_y} D_N \right) \\
M_2 & = & - \frac{4a}{\Rey } \left( \frac{4}{L^2_y} D_N^2 + a^2 I \right) \nonumber \\
& +& {\rm diag}\left( U_\base \right) \left( \frac{4}{L^2_y} D^2_N + 3 a^2I  \right) - 2{\rm i} \omega a I - 
{\rm diag} \left( \frac{\partial^2 U_\base}{\partial y^2} \right).
\end{eqnarray}
The discretization of (\ref{eq:PSE}) in $ x $ is then done as in 
\cite{BertolottiHerbertSpalart1992} by using second
order central finite differences:
\be
\left( M_0 + M_1 \right) \frac{1}{2} \left( \phi_{j+1} + \phi_{j} \right)
+ M_2 \frac{1}{\Delta \xi } \left( \phi_{j+1} - \phi_{j} \right) = 0. 
\label{eq:PSEdiscrete}
\ee
As in \cite{BertolottiHerbertSpalart1992}, the operators $ M_0 $, $ M_1 $ and
$ M_2 $ in (\ref{eq:PSEdiscrete}) are evaluated using the value 
of $ a $ at the midpoint, namely $ a_{mid} = ( a_j + a_{j+1} )/2 $. Since equation (\ref{eq:PSEdiscrete}) is nonlinear in $ a_{j+1} $,
it is solved iteratively together with the constraint (\ref{eq:psiConstraint}).
A first guess $ a^0_{j+1} $ is given by $ a_j $, this allows us to obtain
an approximation $ \phi^0_{j+1} $ at the node $ \xi_{j+1} $ by
solving (\ref{eq:PSEdiscrete}). However, $ \phi^0_{j+1} $ does not
necessarily obey (\ref{eq:psiConstraint}). In order to obtain a new
value $ a^1_{j+1} $, we solve the following equation for the
complex number $ \kappa $:
\be
\alpha \kappa \overline{\kappa} + \beta \kappa + \gamma \overline{\kappa} + \delta = 0.
\ee
This equation is the discrete form of (\ref{eq:psiConstraint}) and
the coefficients are given by
\begin{eqnarray}
\alpha &=& \frac{1}{2\left(\xi_{j+1}-\xi_j\right)}\int_0^{L_y} \, dy \frac{\partial \phi^0_{j+1}}{\partial y}
\overline{\frac{\partial \phi^0_{j+1}}{\partial y}} \\
\beta &=& \frac{1}{2\left(\xi_{j+1}-\xi_j\right)}\int_0^{L_y} \, dy \frac{\partial \phi^0_{j+1}}{\partial y}
\overline{\frac{\partial \phi_{j}^0}{\partial y}} \\
\gamma &=& \frac{1}{2\left(\xi_{j+1}-\xi_j\right)}\int_0^{L_y} \, dy \overline{\frac{\partial \phi^0_{j+1}}{\partial y}}
{\frac{\partial \phi^0_{j}}{\partial y}} \\
\delta &=& \frac{1}{2\left(\xi_{j+1}-\xi_j\right)}\int_0^{L_y} \, dy \frac{\partial \phi^0_{j}}{\partial y}
\overline{\frac{\partial \phi^0_{j}}{\partial y}} 
\end{eqnarray}
The differentiation with respect to $ y $ is as before done 
by means of the differentiation matrix $ D_N $, 
whereas the integration with respect to $ y $ is done by
means of a trapezoidal quadrature rule. Once the root $ \kappa $ is 
found we pose:
\be
a^1_{j+1} = a_{j+1}^0 + \frac{2}{\xi_{j+1} - \xi_j} \ln \kappa. 
\ee
With this new result for $ a_{j+1} $, we solve again (\ref{eq:PSEdiscrete})
and repeat the procedure until convergence, which is obtained 
in our case when 
\be
| \frac{2}{\xi_{j+1} - \xi_j} \ln \kappa | < 10^{-8}. 
\ee
The initial condition $ a_0 $ and $ \phi_0 $ is obtained by means of
equation (26) in \cite{BertolottiHerbertSpalart1992}, which is an eigenvalue
problem similar to the Orr-Sommerfeld equation (\ref{eq:OSE}). 
The program solving the parabolic stability equation has been written in 
MATLAB code and verified carefully. The results of this verification are
presented in appendix \ref{sec:appendixPSE}. There are three parameters
governing the numerical accuracy of the method. These are given by
\begin{eqnarray}
 N_{\pse}, & & \mbox{the number of Chebyshev collocation points in $ y$},\\
 L_y, & & \mbox{the extension of the domain in $ y $ and}\\
 \Delta \xi, & & \mbox{the discretization in $ \xi $ direction}.
\end{eqnarray}
Since all physical quantities are slowly varying in $ \xi $, the
parabolic stability equation solver is relatively insensitive to the
discretization $ \xi $. However, the number of Chebyshev collocation
points $ N_{\pse} $ and the extension of the domain $ L_y $ are important
parameters to the method. The extension $ L_y $ is as for the
Orr-Sommerfeld equation solver determined by formula \ref{eq:extensionPSE}. 
The determination of $ N_\pse $ is done by convergence tests,
cf. appendix \ref{sec:appendixPSE}.
From these tests we found that choosing $ N_\pse = 180 $ allows us
to be on the safe side concerning the accuracy of the present results. This
value has been used for all simulations throughout the present treatise. 
For the computation of a neutral curve, such as the one displayed
in figure \ref{fig:domainDelta0.000475Convergence}, the minimum or maximum
of equation (\ref{eq:stabilityRegionPSE}) is found when the change in $ \xi $
is less than $ 10^{-5} $. For the amplification plots, such as the ones
in section \ref{sec:results}, it is enough to mention
that the plotting accuracy in $ \xi $ needed to produce
a smooth figure is by far more stringent than the numerical 
accuracy needed to make the error contribution by $ \Delta \xi $ 
subdominant to the error contribution by $ N_\pse $.

\subsection{Legendre-Galerkin spectral element Navier-Stokes solver}
\label{sec:NavierStokesSolver}

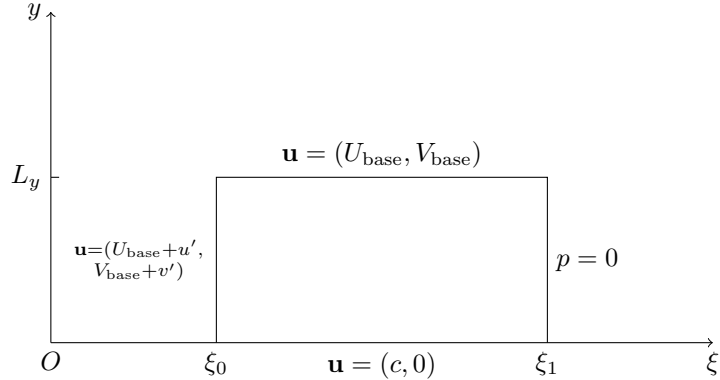
\begin{figure}
\centering
\begin{tikzpicture}[xscale=2.2,yscale=2.2]
  \path (0,0) node[below]{$ O $} coordinate (O);
  \path (1,0) coordinate (x0);
  \path (3,0) coordinate (x1);
  \path (1,1) coordinate (x2);
  \path (3,1) coordinate (x3);
  \path (0,1) coordinate (y1); 
  
  \draw [->] (O) -- (4,0) node [below] {$ \xi $};
  \draw (y1) -- ++(0.05,0);
  \draw  (O) -- (y1) node [left] {$ L_y $};
  \draw [->] (y1) -- (0,2) node [left] {$ y $};
  \draw (x0) node[below]{$ \xi_0 $} --(x2)--(x3)--(x1) node[below] {$ \xi_1$ };
  \node [left] at (1,0.5) {$ \mathbf{u} = (U_\base + u', \atop V_\base + v') $};
  \node [below] at (2.,0.) {$ \mathbf{u}= (c, 0) $};
  \node [above] at (2.,1.) {$ \mathbf{u}= (U_\base, V_\base) $};
  \node [right] at (3.,0.5) {$ p = 0 $};

\end{tikzpicture}

\caption{Set up for the Navier-Stokes solver}
\label{fig:NSsetUp}
\end{figure}

\begin{figure}
\centerline{
\includegraphics[width=\linewidth]{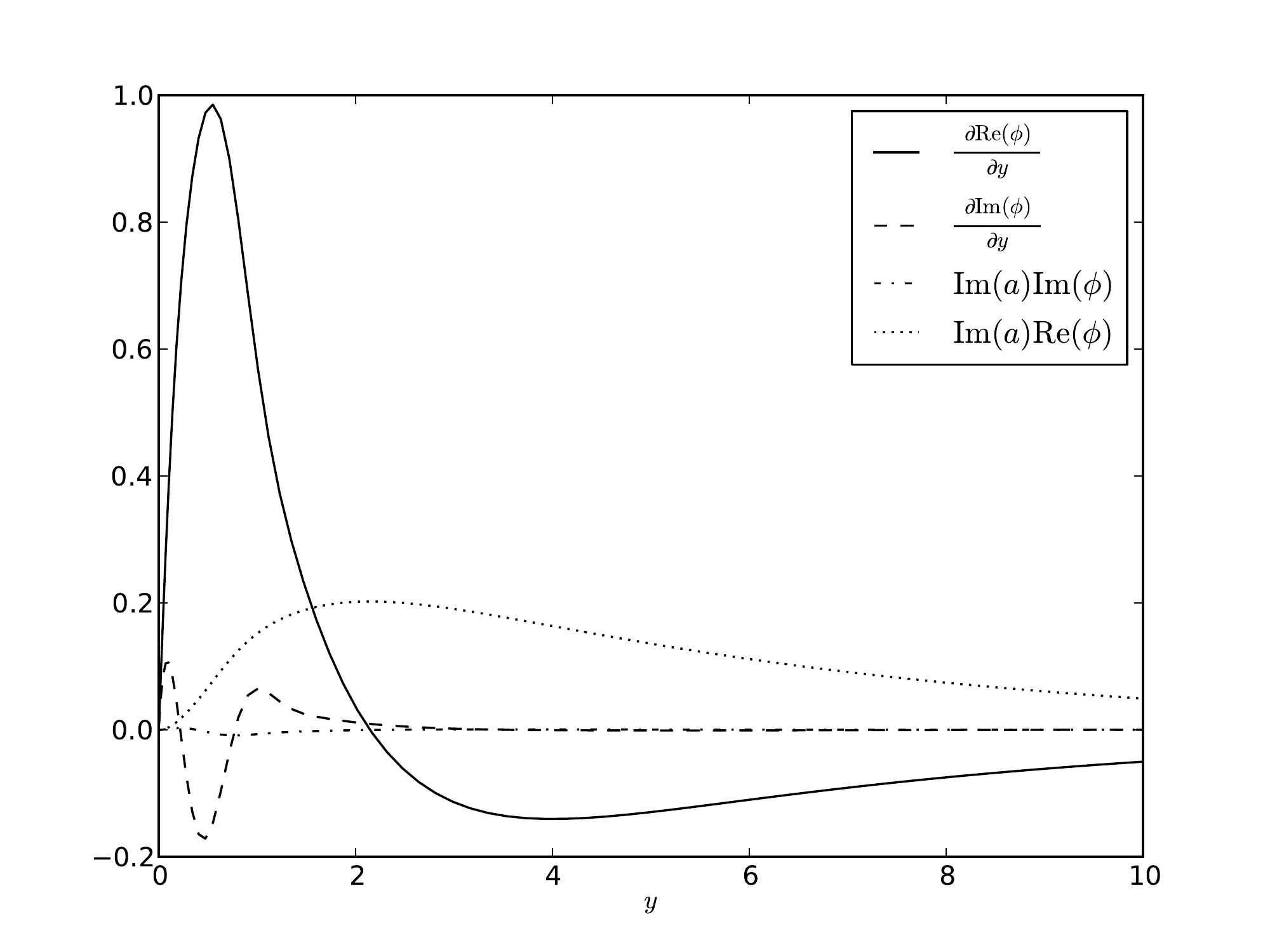}
}
\caption{Profiles of the perturbation used for the Navier-Stokes
solver, cf. equations (\ref{eq:initialConditionNS1}) and (\ref{eq:initialConditionNS2}). The parameters are $ \epsilon =0.4 $, $ \delta = 4.75 \cdot 10^{-4} $ 
and $ \omega = 0.22 $ and the profile was taken at position $ \xi = -0.2375 $. The profiles are only shown to a value of the ordinate of $ y = 10 $. However, for the present case
$ L_y = 45.5 $, a value at which the profiles have decayed sufficiently. The
scaling for the $ y $ axis is given by (\ref{eq:scaling2}).}
\label{fig:TSwave}

\end{figure}

Results obtained by the Orr-Sommerfeld equation solver
and the parabolic stability
equation solver described in 
subsections \ref{sec:OrrSommerfeldSolver} 
and \ref{sec:ParabolicStabilityEquationSolver}, 
respectively, are compared to direct numerical simulations
using the spectral element Navier-Stokes solver NEK5000 which
\cite{nek5000-web-page} developed at the Argon National
Laboratory. The solver is freely available. 
Since control
of the accuracy is crucial to obtain correct growth rates
of the Tollmien-Schlichting waves, a spectral method
was preferred to a low order method such as the one used in
\cite{VittoriBlondeaux2008,VittoriBlondeaux2011}. The
NEK5000 solver is based on a Galerkin formulation of the
Navier-Stokes equations (\ref{eq:NS1}-\ref{eq:NS3}). The 
method uses a standard P/P-2 formulation based 
at the Gau{\ss} Lobatto Legendre nodes for $ u $ and $ v $ 
and the Gau{\ss} Legendre nodes for $ p $ 
\cite{CanutoHussainiQuarteroniZang1993}. 
Convective and diffusive parts are advanced in time by a splitting
scheme. Nonlinear terms are integrated using Orszag's 3/2 rule. 
The pressure is solved by an Uzawa algorithm. Since the
domain is rectangular a fast tensor product solver is used
which increases the efficiency dramatically. More details on
the implementation can be obtained in the documentation
in \cite{nek5000-web-page}. \\
In the present treatise we use for the direct numerical 
simulation an approach akin to the one developed by \cite{Fasel1976}. 
The set up used is 
sketched in figure \ref{fig:NSsetUp}. The domain is rectangular
and aligned with the lower boundary of the computational
domain. Since the frame of reference is moving with the solitary wave,
the wall has a velocity $ (u = c ,v = 0) $, given by the speed of the solitary wave. The upper boundary is
situated at $ y = L_y $, where $ L_y $ is given by (\ref{eq:extensionPSE}). 
The velocity needed for the velocity boundary condition
at the upper boundary is obtained by
$ (U_\base ,V_\base) $ computed in beforehand by means of 
the boundary layer solver. The right boundary of the domain at $ \xi = \xi_1 $ 
is a simple
outflow boundary with the condition $ p = 0 $. The left boundary condition is
a velocity inlet at $ \xi = \xi_0 $, where
we in addition to the base flow add a Tollmien-Schlichting wave $ (u',v')$. The velocity $(u,v)$ at
the inflow is thus given by:
\be
u = U_\base + u' \quad v = V_\base + v', \label{eq:inflowNS}
\ee
where $ U_\base $ and $ V_\base $ is the boundary layer flow and computed
by the boundary layer solver in subsection \ref{sec:boundaryLayerSolver}. 
This approach to introduce the perturbation in the direct numerical
simulation was first developed by \cite{Fasel1976}.
In the present case, 
the perturbation $ (u',v') $ is computed by means of the parabolic
stability equation:
\begin{eqnarray}
u' & = & A_{0} \left( \frac{\partial \Real \left( \phi \right) }{\partial y} \cos(
\Imag(a) \xi_0 - \omega t)
- \frac{\partial \Imag \left( \phi \right) }{\partial y} \sin( \Imag(a) x_0 - \omega t) \right)  \label{eq:initialConditionNS1}\\
v'& = & A_{0} \Imag(a) \left( \Real \left( \phi \right) \sin (\Imag(a) x_0 - \omega t)  + \Imag\left( \phi \right) \cos(
\Imag(a) \xi_0 - \omega t) \right) \label{eq:initialConditionNS2}
\end{eqnarray}
where $ \phi $ and $ a $ are given by equation (\ref{eq:PsiPSE}).
The amplitude is
controlled by $ A_{0} $, which is set to $ 5\cdot 10^{-4} $ 
for all simulations
in order to make the perturbation small with respect to the mean flow,
which is of order $ 1 $. An example of
$ \phi $ obtained by means of the parabolic stability equation solver
can be seen in figure \ref{fig:TSwave}. We remark that we neglected in
(\ref{eq:initialConditionNS1}) and (\ref{eq:initialConditionNS2}) the terms due
to the growth rate, the real part of $ a $. 
However, this is not crucial, 
since it is small compared to the remaining parts. Another way of introducing
$ (u',v') $ at the inflow would be to use only:
\begin{eqnarray}
u' & = & A_{0} \frac{\partial \Real \left( \phi \right) }{\partial y} \cos(
\Imag(a) \xi_0 - \omega t) \label{eq:initialConditionNS1Fasel}, \\
v'& = & A_{0} \Imag(a) \Real \left( \phi \right) \sin(
\Imag(a) \xi_0 - \omega t) \label{eq:initialConditionNS2Fasel},
\end{eqnarray}
as was done in \cite{Fasel1976}. By anticipating a few results
from section \ref{sec:results}, we tried this approach and
it lead to a 
wrong amplification of the Tollmien-Schlichting wave in a region
behind the inflow as it did in \cite{Fasel1976}. 
In the rest of the computational domain the 
amplification was, however, correct.
Presumably, the Tollmien-Schlichting wave introduced
at the inflow
by means of (\ref{eq:initialConditionNS1Fasel}) and 
(\ref{eq:initialConditionNS2Fasel}) needs to develop to the
correct shape during a few wavelengths. This behaviour has been utilized
by \cite{BertolottiHerbertSpalart1992} for their direct numerical 
simulations. They used a generic shape mimicking a Tollmien-Schlichting wave
to introduce the perturbation which then over a few wavelengths developed to 
the correct solution. 
As long as the frequency $ \omega $ of the perturbation generating the
Tollmien-Schlichting wave in the direct numerical simulation
is the same, the amplification of the Tollmien-Schlichting waves
a few wavelengths downstream will 
be little affected by the details in its introduction. \\
The degree $P $ of the
polynomials and the number $ N_x, N_y $ of elements in 
$ \xi $ and $ y $ direction, respectively,
are
important for the numerical accuracy of the simulations. 
A first verification of the Navier-Stokes solver consists in
testing the convergence of the solver when setting $ u' = v' = 0$. 
In this case the results can be compared to
the results computed by the boundary layer equations solver,
cf. section \ref{sec:boundaryLayerSolver}. This verification
is presented in appendix \ref{sec:appendixNS}. 
The length $ L_y $ is as for the Orr-Sommerfeld and the 
parabolic stability equation solver
determined by the width of the Tollmien-Schlichting
wave, equation (\ref{eq:extensionPSE}). 
Increasing the degree $P $ of the polynomials gives better 
agreement between the amplification computed by the
Navier-Stokes solver and by the parabolic stability equation 
solver. For the results in section \ref{sec:results}, we chose
to work with $ P = 11 $. Although the boundary layer flow is slowly 
changing in $ \xi $, the Tollmien-Schlichting wave-length is short, which
makes it necessary to have similar spatial resolutions in
$ \xi $ and $ y $. For the results presented 
in section \ref{sec:results}, we used $ N_x = 300 $ elements in
$ \xi $ direction for the cases with $ \delta = 8 \cdot 10^{-4}$ and $ \delta = 4.75 \cdot 10^{-4} $ 
and $ N_x = 600 $ for the case with $ \delta = 10^{-4}$ since its domain
has twice the size in $ \xi $ as compared to the other cases. 
In the $ y $ direction, $ N_y = 12 $ elements were used in all simulations. 

\clearpage
\section{Results} \label{sec:results}

The results of the present analysis are divided into two parts. First,
we discuss the linear and nonlinear boundary layer solutions in
\cite{LiuParkCowen2007} in connection with the erroneous formula
presented in \cite{LiuOrfila2004}, cf. subsection \ref{sec:boundaryLayerLiu}.
Second,
the general stability properties of the boundary layer flow
under a solitary wave are presented in subsection \ref{sec:stabilityDomains}.

\subsection{Comparisons between linear and nonlinear boundary layer solutions}
\label{sec:boundaryLayerLiu}

\begin{figure}
\centerline{
\includegraphics[width=0.5\linewidth]{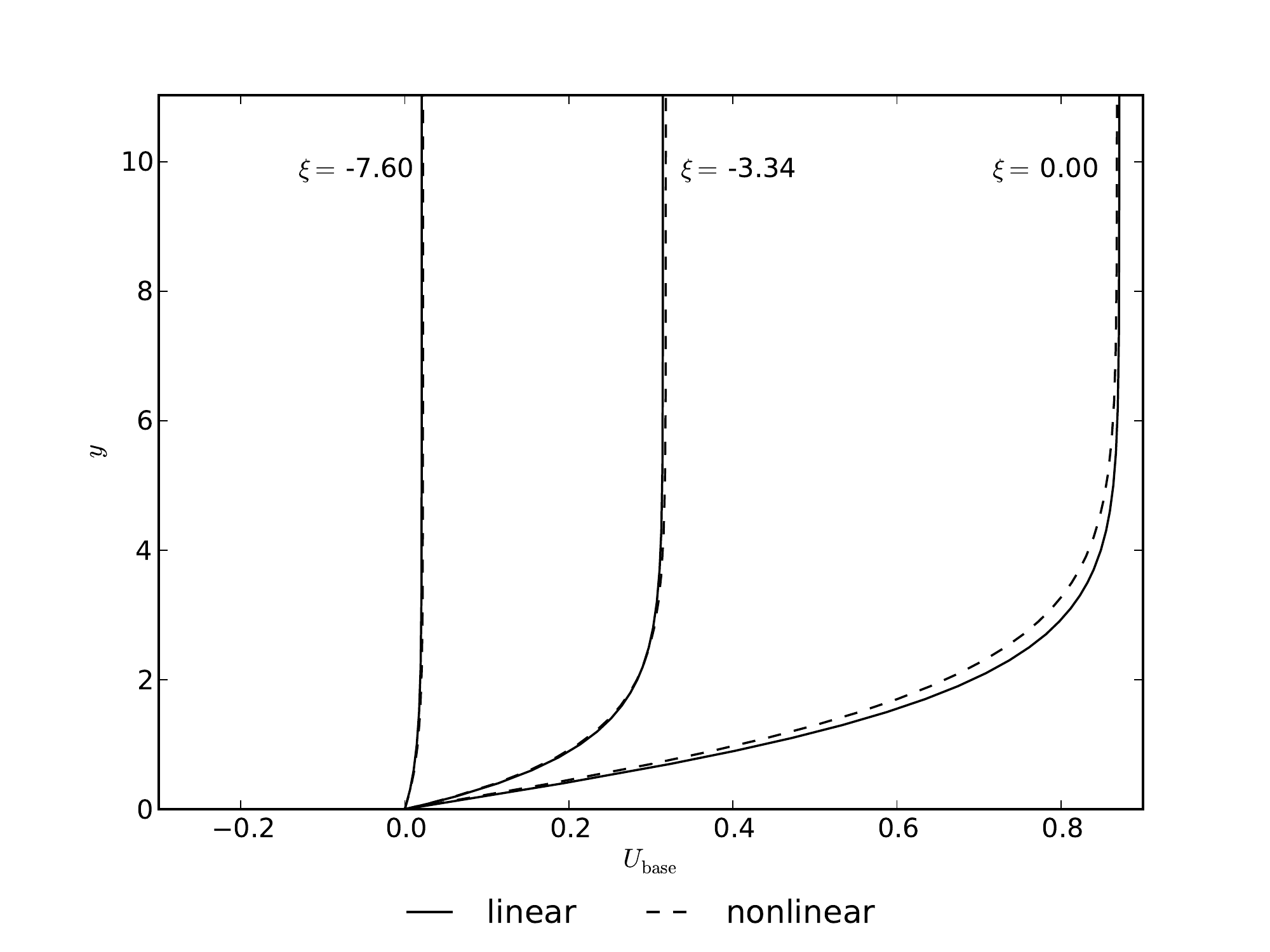}
\includegraphics[width=0.5\linewidth]{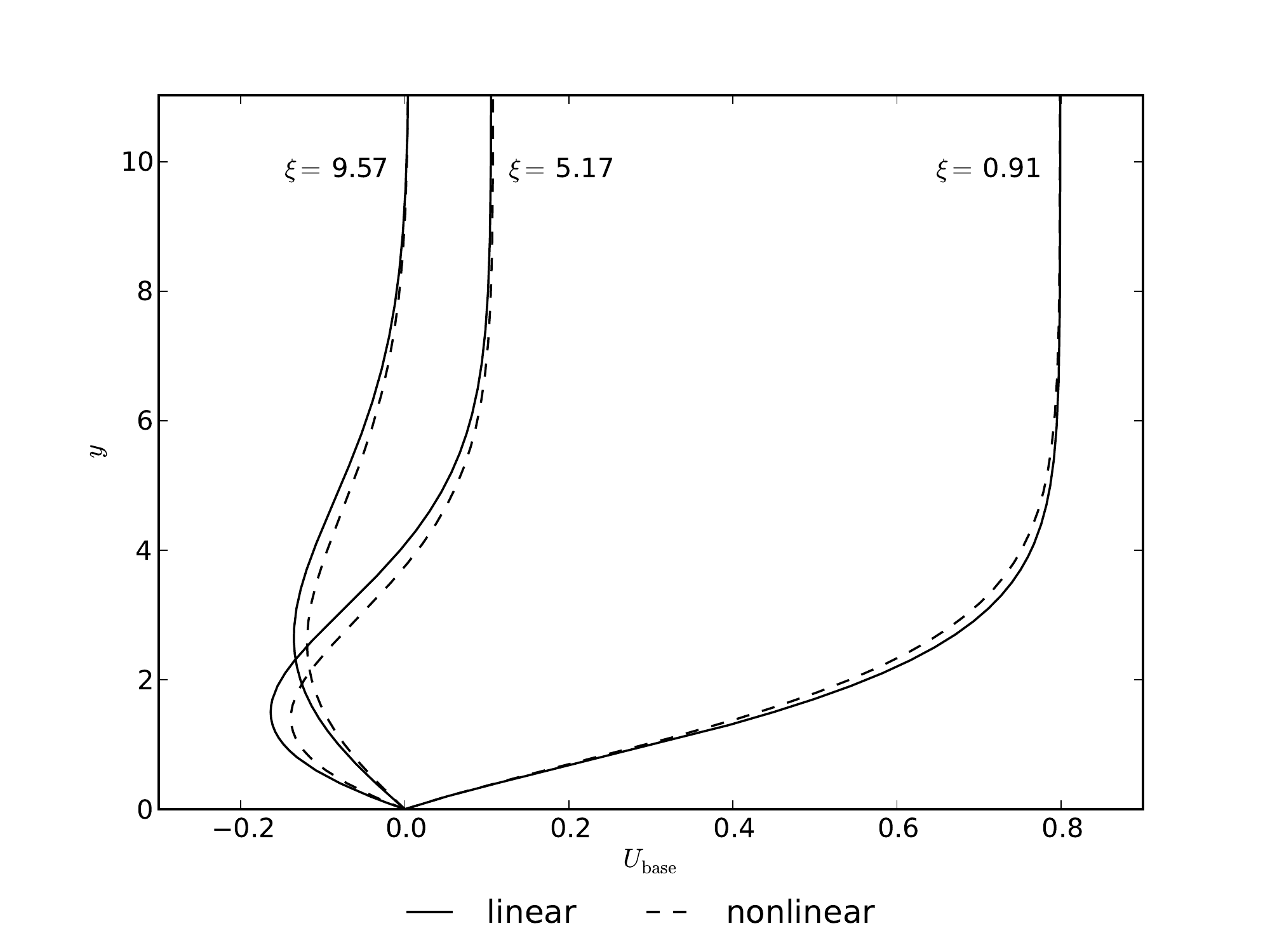}
}
\caption{Profiles of the horizontal velocity for the case $ \epsilon = 0.2 $
taken at different values of $\xi$. 
The linear profile is computed by means of equation (\ref{eq:linearLiu}),
whereas the nonlinear profile is computed using 
equations (\ref{eq:BL1}) and (\ref{eq:BL2}). The scaling is given by
(\ref{eq:scalingLiu1}) and (\ref{eq:scalingLiu2}).}
\label{fig:liuProfileEpsilon0.2}
\end{figure}

\begin{figure}
\centerline{
\includegraphics[width=0.5\linewidth]{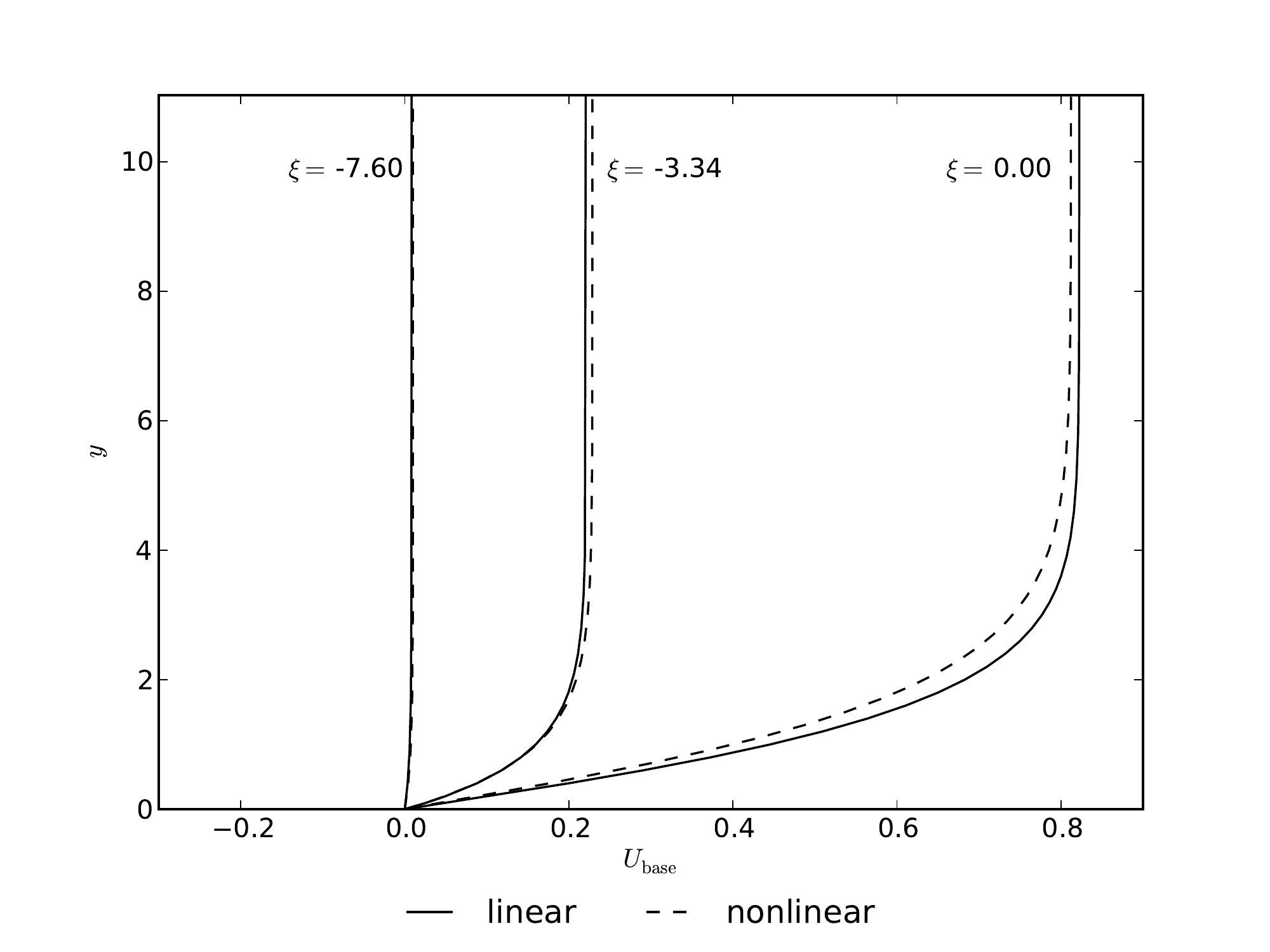}
\includegraphics[width=0.5\linewidth]{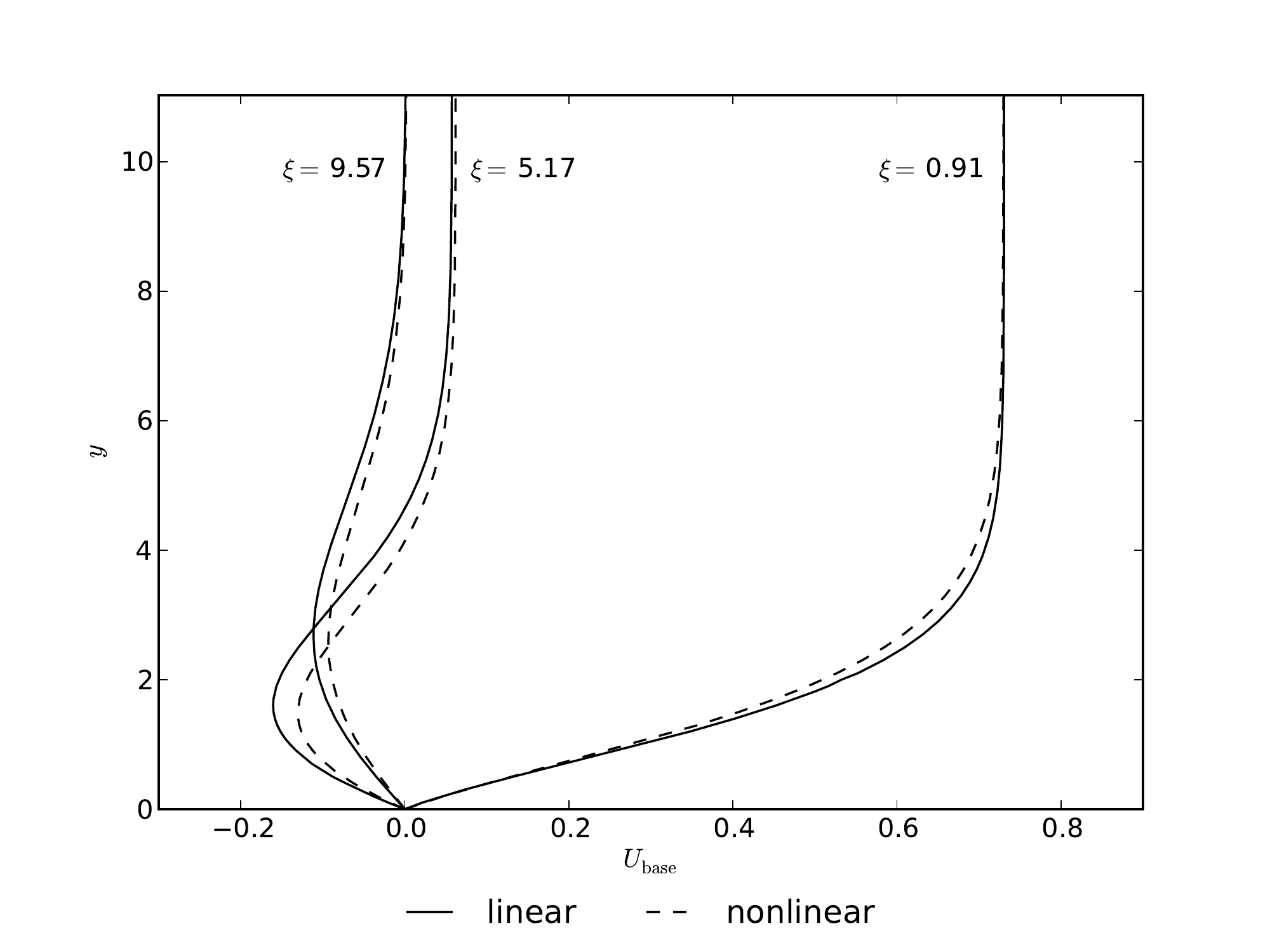}
}
\caption{Profiles of the horizontal velocity for the case $ \epsilon = 0.3 $
taken at different values of $\xi$. 
The linear profile is computed by means of equation (\ref{eq:linearLiu}),
whereas the nonlinear profile is computed using 
equations (\ref{eq:BL1}) and (\ref{eq:BL2}). The scaling is given by
(\ref{eq:scalingLiu1}) and (\ref{eq:scalingLiu2}).}
\label{fig:liuProfileEpsilon0.3}
\end{figure}

\begin{figure}
\centerline{
\includegraphics[width=0.5\linewidth]{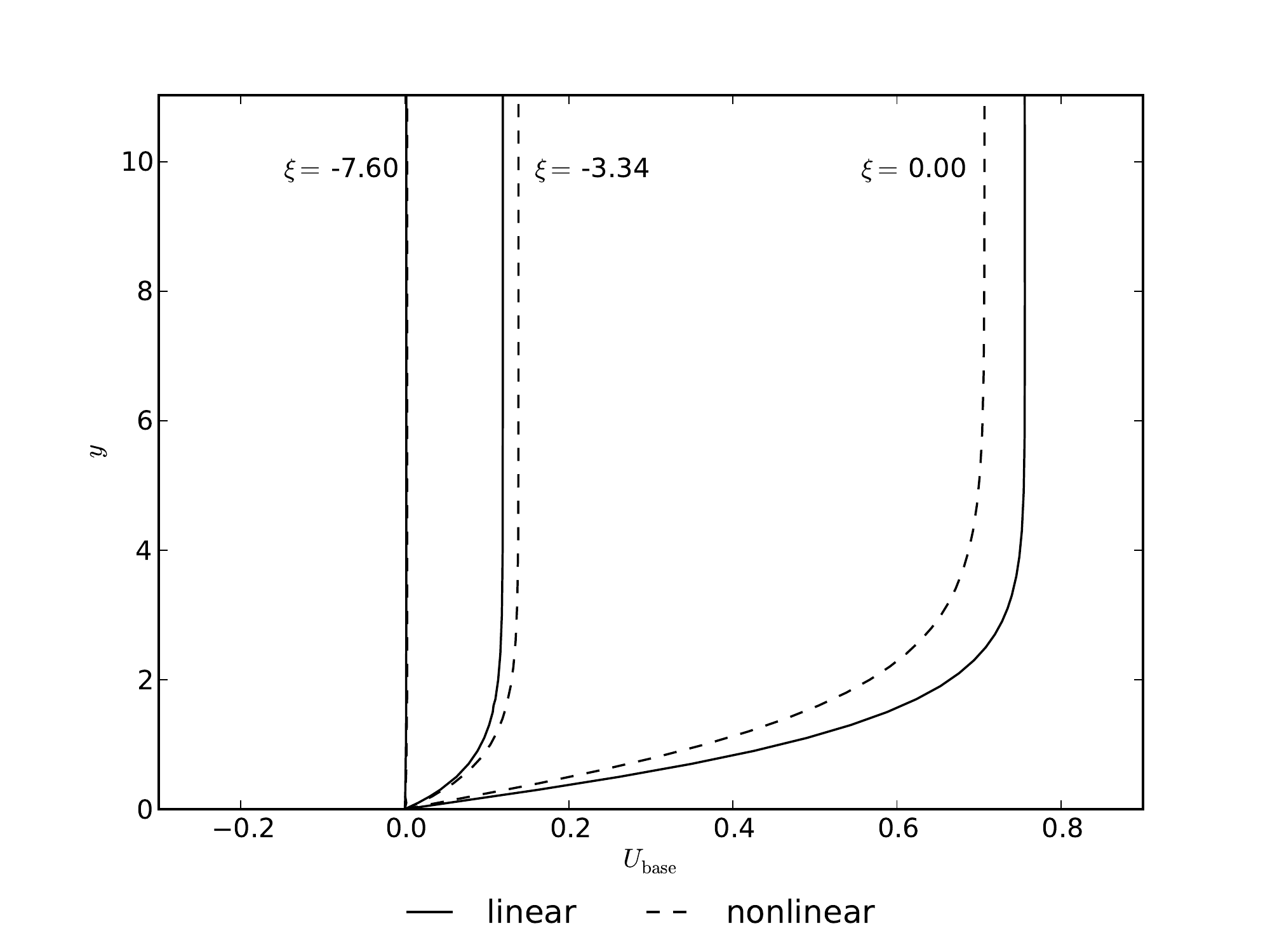}
\includegraphics[width=0.5\linewidth]{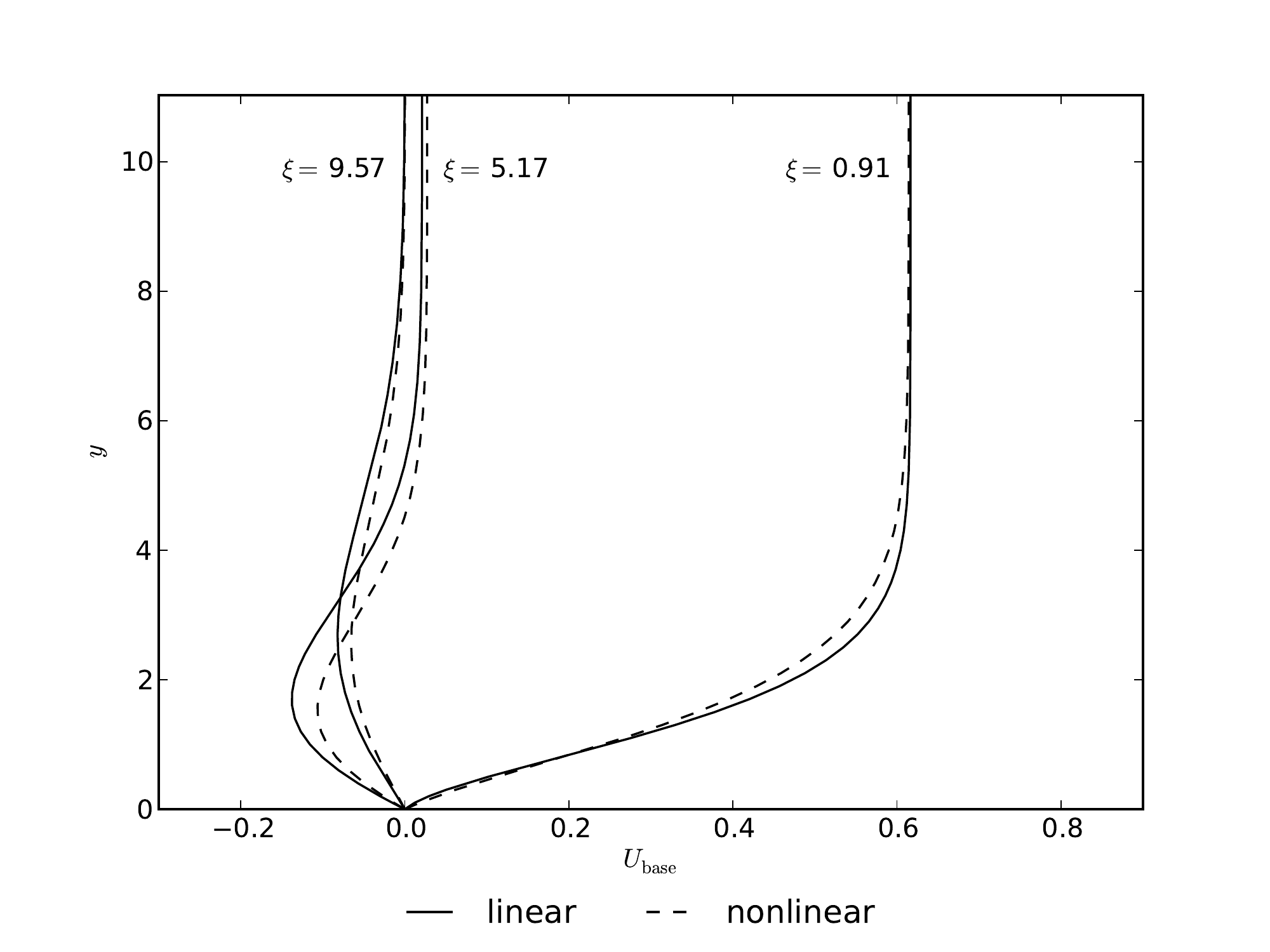}
}
\caption{Profiles of the horizontal velocity for the case $ \epsilon = 0.5 $
taken at different values of $\xi$. 
The linear profile is computed by means of equation (\ref{eq:linearLiu}),
whereas the nonlinear profile is computed using 
equations (\ref{eq:BL1}) and (\ref{eq:BL2}). The scaling is given by
(\ref{eq:scalingLiu1}) and (\ref{eq:scalingLiu2}).}
\label{fig:liuProfileEpsilon0.5}
\end{figure}

In \cite{LiuOrfila2004} the parameter corresponding to $ \delta $ is
denoted by $ \alpha $ and includes a parameter controlling
the dispersion of the solitary wave. Dispersion is less of an issue for
the analysis of the boundary layer stability, such that
we follow the parameter choice by \cite{VittoriBlondeaux2008}. 
Changes in the boundary layer flow
along the horizontal direction appear on a scale of $ h_0 $,
whereas changes in vertical direction happen on a scale of $ \delta^* $.
For this parameter choice the corresponding
inhomogeneous scaling in \cite{LiuOrfila2004} would be given by
the following expressions.
\be
x = \frac{x^*}{h_0}, \quad y = \frac{y^*}{\delta h_0}. \label{eq:scalingLiu1}
\ee
The velocity components are scaled accordingly:
\be
u = \frac{u^*}{\epsilon \sqrt{gh_0}}, \quad v = \frac{v^*}{\epsilon \delta \sqrt{gh_0}}. \label{eq:scalingLiu2}
\ee
The expansion 
given in equations (2.7) and (2.8) in \cite{LiuOrfila2004},
may then be written:
\begin{eqnarray}
u &=& \frac{\partial \tilde{\Phi}}{\partial x} + u_0^r + \delta u_1^r + \ldots , \label{eq:liuExpansion1}\\
v &=& \frac{\partial \tilde{\Phi}}{\partial y} + v^r_0 + \delta v_1^r + \ldots,
\label{eq:liuExpansion2}
\end{eqnarray}
where the superscript $ ^r $ means that the velocity field is
not irrotational. We remark that $ \tilde{\Phi} $ is a different
mathematical object than the flow potential in (\ref{eq:inviscidFlow}). 
The potential $ \tilde{\Phi} $ represents the whole velocity field
in the bulk of the fluid including the entrainment velocity 
induced by the viscous boundary layer, as explained in \cite{LiuOrfila2004}.
In addition the scaling of $ \tilde{\Phi} $ is different than
the one in  (\ref{eq:inviscidFlow}). 
Introducing this expansion into 
the Navier-Stokes equations (\ref{eq:NS1}-\ref{eq:NS2}) and neglecting
terms of order $ \delta $, we are given a set of equations for $ u_0^r $ and
$ v_0^r $:
\begin{eqnarray}
\frac{\partial}{\partial x} u^r_0 + \frac{\partial}{\partial y} v^r_0 & = & 0, 
\label{eq:liuOrfila1} \\
\frac{\partial}{\partial t} u^r_0 
+ \epsilon \Big( \frac{\partial}{\partial x} \tilde{\Phi} \frac{\partial}{\partial x} u_0^r
+ u_0^r \frac{\partial^2}{\partial x^2} \tilde{\Phi} + u_0^r \frac{\partial}{\partial x} u_0^r & & \nonumber \\
+\quad  \frac{\partial}{\partial y} \tilde{\Phi} \frac{\partial}{\partial y} u_0^r 
+ v_0^r \frac{\partial^2}{\partial y \partial x} \tilde{\Phi} + v_0^r 
\frac{\partial}{\partial y} u_0^r \Big) & = &
 \frac{1}{2} \frac{\partial^2}{\partial y^2} u_0^r.
\label{eq:liuOrfila2}
\end{eqnarray}
From equation (\ref{eq:liuOrfila2}), we observe that in 
equation (2.11) in \cite{LiuOrfila2004} the terms containing $ \tilde{\Phi} $
have been omitted. This is not so much of an issue for
the work in \cite{LiuOrfila2004}, since \cite{LiuOrfila2004} 
continued their analysis with the linearized equations in $ \epsilon $.
However, for the investigation 
of the boundary layer flow under a solitary wave in 
\cite{LiuParkCowen2007} it is of importance. The entire 
part on the nonlinear boundary layer solution
in \cite{LiuParkCowen2007} is based on equation (2.11) in \cite{LiuOrfila2004}
and needs therefore to be modified. The
conclusions for the nonlinear boundary layer
solution drawn on the basis of this analysis are not correct.
In particular their statements that
for large values of $ \epsilon $ 
the differences in the linear and nonlinear boundary layer solutions are 
'not very significant' and the profiles are 'surprisingly close' 
need to be reconsidered. The present subsection shall elucidate
this issue. \\
Opposed to the nonlinear boundary layer solution, the linear
boundary layer solution 
for the horizontal velocity component in \cite{LiuParkCowen2007} is
correct and given by:
\be
u_0^r(x,y,t) = - \frac{y}{\sqrt{2\pi}} 
\int \limits_0^t \frac{\partial \Phi}{\partial x}( x, 0, \tau/2)
\frac{ e^{-\frac{y^2}{2 ( t - \tau )}}}{ \sqrt{ \left( t - \tau \right)^3 }}
\, d\tau,  \label{eq:linearLiu}
\ee
where $ \Phi $ is the inviscid base flow as in
equation (\ref{eq:inviscidFlow}). 
However, the boundary conditions for the linear wall normal 
velocity component in \cite{LiuParkCowen2007} are wrong. 
As for the Blasius boundary layer, the boundary layer will
displace fluid and therefore the condition
\be
v_0^r \rightarrow 0 \quad \mbox{for} \quad y \rightarrow \infty 
\label{eq:boundaryConditionLiu}
\ee
in \cite{LiuParkCowen2007} is not correct and $ v_0^r $ cannot
satisfy the impermeability condition $ v_0^r = 0 $ at the 
wall and condition (\ref{eq:boundaryConditionLiu}) at the same time. 
The condition (\ref{eq:boundaryConditionLiu}) is, however, 
correct in \cite{LiuOrfila2004}, since there, $ \partial \tilde{\Phi}/\partial y $
accounts for the displacement of fluid by the boundary layer. 
The solution for $ v_0^r $ given by
equation (2.20) in \cite{LiuParkCowen2007} needs therefore to be
replaced by 
\be
v_0^r (x, y, t) = -\frac{1}{\sqrt{2\pi}}
\int \limits_0^t \frac{\partial^2 \Phi}{\partial x^2} (x,0,\tau/2)
\frac{1}{ \sqrt{ \left( t - \tau \right) }} 
\left(  e^{-\frac{y^2}{2 ( t - \tau )}} - 1 \right)
\, d\tau. \label{eq:normalComponentLiu}
\ee
\cite{LiuParkCowen2007} reported excellent agreement between 
their experimental results and their theoretical analysis. In
particular they presented profiles for the case $ \epsilon = 0.2 $
and $ \delta = 4.4\cdot 10^{-3}$ at different locations $ \xi $. 
The values of $ \xi $ in their scaling (i.e. $ \xi = 0.5 $, $ 0.22 $,
$ -0.06 $, $ -0.34 $, $ -0.63 $) correspond in the present scaling
to the values $ \xi = -7.6 $, $ -3.34 $, $ 0.91 $, $ 5.17 $ and $ 9.57 $. 
In figures \ref{fig:liuProfileEpsilon0.2},
\ref{fig:liuProfileEpsilon0.3}
and \ref{fig:liuProfileEpsilon0.5}, we plotted the profiles
at these locations and under the crest $ \xi = 0 $ using
the linear solution, equation (\ref{eq:linearLiu}),
and the nonlinear solution by the present boundary layer solver,
cf. subsection \ref{sec:boundaryLayerSolver}, for
the cases $ \epsilon = 0.2 $, $ 0.3 $ and $ 0.5 $ respectively. 
As can be observed from these figures,  qualitatively the 
linear and nonlinear profiles
are displaying the same behavior for all
values of $ \epsilon $. However, quantitative differences arise
already for a value of $ \epsilon = 0.2 $. The difference between
the linear and nonlinear profiles is 
largest for the position $ \xi = 5.17 $. This is not surprising,
since the flow in the boundary layer reverses its direction here. 
Convection has an influence in the sense that it softens the
linear effect such that the profiles become less extreme. 
This effect can also be seen in the experimental results
presented in figure 9 in \cite{LiuParkCowen2007}, where the
experimental profile for $ \xi = 5.17 $ ($\xi=-0.34$ in their scaling)
displays a softer bend than the nonlinear profile by \cite{LiuParkCowen2007}. 
Increasing the amplitude leads to increasing differences between
the linear and the nonlinear solutions. As a measure of the
difference, the maximum difference between each profile can 
be computed and compared to $ \epsilon $. This will result
into an error of $ 4 \%  $ for the case $ \epsilon = 0.2 $,
$ 5\% $ for $ \epsilon = 0.3 $ and 
$ 10 \% $ for $ \epsilon = 0.5 $. This is, however, a
rather coarse estimation of the error, but gives a
first indication on which differences to expect when comparing results
based on linear or nonlinear profiles. Although,
\cite{LiuParkCowen2007} mentioned that they 
performed experiments for $ \epsilon = 0.3 $, they
did not present results for this case. 

\subsection{Linear stability of the boundary layer flow
under a solitary wave} \label{sec:stabilityDomains}

\begin{figure}
  \centerline{\includegraphics[width=\linewidth]{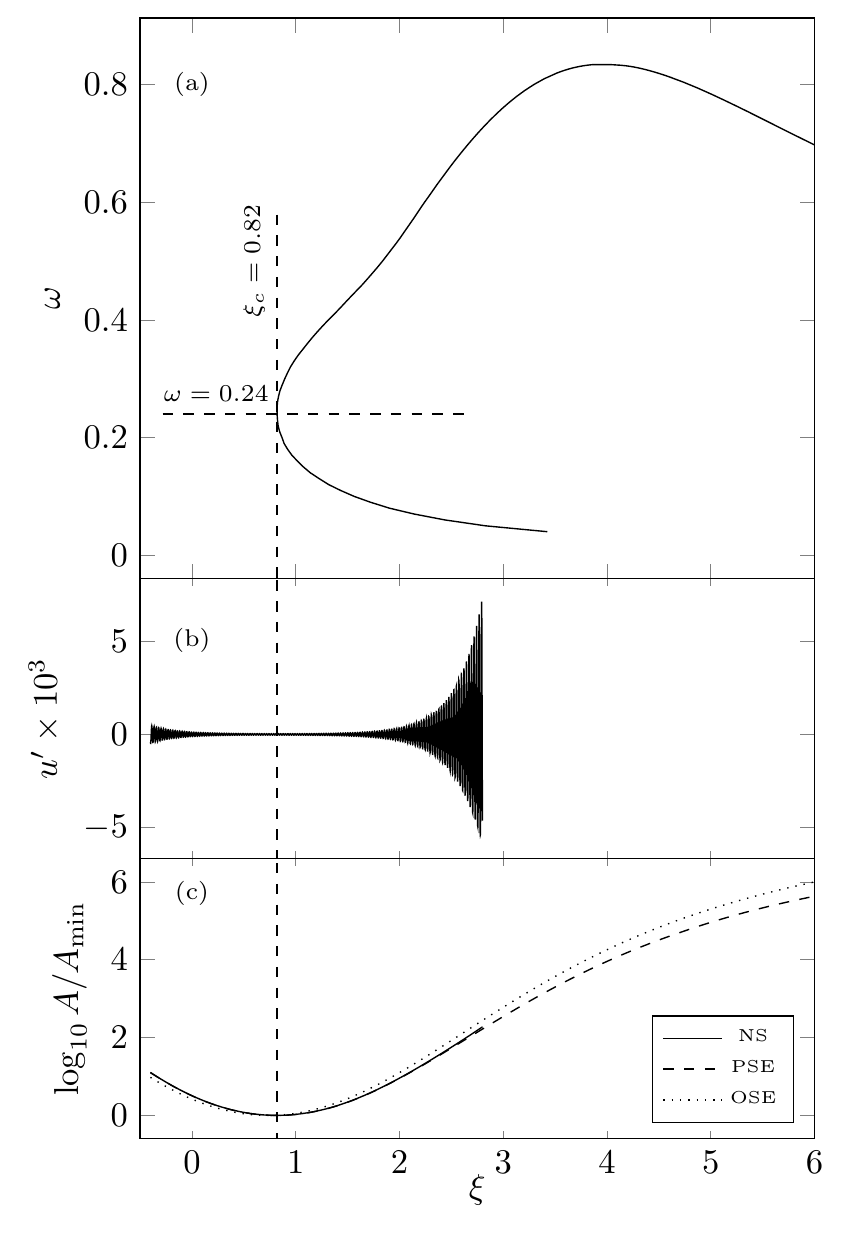}}
  \caption{(a) Stability domain for the case $ \epsilon = 0.4 $ and
$ \delta = 8 \cdot 10^{-4} $.
(b) Horizontal perturbation $ u' $ of the horizontal velocity
component recorded at a distance $ y = 0.4943 $ from the wall
for a Tollmien-Schlichting wave of $ \omega = 0.24 $ computed
by means of the Navier-Stokes solver. (c) Amplification of the
Tollmien-Schlichting wave with $ \omega = 0.24 $
computed by means of the Orr-Sommerfeld
equation solver (OSE), the parabolic stability equation solver (PSE)
and the Navier-Stokes solver (NS). Here and in the subsequent figures: $ \xi $ is scaled by (\ref{eq:scaling2}), while $ \omega $ by means of (\ref{eq:scaling2time}). }
\label{fig:plot1NavierStokes}
\end{figure}

\begin{figure}
  \centerline{\includegraphics[width=\linewidth]{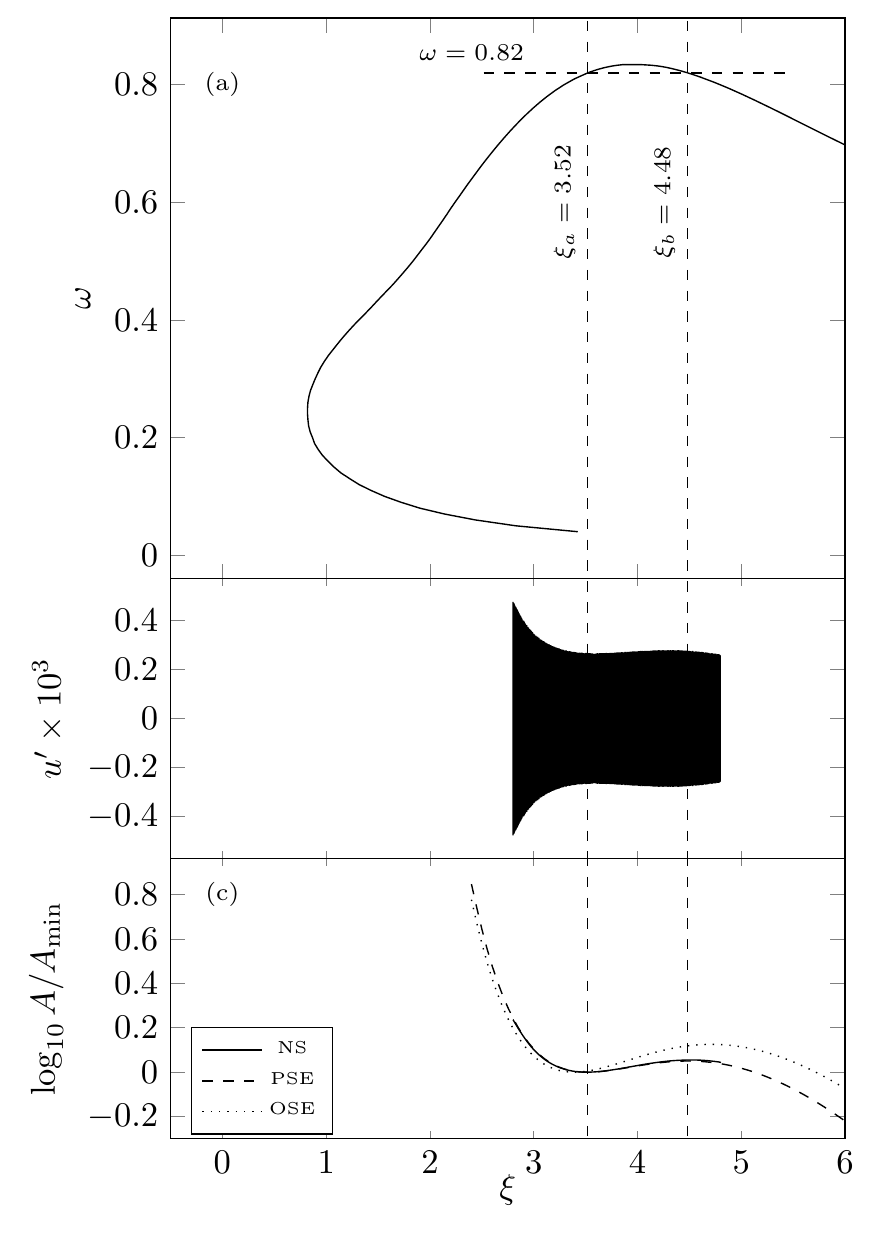}}
  \caption{(a) Stability domain for the case $ \epsilon = 0.4 $ and
$ \delta = 8 \cdot 10^{-4} $.
(b) Horizontal perturbation $ u' $ of the horizontal velocity
component recorded at a distance $ y = 0.4943 $ from the wall
for a Tollmien-Schlichting wave of $ \omega = 0.82 $ computed
by means of the Navier-Stokes solver. (c) Amplification of the
Tollmien-Schlichting wave with $ \omega = 0.82 $
computed by means of the Orr-Sommerfeld
equation solver (OSE), the parabolic stability equation solver (PSE)
and the Navier-Stokes solver (NS).}
\label{fig:plot2NavierStokes}
\end{figure}

\begin{figure}
  \centerline{\includegraphics[width=\linewidth]{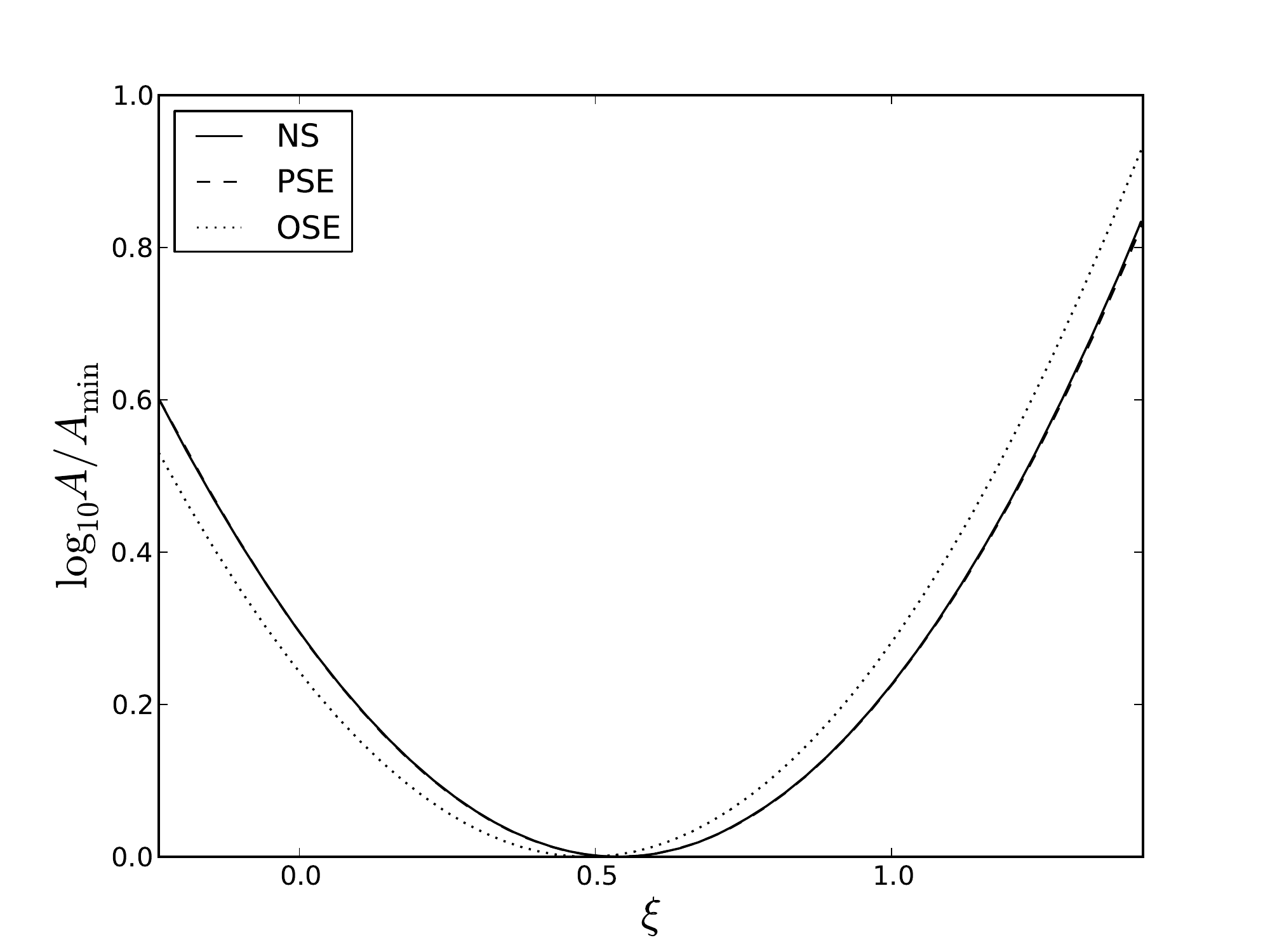}}
  \caption{Amplification of the Tollmien-Schlichting wave
for the case $ \epsilon = 0.4 $, $\delta = 4.75\cdot 10^{-4} $ and
$ \omega = 0.22 $. The amplification has been computed
by means of the Orr-Sommerfeld equation solver (OSE), the
parabolic stability equation solver (PSE) and the Navier-Stokes solver (NS).}
\label{fig:amplificationDelta0.000475}
\end{figure}

\begin{figure}
  \centerline{\includegraphics[width=\linewidth]{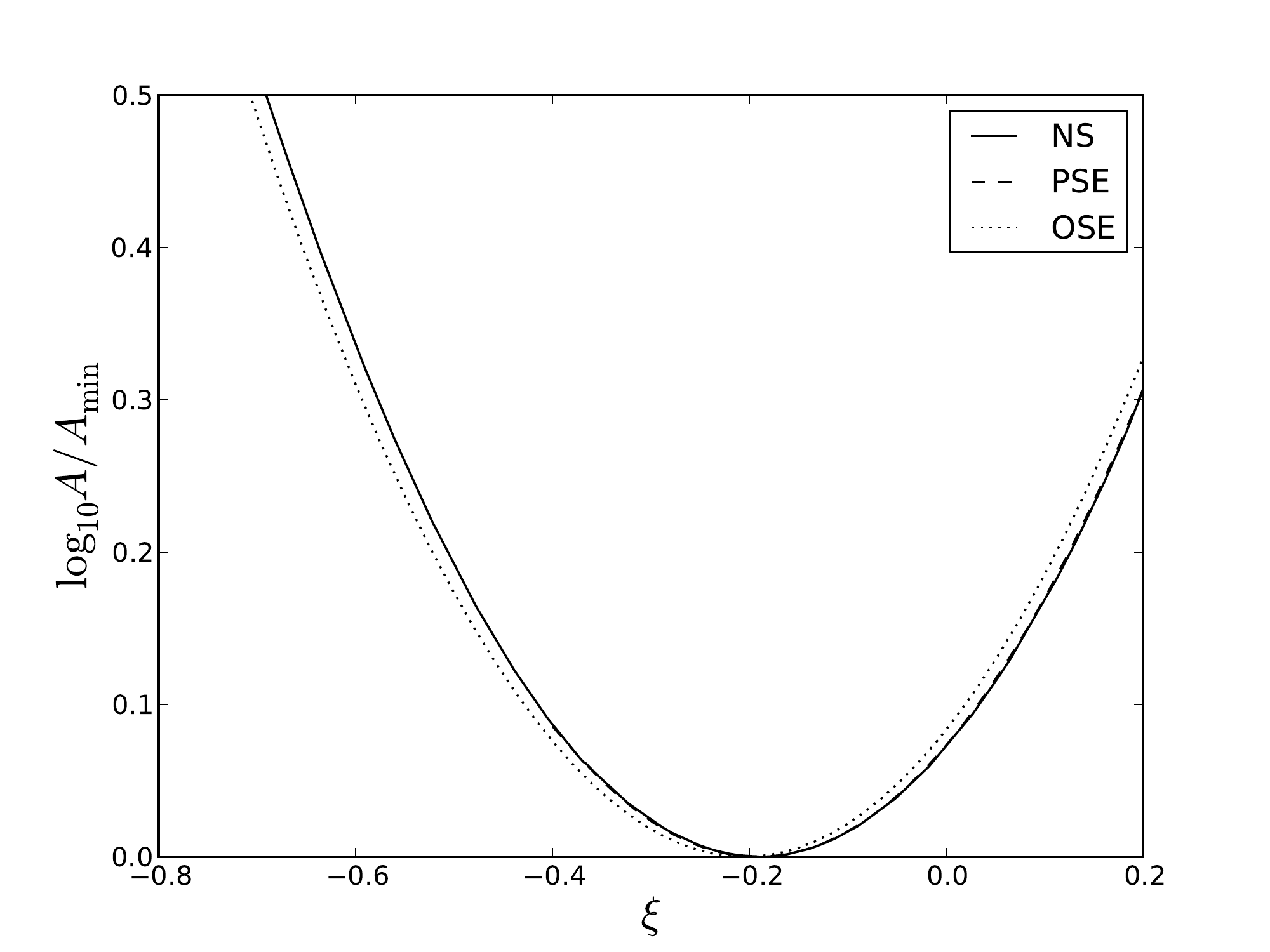}}
  \caption{Amplification of the Tollmien-Schlichting wave
for the case $ \epsilon = 0.4 $, $\delta = 10^{-4} $ and
$ \omega = 0.185 $. The amplification has been computed
by means of the Orr-Sommerfeld equation solver (OSE), the
parabolic stability equation solver (PSE) and the Navier-Stokes solver (NS).}
\label{fig:amplificationDelta0.0001}
\end{figure}

\begin{figure}
  \centerline{\includegraphics[width=\linewidth]{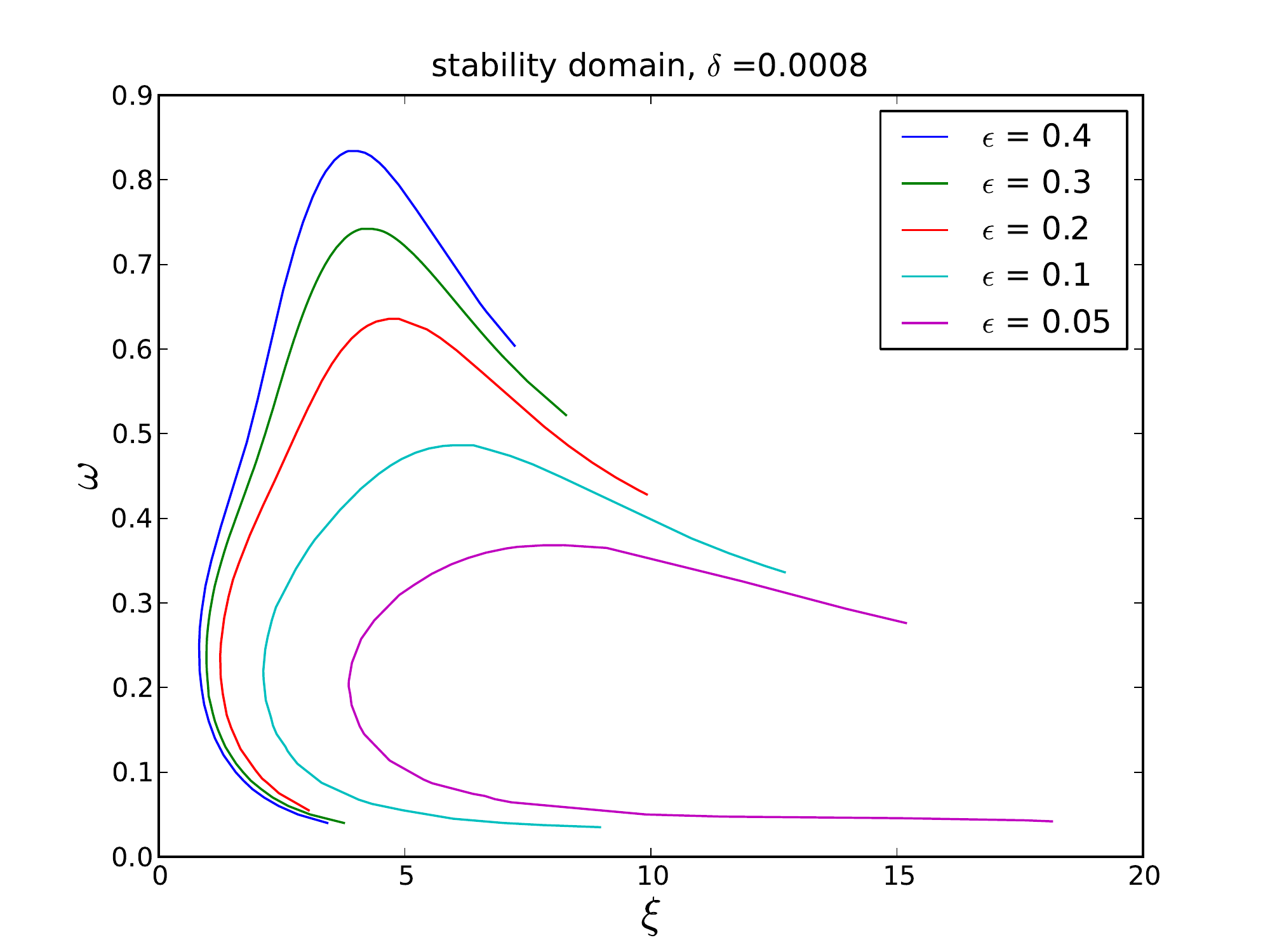}}
  \caption{Stability domain for $ \delta = 8\times 10^{-4} $. The region bounded by the curves is the unstable region.}
\label{fig:domainDelta0.0008}
\end{figure}

\begin{figure}
  \centerline{\includegraphics[width=\linewidth]{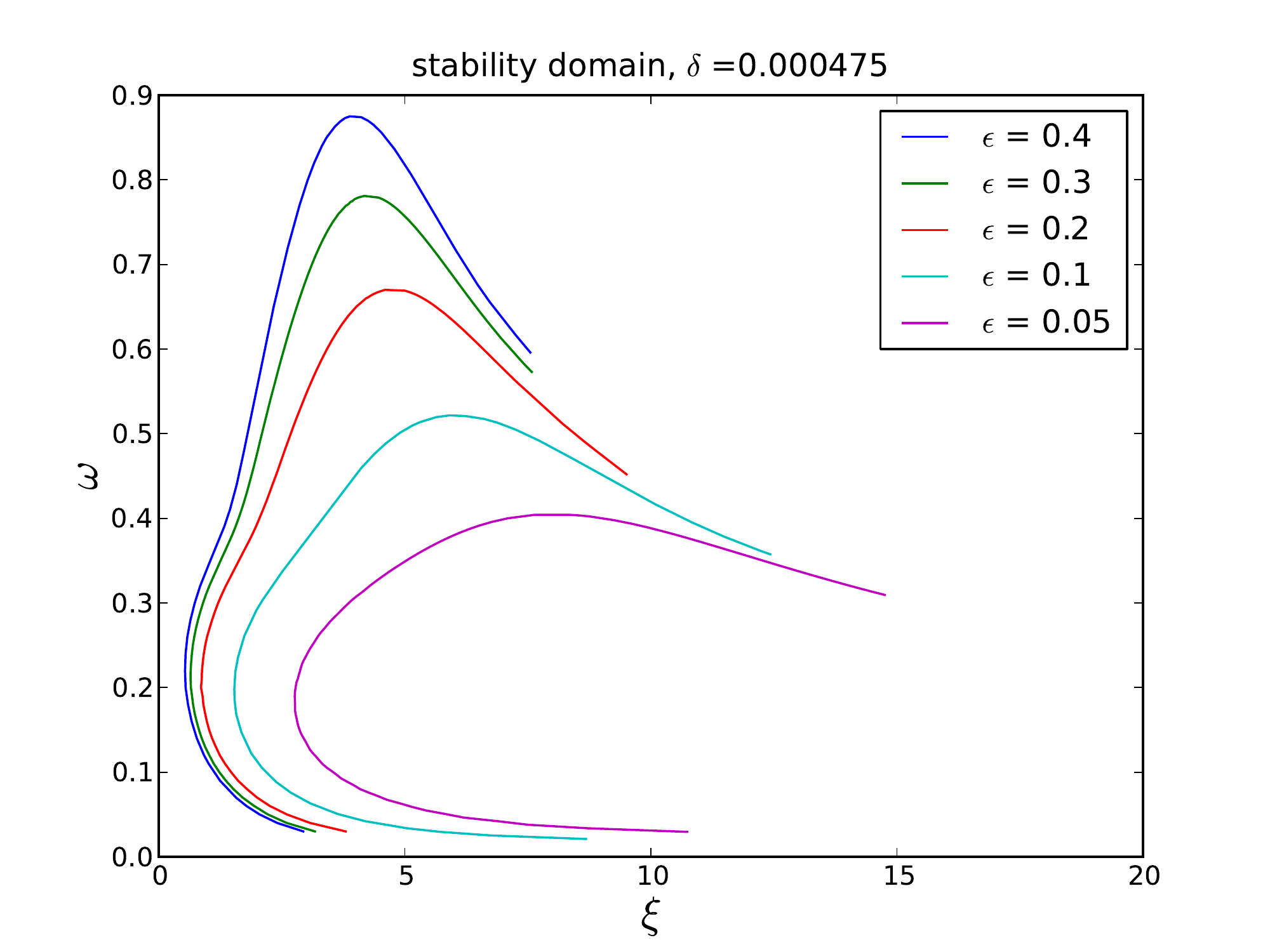}}
  \caption{Stability domain for $ \delta = 4.75\times 10^{-4} $. The region bounded by the curves is the unstable region.}
\label{fig:domainDelta0.000475}
\end{figure}

\begin{figure}
  \centerline{\includegraphics[width=\linewidth]{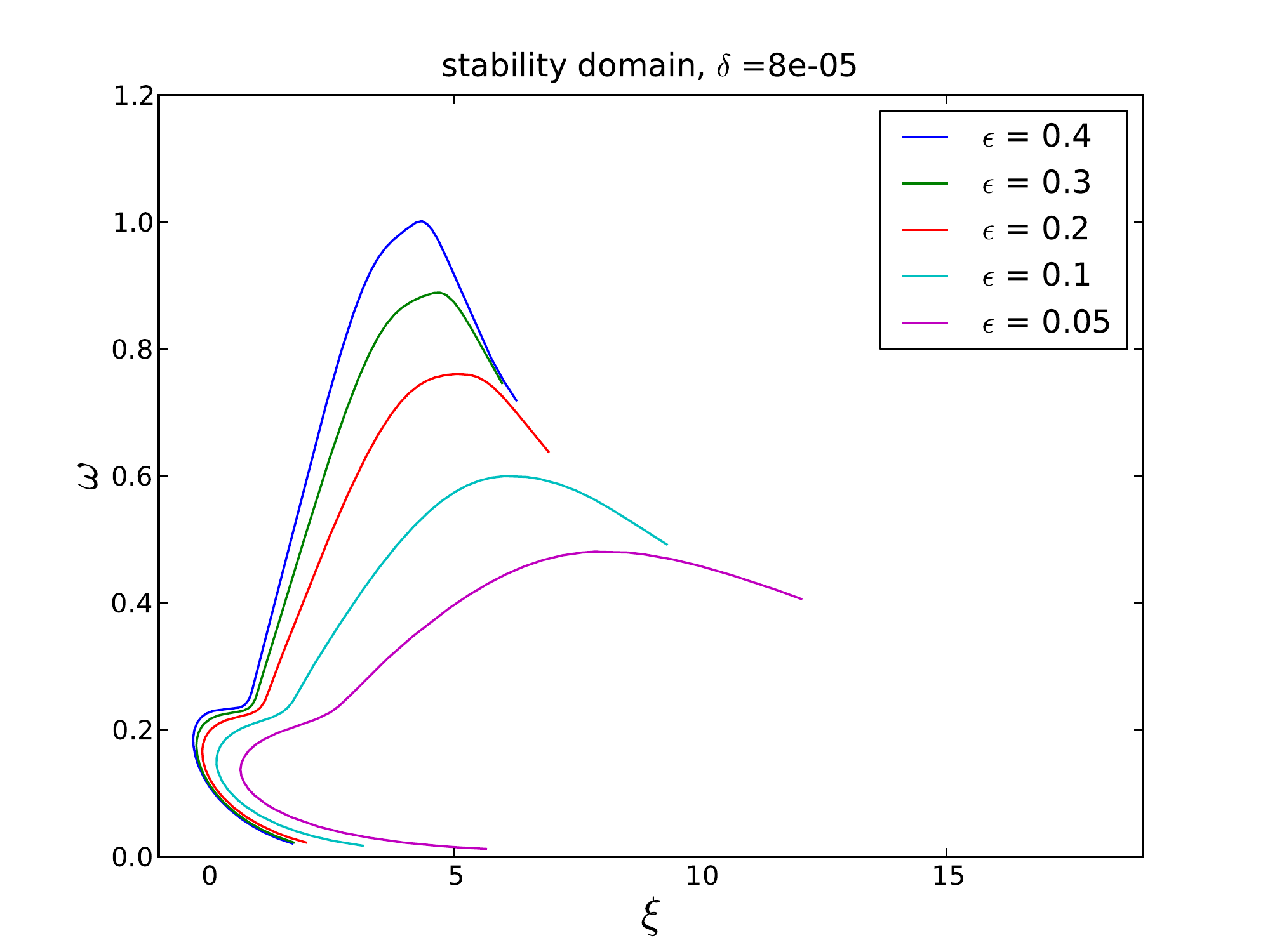}}
  \caption{Stability domain for $ \delta = 8\times 10^{-5} $. The region bounded by the curves is the unstable region.}
\label{fig:domainDelta8e_5}
\end{figure}

\begin{figure}
  \centerline{\includegraphics[width=\linewidth]{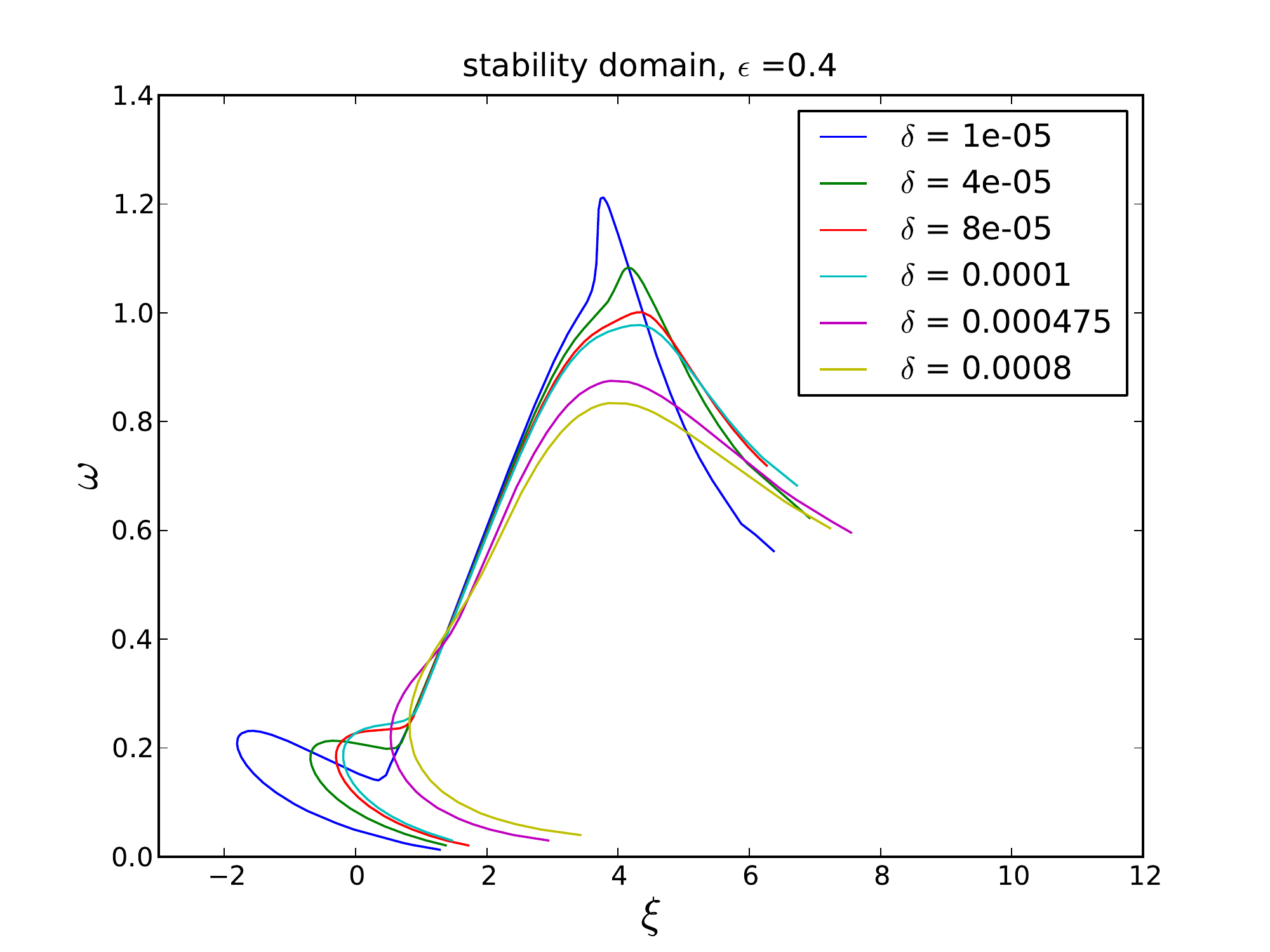}}
  \caption{Stability domain for $ \epsilon = 0.4 $. The region bounded by the curves is the unstable region.}
\label{fig:domainEpsilon0.4}
\end{figure}

\begin{figure}
  \centerline{\includegraphics[width=\linewidth]{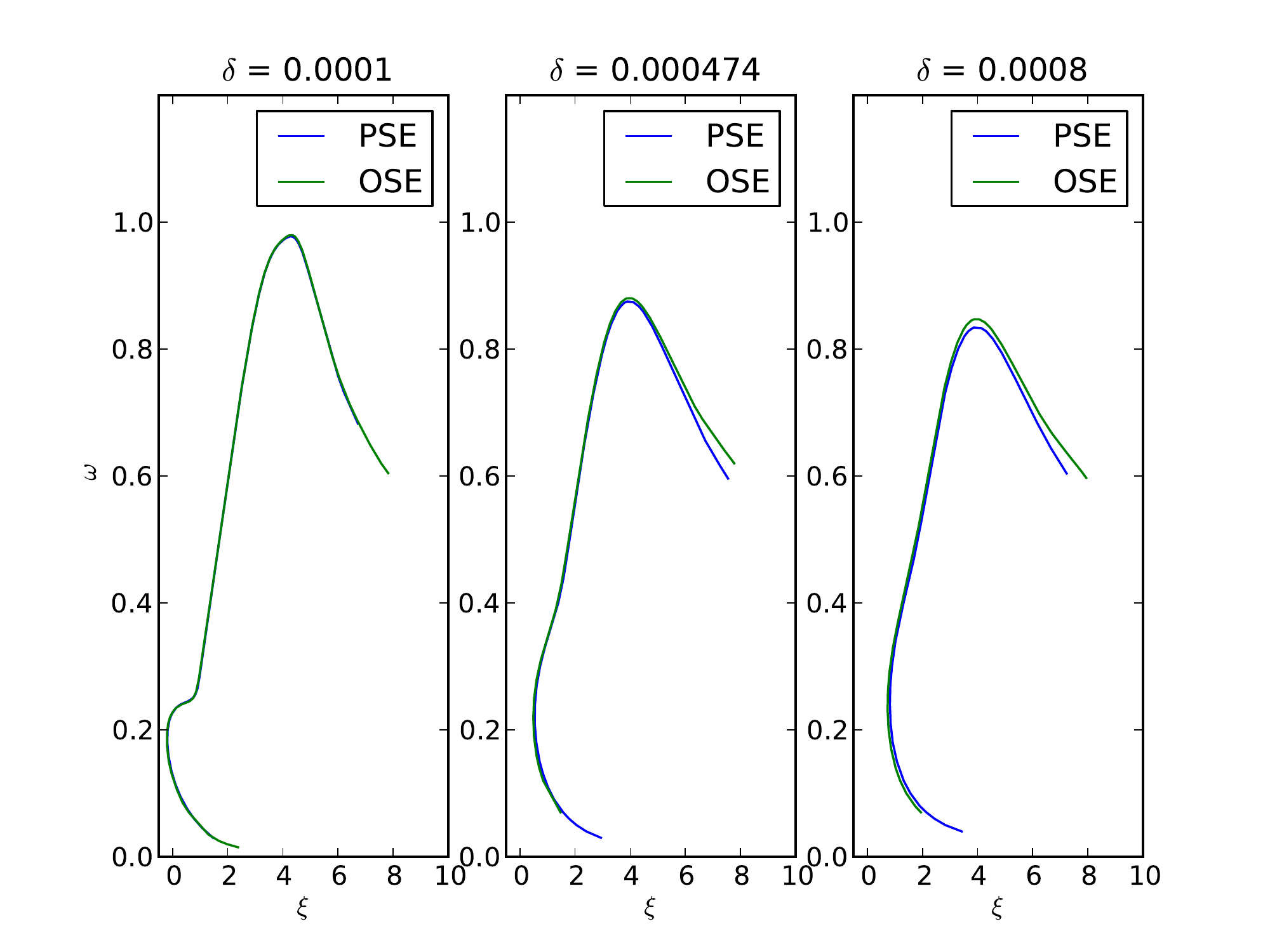}}
  \caption{Comparison of the unstable region for the cases
$ \epsilon = 0.4 $ and $ \delta = 8\cdot10^ {-4}, 4.75\cdot10^ {-4},10^ {-4} $
computed by means of the Orr-Sommerfeld equation (OSE) 
and the parabolic stability equation (PSE).}
\label{fig:OSEvsPSE}
\end{figure}

\begin{figure}
  \centerline{\includegraphics[width=\linewidth]{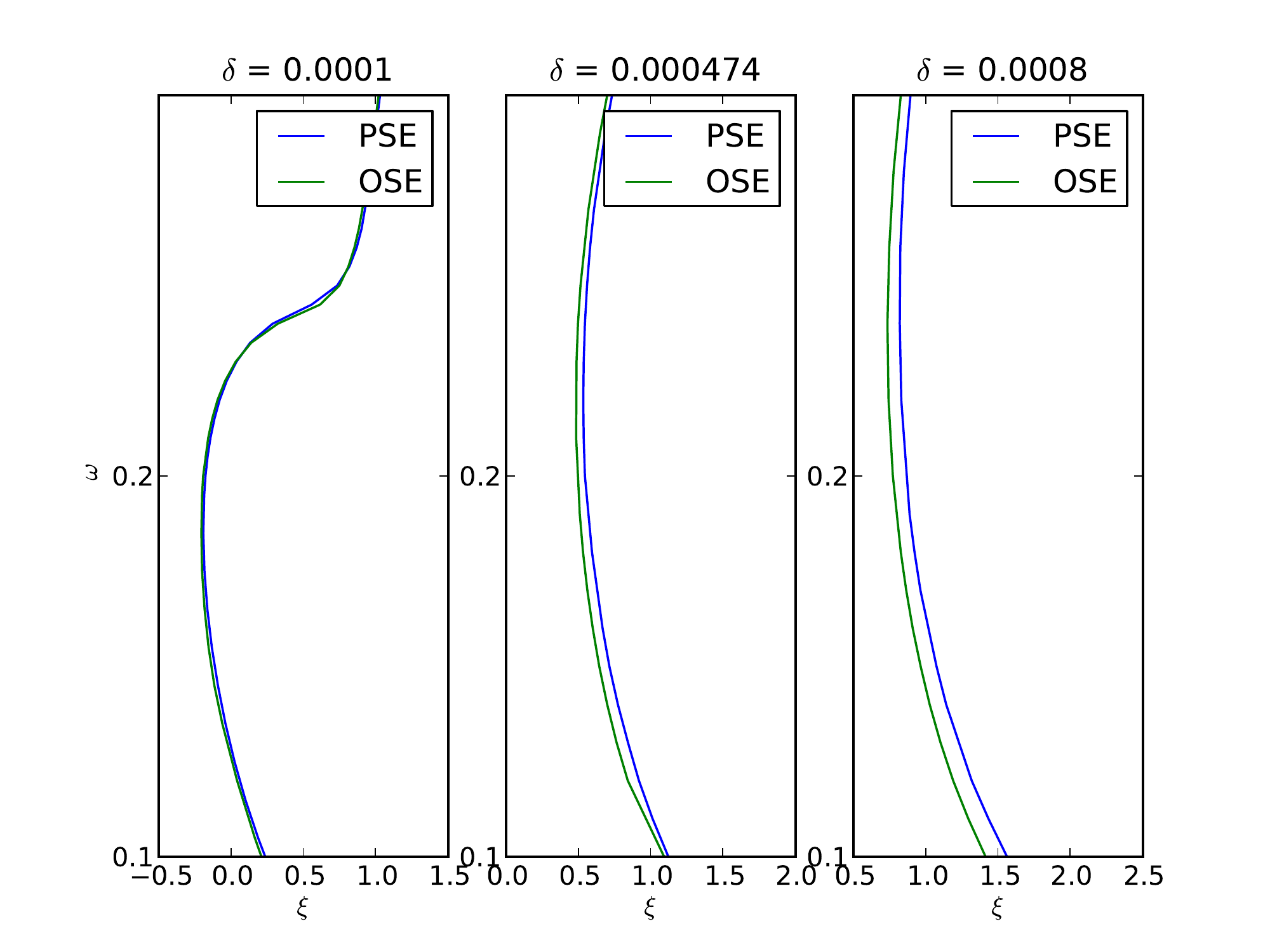}}
  \caption{Zoom onto parts of the unstable regions displayed in 
figure \ref{fig:OSEvsPSE}.}
\label{fig:OSEvsPSEzoom}
\end{figure}

\begin{figure}
  \centerline{\includegraphics[width=\linewidth]{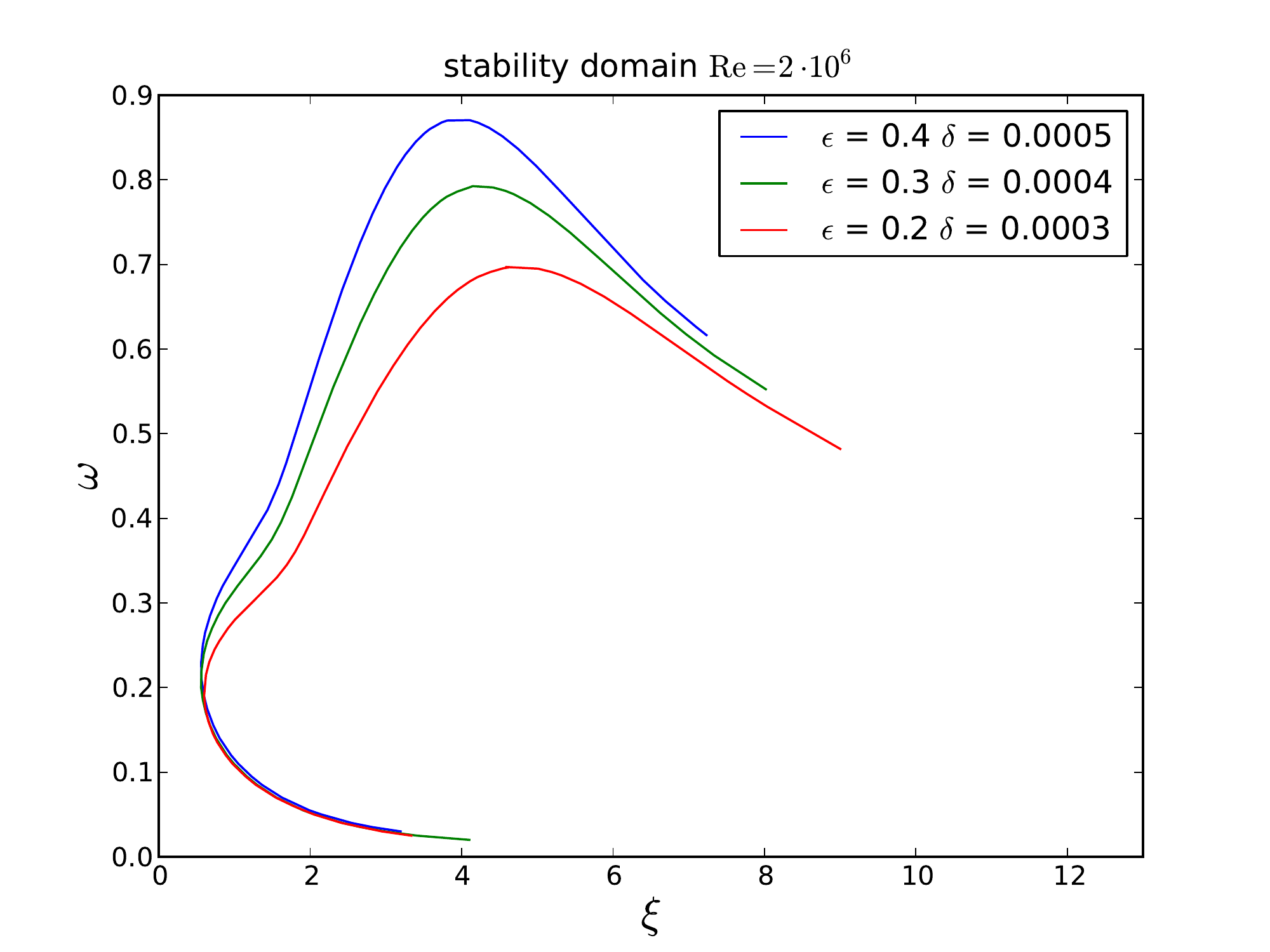}}
  \caption{Stability domain for $ \Rey_\summer = 2 \cdot 10^6 $. The
amplitude $ \epsilon $ and the parameter $ \delta $ are chosen
in such a way that $ \Rey_\summer $ is approximately equal to $ 2 \cdot 10^6 $.}
\label{fig:domainReSummer}
\end{figure}

\begin{figure}
  \centerline{\includegraphics[width=\linewidth]{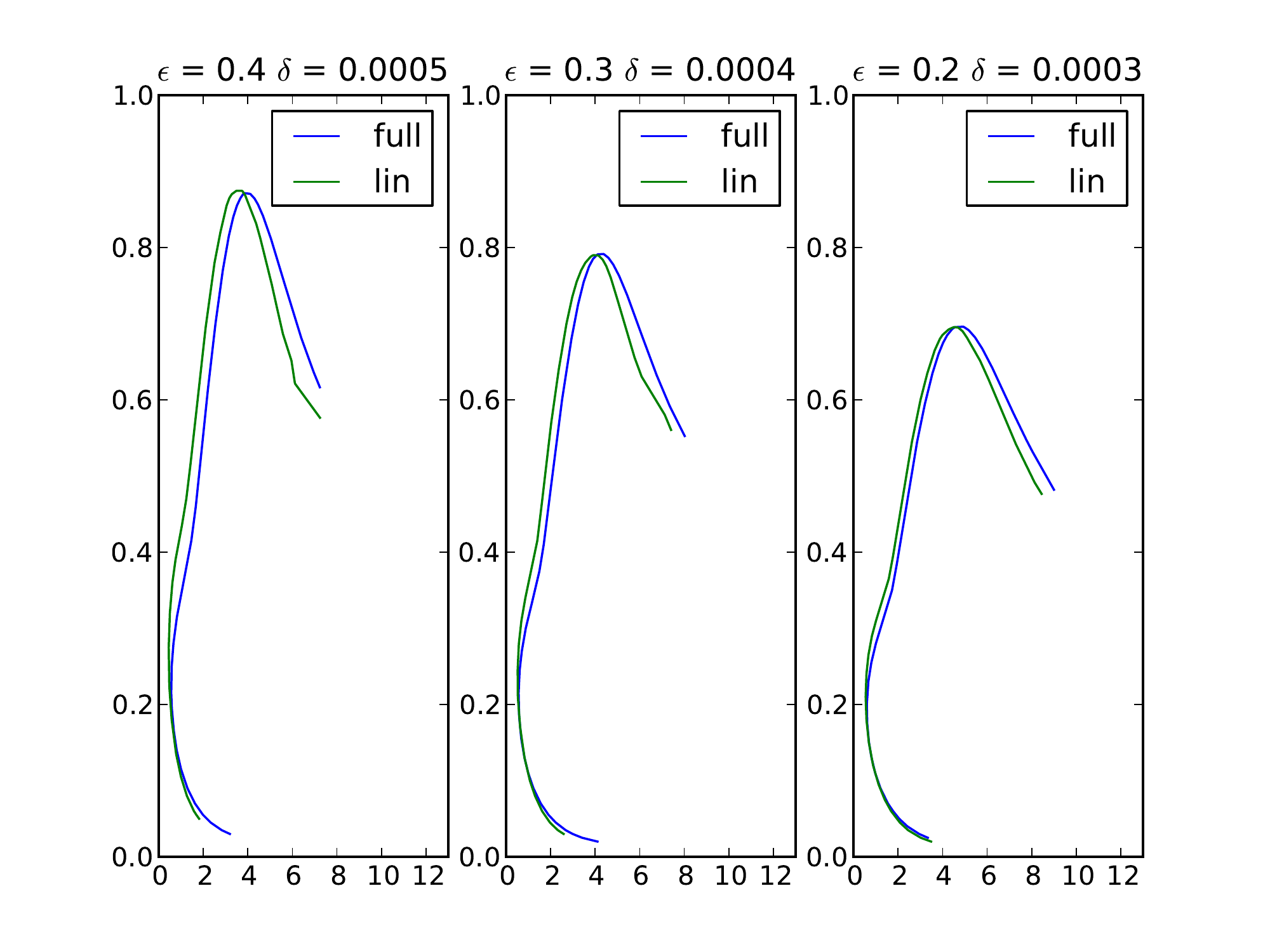}}
  \caption{Stability domains for the linearized boundary layer profile 
by \cite{LiuParkCowen2007}
and the nonlinear profile based on the full potential solution,
 for $ \Rey_\summer = 2 \cdot 10^6 $. The
amplitude $ \epsilon $ and the parameter $ \delta $ are chosen
in such a way that $ \Rey_\summer $ is approximately equal to $ 2 \cdot 10^6 $.}
\label{fig:domainLinearReSummer}
\end{figure}

\begin{figure}
  \centerline{\includegraphics[width=\linewidth]{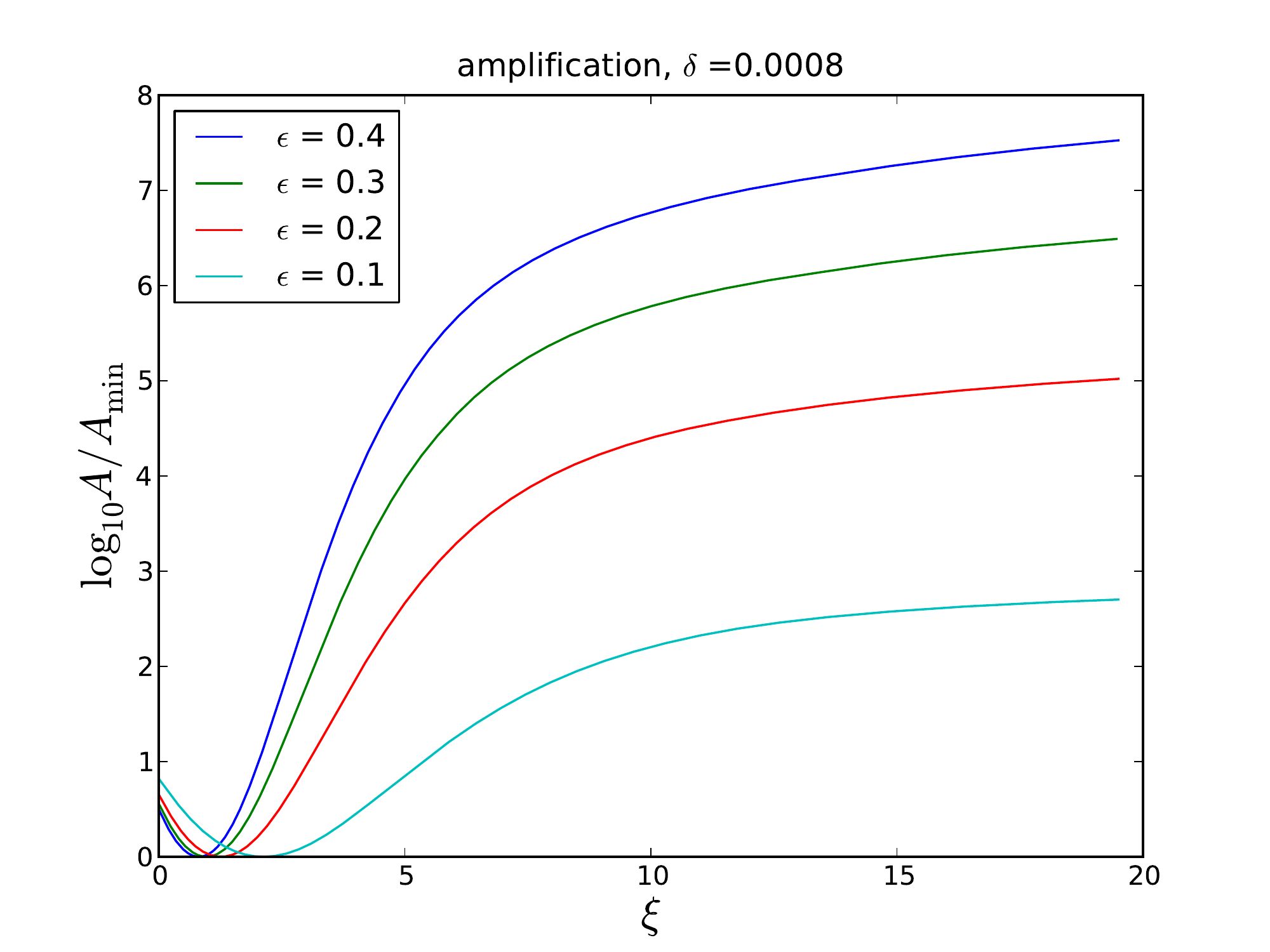}}
  \caption{Amplification of Tollmien-Schlichting waves for the cases
    listed in table \ref{tab:amplification} using the parabolic stability equation solver.}
\label{fig:amplificationDelta0.0008}
\end{figure}

\begin{figure}
  \centerline{\includegraphics[width=\linewidth]{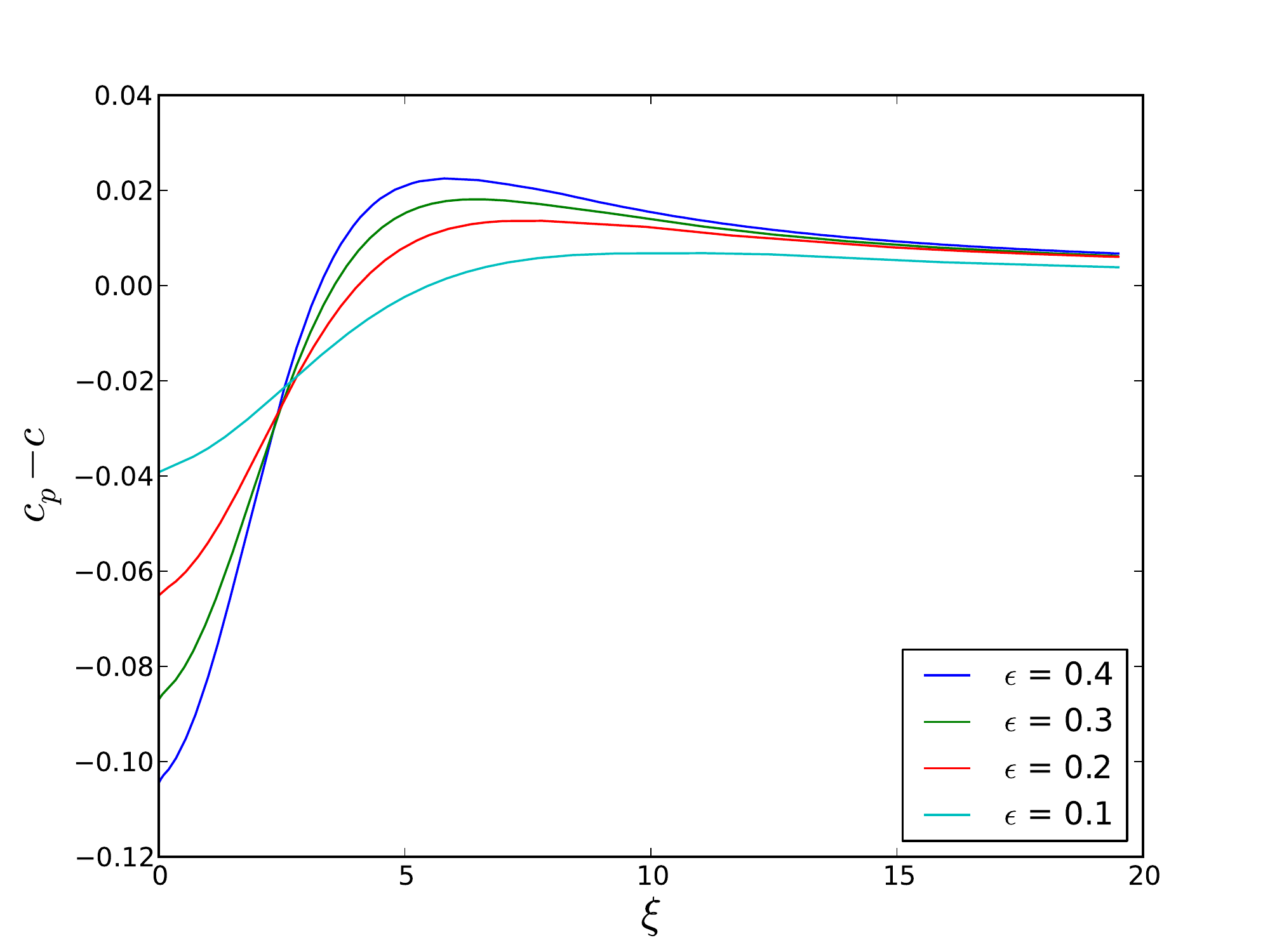}}
  \caption{Absolute phase speed $ c_p - c $ of 
 Tollmien-Schlichting waves for $ \delta = 8 \cdot 10^{-4} $.
 The parameters of the Tollmien-Schlichting waves are given 
in table \ref{tab:amplification}.}
\label{fig:phaseSpeed}
\end{figure}

In the present treatise, three different methods have been chosen
to investigate the linear stability of the boundary layer flow
under a solitary wave. The methods can be ranged by physical 
complexity, meaning the amount of physical effects their model
includes. On a first level, we have the Orr-Sommerfeld equation
solver, neglecting the nonparallel
effects of the flow and nonlinear interactions
of the perturbation. The parabolic stability equation,
on the other hand, belongs as the Orr-Sommerfeld equation to
the class of model equations, but includes effects of nonparallelism.
However, also here nonlinear interactions of the perturbation
are neglected.  
The Navier-Stokes solver is on the highest level, 
based on first principles for an incompressible Newtonian fluid. 
It accounts for all physical effects in the flow. As the amplitude
of the perturbation is considered small, the parabolic stability
equation solver and the Navier-Stokes solver are expected to produce
similar results, since they both contain all the essential physics
of the problem. This has been observed by \cite{BertolottiHerbertSpalart1992}
and \cite{JoslinStreettChang1993} in their stability analysis
of the Blasius boundary layer. \\
Figure \ref{fig:plot1NavierStokes} shows three plots
for the case $ \epsilon = 0.4 $ and $ \delta = 8\cdot10^{-4} $.
In figure \ref{fig:plot1NavierStokes}(a), 
the stability domain of the flow, computed using the 
parabolic stability equation solver, has been plotted
by means of criterion (\ref{eq:stabilityRegionPSE}).
For the region in $ (\xi,\omega) $
bounded by the neutral curve, Tollmien-Schlichting
waves start to grow and may destabilize the flow. 
For the region outside of the neutral curve, the flow
is stable and perturbations decay. At the neutral curve
there is neither growth nor decay of the perturbations. 
The position $ \xi_c $ leftmost on the neutral curve is 
called the critical position, since for $ \xi > \xi_c $
perturbations are expected to grow. We now might pick a particular
frequency $ \omega $ and follow the evolution of
a Tollmien-Schlichting wave with this frequency. As a matter
of fact the discrete spectrum of the Orr-Sommerfeld equation
and the initial condition for the parabolic stability equation solver
allows for an infinity of such Tollmien-Schlichting waves.
Among these Tollmien-Schlichting waves, we always 
choose the most destabilizing one, meaning the one with
maximum real part of $ a $. To ease the discussion, we 
shall call the most destabilizing Tollmien-Schlichting wave simply the 
Tollmien-Schlichting wave in the following. 
In figure \ref{fig:plot1NavierStokes}(b), we have chosen
the Tollmien-Schlichting wave with the 
critical frequency $ \omega_c = 0.24 $. 
Dumping the velocity field $ u(\xi,y,t) $ from the Navier-Stokes
solver at a specific point in time $ t $, allows us to compute the 
perturbation velocity $ u'(\xi,y,t) $ by subtracting the base flow
$ U_\base(\xi,y) $ from the velocity field:
\be
u' = u - U_\base. 
\ee
In figure \ref{fig:plot1NavierStokes}(b), a slice of $ u' $ at
$ y = 0.4943 $ is plotted. We clearly observe a sinusoidal wave
which decays until it reaches the critical position $ \xi_c = 0.82 $ 
and then starts to grow, albeit sllowly at the beginning. 
This qualitative picture can be
analyzed further. Using the solution $ u(x,y,t) $ by
the Navier-Stokes solver, we compute the amplitude of 
the Tollmien-Schlichting wave by
means of the envelope of $ u' $ at its maximum, i.e.:
\be
A(\xi) = \max_{t,y} | u'(\xi,y,t) |. \label{eq:DNSAmplification}
\ee
This envelope can then be compared to the resulting amplifications
using the parabolic stability equation,
and the Orr-Sommerfeld equation,
computed by means of equations (\ref{eq:amplificationPSE})
and (\ref{eq:ampliOSE}), respectively. The result of this
comparison is depicted in figure \ref{fig:plot1NavierStokes}(c).
All three methods predict first a decay of the Tollmien-Schlichting wave
followed by growth. The results by the 
Navier-Stokes solver and by the parabolic stability equation
solver agree remarkably well. This corresponds to 
the good agreement between the Navier-Stokes solver and the parabolic
stability
equation solver reported for the Blasius boundary layer 
mentioned above \cite{BertolottiHerbertSpalart1992,JoslinStreettChang1993}. 
The results by the Orr-Sommerfeld solver on the other 
hand display 
an earlier growth of the Tollmien-Schlichting wave compared
to the other two solvers. This indicates that nonparallel effects are 
significant and lead to quantitative differences.\\
Instead of
choosing the critical frequency $ \omega_c $, 
we might choose any other frequency,
say $ \omega = 0.82 $, cf. figure \ref{fig:plot2NavierStokes}. As can
be seen from figure \ref{fig:plot2NavierStokes}(a), the Tollmien-Schlichting
wave for $ \omega = 0.82 $ crosses the neutral curve twice. It enters
the unstable region at $ \xi_a = 3.52 $ and leaves it at $ \xi_b = 4.48 $. 
The wave is thus expected to first decay until $ \xi_a $, to 
grow subsequently until $ \xi_b $, and to decay again. This can also be
seen by $ u' $ computed by means of the Navier-Stokes solver in
figure \ref{fig:plot2NavierStokes}(b). All three methods predict
first decay of the wave, followed by growth and then decay again, cf.
figure \ref{fig:plot2NavierStokes}(c). 
The amplification computed by means of the Navier-Stokes solver
and by means of the parabolic stability equation solver agree 
again remarkably well. As before the Orr-Sommerfeld solver predicts
a somewhat different amplification of the Tollmien-Schlichting wave. \\
Also for other cases, a remarkably good agreement between
Navier-Stokes results and results by the parabolic stability equation solver
can be observed. In figures \ref{fig:amplificationDelta0.000475} and 
\ref{fig:amplificationDelta0.0001}, we displayed the amplification
by the three methods for the cases $ \epsilon = 0.4 $, $ \delta = 4.75 \cdot 10^{-4} $, $ \omega = 0.22 $ and $ \epsilon = 0.4 $, $ \delta = 10^{-4} $, $ \omega = 0.185 $, respectively. The chosen frequencies correspond to the
critical cases. As before the amplification by the Orr-Sommerfeld
solver displays an earlier growth of the instability. However, for
decreasing $ \delta $, the difference becomes smaller. This is
not surprising, since the vertical velocity component
in the boundary layer scales
like $ \delta $ and so do the nonparallel effects. For the 
$ \epsilon = 0.4 $, $ \delta = 10^{-4} $, $ \omega = 0.185 $ case,
the critical position $ \xi_c $ is in the acceleration region of 
the flow ($ \xi < 0 $), as can be seen from figure 
\ref{fig:amplificationDelta0.0001}, we shall discuss this
feature in a few lines below. The good agreement between 
Navier-Stokes solver and parabolic stability 
equation solver encourages us to continue the present stability
analysis mainly by means of the parabolic stability equation solver. \\
In figures \ref{fig:domainDelta0.0008}, \ref{fig:domainDelta0.000475},
and \ref{fig:domainDelta8e_5}
the stability domains in the $ (\xi, \omega) $ plane 
for the values of $ \delta = 8\cdot 10^{-4},4.75 \cdot 10^{-4}$ and $ 8\cdot 10^{-5} $, respectively, 
are displayed for different values of $ \epsilon $. 
We observe that for the values of $ \delta = 8\cdot 10^{-4} $ and $ \delta =
4.75 \cdot 10^{-4} $ 
all unstable regions are situated completely in the decelerating region
of the flow. This is the region
where the pressure gradient $ dp^{\rm ext}/d\xi $
is favorable for instability. Although the method
for linear stability
by \cite{BlondeauxPralitsVittori2012} is 
derived for unsteady flows and
growth in time,
their unstable regions in the  $ (\xi,k) $ plane, where $ k $ is the chosen
wavenumber, are also entirely situated in the deceleration region,
which supports the present result. 
As mentioned by \cite{SumerJensenSorensenFredsoeLiuCarstensen2010}, 
instability can be expected in this region of the 
flow, since the profile $ U_\base $ displays an inflection point 
behind the crest, i.e. in the domain $ \xi > 0 $. Rayleigh's inflection point
theorem is, however, not entirely applicable for the present case, 
since non-parallel effects are not negligible and viscosity
plays an important role for the growth of the perturbation. 
Not surprisingly, the regions become larger for increasing amplitude 
$ \epsilon $ and decreasing $ \delta $. Viscosity is thus a stabilizing factor.
For $ \delta = 10^{-4} $ and smaller, 
cf. figure \ref{fig:domainDelta8e_5}, we observe that the unstable region
even grows beyond the line $ \xi = 0 $ into the region of accelerated flow.
Figure \ref{fig:domainEpsilon0.4} shows how the stability domain evolves
for $ \epsilon = 0.4 $, when $ \delta $ is decreased. 
The unstable domain forms a 'tongue', for lower values of $ \omega $, reaching into the $ \xi < 0 $ region for decreasing $ \delta $. 
This 'tongue' is probably of viscous nature, since Rayleigh's stability
 criterion (although only valid for strictly parallel flows) 
does not allow for instability in this region. As such the form of this tongue
is reminiscent of the unstable region of the Blasius boundary layer \cite{DrazinReid1981}. There is reason to believe that the instability mechanism in this case
is similar to the one of the Blasius boundary layer
\cite{BainesMujumdarMitsudera1996}. \\
In order to estimate the significance of nonparallel effects, we computed the
stability domains for the present boundary layer flow using the Orr-Sommerfeld
equation for the
cases $ \epsilon = 0.4 $ and $ \delta = 8\cdot 10^{-4} $, $4.75 \cdot 10^{-4} $,
$ 10^{-4} $. The results, in comparison to the ones by the parabolic stability
equation, are plotted in figures \ref{fig:OSEvsPSE} and \ref{fig:OSEvsPSEzoom}. 
As also observed for individual frequencies above,
the Orr-Sommerfeld equation predicts an earlier growth of
the instability. As before, the difference between the unstable regions computed
by means of the Orr-Sommerfeld and the parabolic stability equation
becomes, however, smaller for smaller $ \delta $.
Although there
are differences,
the results by the two methods are still in good agreement, which supports
the correctness of the present approach. \\
\cite{SumerJensenSorensenFredsoeLiuCarstensen2010} 
observed that irregular signals in the boundary layer can appear 
in front of the crest for higher Reynolds numbers $ \Rey_\summer $. In particular they 
presented the case $ \Rey_\summer = 2 \cdot 10^6 $, cf. figure 10 (d) in 
\cite{SumerJensenSorensenFredsoeLiuCarstensen2010}, in which 
instabilities are observable for $ \xi < 0 $. 
In figure \ref{fig:domainReSummer},
the stability domains for different 
values of $ \epsilon $ and $ \delta $ are plotted. 
The values of $ \epsilon $ and $ \delta $ are chosen such that the cases
correspond approximately to
 $ \Rey_\summer = 2 \cdot 10^6 $. 
In order to convert between $ (\epsilon,\delta) $ and $ \Rey_\summer $, we use the conversion formula given in 
\cite{VittoriBlondeaux2011}:
\be
\Rey_\summer = \frac{4}{\sqrt{3}}{\frac{ \epsilon^{3/2}}{\delta^2}}. 
\ee
From figure \ref{fig:domainReSummer}, we observe that the unstable
region is not yet
crossing the line $ \xi = 0 $ for different values of $ \epsilon $ and
$ \delta$ corresponding to $ \Rey_\summer = 2 \cdot 10^6 $. 
\cite{SumerJensenSorensenFredsoeLiuCarstensen2010} 
suspected turbulent spots to appear before $ \xi < 0 $ 
for this Reynolds number. 
This is not only in contrast 
to the present results indicating that Tollmien-Schlichting
waves start growing only after passage of the crest
for this case, 
but also to the works by 
\cite{VittoriBlondeaux2008,VittoriBlondeaux2011} who found that
for even bigger Reynolds numbers $ \Rey_\summer $, the region of instability
is always located behind the crest. There are qualitative 
distinctions in the different analyses which need to be mentioned. 
\cite{SumerJensenSorensenFredsoeLiuCarstensen2010}, 
\cite{VittoriBlondeaux2008,VittoriBlondeaux2011} and \cite{BlondeauxPralitsVittori2012}
considered temporal
growth of instabilities, whereas in the present discussion we focus
on spatial growth. 
One important distinction between the three different works is
the velocity profile used. The velocity profile in the 
boundary layer in the experiments by 
\cite{SumerJensenSorensenFredsoeLiuCarstensen2010} 
and in the simulations in \cite{BlondeauxPralitsVittori2012}
followed from a free stream flow in the form of 
a simple $ {\rm sech^2}( \omega t) $ profile. 
\cite{VittoriBlondeaux2008,VittoriBlondeaux2011},
on the other hand, invoked Grimshaw's solution \cite{Grimshaw1971}
for the outer flow,
which is better than the $ {\rm sech}^{2}( \omega t) $ profile,
but still deviates markedly from the exact one for higher
amplitudes.
Even more important, in all these refences
the approximation of spatially uniform free
stream flow was made, which corresponds to the
linear boundary layer solution by \cite{LiuParkCowen2007},
equation (\ref{eq:linearLiu}), since the nonlinear term vanishes. In
the present discussion the velocity profile is the result
of the nonlinear boundary layer equations (\ref{eq:BL1}-\ref{eq:BL3}) using
a fully nonlinear solution of the potential equations.
As shall be discussed below,
\cite{VittoriBlondeaux2008,VittoriBlondeaux2011} (and
also \cite{SumerJensenSorensenFredsoeLiuCarstensen2010} in their experiments)
did not directly control the amplitude of the perturbation which might
lead to a retarded appearance of the instability in their simulations. 
We doubt, however, that this is the reason for the discrepancy between
the results of \cite{VittoriBlondeaux2008,VittoriBlondeaux2011} and
\cite{SumerJensenSorensenFredsoeLiuCarstensen2010} for the 
$ \Rey_\summer = 2 \cdot 10^6 $ case. In figure \ref{fig:domainLinearReSummer}, 
we plotted together with the results based on the full potential
solution, stability domains for the linearized boundary layer profile by
\cite{LiuParkCowen2007},
equation (\ref{eq:linearLiu}) with the correct formula for the
normal velocity component, equation (\ref{eq:normalComponentLiu})
by means of the 
parabolic stability equation solver.
As can be seen the unstable region starts earlier for the linearized
profile than for the fully nonlinear profile. In addition the
critical frequency $ \omega_c $ is different than for the 
nonlinear case. Nevertheless, it does not
extend into the acceleration region of the flow. Therefore, we believe
that the discrepancy between the results by \cite{VittoriBlondeaux2008,VittoriBlondeaux2011} and \cite{SumerJensenSorensenFredsoeLiuCarstensen2010} for this case might be  due to some disturbance in 
the experiments for higher $ \Rey_\summer $. 
We do, however, support the general observation by 
\cite{SumerJensenSorensenFredsoeLiuCarstensen2010}, that instability
can occur in the acceleration region of the flow. 
This occurs, however, for higher Reynolds numbers 
$ \Rey_\summer $ than theirs. In figure \ref{fig:domainEpsilon0.4},
the unstable region for the case $ \epsilon = 0.4$, $ \delta = 10^{-4} $ 
enters the acceleration region $ \xi < 0 $.
This case corresponds to a Reynolds number of $ \Rey_\summer = 6 \cdot 10^7 $,
which is more than an order of magnitude larger than the result
by \cite{SumerJensenSorensenFredsoeLiuCarstensen2010}. \\
We found that for all values
of $ \epsilon $ and $ \delta $ considered the boundary layer flow
displays regions of growth of instabilities, confirming the
results by \cite{BlondeauxPralitsVittori2012}. As also mentioned by
\cite{BlondeauxPralitsVittori2012}, this
is, however, 
in contrast to the apparent findings by \cite{SumerJensenSorensenFredsoeLiuCarstensen2010} and \cite{VittoriBlondeaux2008,VittoriBlondeaux2011} which state
that below a critical Reynolds number $ \Rey_\summer $,
the boundary layer profile does not depart from its 
shape given by the boundary layer equations (\ref{eq:BL1}-\ref{eq:BL3}). 
At the critical Reynolds number the references reported, the boundary layer flow
performs a transition from the solution
given by the boundary layer equations (\ref{eq:BL1}-\ref{eq:BL3})
to a laminar flow
with periodic vortex shedding. 
This critical Reynolds number
has been determined by \cite{SumerJensenSorensenFredsoeLiuCarstensen2010}
to have the value $ \Rey_\summer = 2 \cdot 10^5 $, whereas 
\cite{VittoriBlondeaux2011} put this value higher, somewhat below $ 
\Rey_\summer = 5 \cdot 10^5 $, cf. figure 5 in \cite{VittoriBlondeaux2011}. 
As reported, a second transition from laminar vortex shading to turbulent flow
arises for higher values of $ \Rey_\summer $. In the present discussion,
we focus on the first transition where the boundary layer 
flow solution, given by equation (\ref{eq:BL1}-\ref{eq:BL3}), becomes unstable.
We do, however, not make any prediction on what will happen after transition.
Linear stability cannot predict what happens
once the flow turns unstable. 
The flow might become turbulent, it might go
over to a different laminar regime or as we shall see it might also 
continue to 'display' the original solution given by the boundary layer
equations (\ref{eq:BL1}-\ref{eq:BL3}). \\
Since, in the sense of linear stability the flow is always unstable,
the question is rather when do these instabilities become visible. 
\cite{BlondeauxPralitsVittori2012} tried to answer this question by looking
at the kinetic energy of the perturbations in their Navier-Stokes solution.
They observed that when fixing $ \delta $ the growth of the
kinetic energy of the perturbations becomes more important
for increasing $ \epsilon $ until for a critical $ \epsilon_c $, 
the kinetic energy of the perturbations cannot be neglected anymore. 
They therefore concluded that a critical set of parameters $ (\delta_c , \epsilon_c ) $ can be given for which the flow turns unstable, although they admit
that there is a certain arbitrariness in the choice of $ (\delta_c , \epsilon_c ) $. Their analysis is, however, misleading. 
As we shall see, the fact when a flow 
turns visibly unstable depends in a great deal on the initial amplitude
of the perturbation which was uncontrolled in the experiments by 
\cite{SumerJensenSorensenFredsoeLiuCarstensen2010} and also in
the simulations by \cite{VittoriBlondeaux2008,VittoriBlondeaux2011}. 
Visibility of the instability can be measured by the amplification of
the perturbation. As a matter of fact a general perturbation consists
of all possible frequencies $ \omega $ and the amplification of each
mode should be investigated. To simplify the analysis we only consider
the amplification of a Tollmien-Schlichting wave of 
frequency $ \omega_c $ where $ \omega_c $ is
the frequency of the Tollmien-Schlichting
starting to grow at the critical position $ \xi_c $.
The parameter range investigated is $ \delta = 8 \cdot 10^{-4} $ and 
$ \epsilon = 0.1,0.2,0.3,0.4 $. The stability domains
for these cases have been displayed in figure \ref{fig:domainDelta0.0008}
and by going along the neutral curve to its leftmost extremum,
the critical position $ \xi_c $ can be found. The critical parameters
for the Tollmien-Schlichting waves have been listed
in table \ref{tab:amplification}. 
The cases $ \delta = 8 \cdot 10^{-4} $ and 
$ \epsilon = 0.1,0.2,0.3,0.4 $ have also been 
investigated by \cite{VittoriBlondeaux2008,VittoriBlondeaux2011}.
In figure 1 in \cite{VittoriBlondeaux2011}, time profiles 
of the horizontal velocity component at a point in space are plotted. 
In addition, in figure 5 in \cite{VittoriBlondeaux2011} it is clearly
visible that the case $ (\epsilon = 0.1,\delta = 8 \cdot 10^{-4}) $
has been considered stable by both 
\cite{SumerJensenSorensenFredsoeLiuCarstensen2010} and 
\cite{VittoriBlondeaux2011}. The cases $ (\epsilon = 0.3,\delta = 8 \cdot 10^{-4}) $ and $ (\epsilon = 0.4,\delta = 8 \cdot 10^{-4}) $ on the other hand
were found to be unstable by both \cite{SumerJensenSorensenFredsoeLiuCarstensen2010} and 
\cite{VittoriBlondeaux2011}. However, the case 
$ (\epsilon = 0.2,\delta = 8 \cdot 10^{-4}) $ was classified unstable by 
\cite{SumerJensenSorensenFredsoeLiuCarstensen2010}, whereas
\cite{VittoriBlondeaux2011} deemed it stable. \\
In figure \ref{fig:amplificationDelta0.0008} the growth of the
Tollmien-Schlichting wave has been recorded in terms of the amplification
of the perturbation,
cf. equation (\ref{eq:amplificationPSE}), using the parabolic stability
equation solver. At the
chosen point $ \xi = 19.5 $, we measured the amplitude of the
signal and compared it with the minimum amplitude 
at the critical position $ \xi_c $, of the signal which gives us
the amplification of the signal at the above point. These amplifications
are listed in table \ref{tab:amplification}. The value $ \xi = 19.5 $ seemed
reasonable to us, since the maximum extension in time behind the crest used by 
\cite{VittoriBlondeaux2011} for example in figure 1 in \cite{VittoriBlondeaux2011} is 20 which corresponds approximately to a spatial extension between
$ 21 $ and $ 23.6 $. 
If we assume for the time being that the 
Tollmien-Schlichting waves start to roll up into vortices once their
amplitude has grown to a value comparable to the mean flow, of order 1 thus, 
we obtain the result that the initial perturbation at $ \xi_c $ for
the case $ (\epsilon = 0.1,\delta = 8 \cdot 10^{-4}) $
has at least been smaller than $ 10^{-2.7} $ in the experiments by
\cite{SumerJensenSorensenFredsoeLiuCarstensen2010} and the simulations
by \cite{VittoriBlondeaux2008,VittoriBlondeaux2011}. The case 
$ (\epsilon = 0.3,\delta = 8 \cdot 10^{-4}) $ on the other hand
tells us that the perturbation for both \cite{SumerJensenSorensenFredsoeLiuCarstensen2010} and
\cite{VittoriBlondeaux2008,VittoriBlondeaux2011} must have had an
initial amplitude approximately equal or larger than $ 10^{-6.5} $.
Since the experiments turned unstable for $ \epsilon = 0.2 $,
the initial amplitude in the experiments must have been 
approximately equal or larger than $ 10^{-5} $. On the other hand 
the amplitude of the perturbation at the critical position in the 
simulations by \cite{VittoriBlondeaux2008,VittoriBlondeaux2011} 
must have been smaller than $ 10^{-5} $ for this case. In figure
1(b) in \cite{VittoriBlondeaux2011} some wiggles in the temporal
evolution of the horizontal velocity appear at approximately $ t = 6 $ behind the crest for
the case $ (\epsilon = 0.3,\delta = 8 \cdot 10^{-4}) $, using the speed
$ c $ of a solitary wave for this amplitude, this point in time corresponds
to a position behind the crest of $ \xi = 6.83 $. From figure
\ref{fig:amplificationDelta0.0008}, we obtain an amplification of 
$ 10^{5.0} $ at $ \xi = 6.83 $. The same can be done for the
case $ (\epsilon = 0.4,\delta = 8 \cdot 10^{-4}) $, cf. figure 1(c) in
\cite{VittoriBlondeaux2011}, where the wiggles start at 
approximately $ t = 4 $ behind the crest,
corresponding to $ \xi = 4.71 $. The amplification at that position,
taken from figure \ref{fig:amplificationDelta0.0008} is $ 10^{4.7} $
Although the above analysis is relatively crude, since the 
works by \cite{SumerJensenSorensenFredsoeLiuCarstensen2010},
by \cite{VittoriBlondeaux2008,VittoriBlondeaux2011} and the present
one are not comparable in detail, it suggests that at the 
critical position $ \xi_c $ the amplitude of the perturbation 
in the simulations by \cite{VittoriBlondeaux2008,VittoriBlondeaux2011}
was smaller than that of the perturbation in the experiments by 
\cite{SumerJensenSorensenFredsoeLiuCarstensen2010}. 
Since linear stability predicts a strong decay of the
perturbations in the acceleration region, the
initial $ 10^{-4} $ amplitude perturbation, which
\cite{VittoriBlondeaux2008,VittoriBlondeaux2011}
reported to have imposed onto the initial condition before the solitary
wave arrived, should have decayed until the critical position
to values much lower than 
the $ 10^{-5}$ estimated above. In addition, since they introduced white noise the initial amplitude on a Tollmien-Schlichting wave, conceivable as the
result of a normal mode decompositon, would be much less than  $ 10^{-4} $.
According to this, combined with our analysis, they should not have observed
appreciable instabilities for the case in question $ (\epsilon = 0.3,\delta = 8 \cdot 10^{-4}) $.
 Unfortunately,
no sensitivity test on the influence of the initial perturbation was reported in 
\cite{VittoriBlondeaux2008,VittoriBlondeaux2011} and \cite{BlondeauxPralitsVittori2012}. It would be of interest how their results would change
from the published once, if, for instance, no initial seeding or seeding with a certain spectrum had been applied.
However, the Navier-Stokes solver used
in \cite{VittoriBlondeaux2008,VittoriBlondeaux2011} and \cite{BlondeauxPralitsVittori2012}
may have produce a certain level of numerical noise, which might have provided
a sufficient level at $\xi_c$ for instabilities to become visible. 
We find support for this presumption in figure 4 in \cite{BlondeauxPralitsVittori2012}, where the level of kinetic energy 
of the perturbation seems to stay on a stable level of around $ 10^{-8}-10^{-9}
 $ for all cases of $ \epsilon $, before arriving at the critical position. 
This indicates that their code bears a source of numerical noise with
an amplitude of approximately  $ 10^{-4}-10^{-4.5} $,
which combined with an amplification factor of $10^5$, as found above,
well may have caused visible disturbances. Several reasons for the
numerical source of noise are thinkable, such as truncation errors or 
incomplete pressure
solutions. As a conclusion, it needs to be said that in the works of 
\cite{SumerJensenSorensenFredsoeLiuCarstensen2010}, 
\cite{VittoriBlondeaux2008,VittoriBlondeaux2011} and \cite{BlondeauxPralitsVittori2012},
the
triggering mechanism of the instability is not well controlled.
Since the initial amplitude is crucial for the visual appearance of the
perturbation, the meaning of 
classifications such as the one presented in figure 5 in 
\cite{SumerJensenSorensenFredsoeLiuCarstensen2010} or
figure 5 in \cite{VittoriBlondeaux2011} or
the determination of critical parameters $ (\delta_c,\epsilon_c) $
in \cite{BlondeauxPralitsVittori2012} needs to be taken with care. A 
boundary layer flow under a solitary wave considered
to be unstable by criteria proposed by either
\cite{SumerJensenSorensenFredsoeLiuCarstensen2010},
\cite{VittoriBlondeaux2008,VittoriBlondeaux2011} or \cite{BlondeauxPralitsVittori2012},
might in a different setting seem to be stable as long as the 
initial amplitude of the perturbation at the
critical position is smaller than its amplification.
On the other hand, a flow determined to be stable by one 
of the above criteria might display instabilities once the initial
amplitude is equal or larger than its amplification.
In their related study on boundary layers on a $10^\circ$ beach \cite{PedersenLindstromBertelsenJensenSaelevik2013} reported (section IVA) for one case (incident wave with $\epsilon=0.3$) that 
instability occurred in an up-beach position, immediately after an inflection 
point in the boundary layer profile was observed. However, 
this was observed in 3
 out of 4 experiments, and never in a location close to equilibrium shoreline,
where inflection points in the retardation phase were also present. 
In view of the present investigation we presume that the flow in those experiments were  
unstable in most of the retardation phase and that the visible appearance of 
instabilities
was due to the integrated amplification factor and the level of disturbances, which in this case may stem from  
particle seeding (for PIV measurements), contact point dynamics and residual motion from preceding experiment. In fact, unless the disturbances is actively controlled we 
do not believe that experimental repeatability for a flow transition of this type can be obtained.\\
A quantity of interest, also investigated in 
\cite{BlondeauxPralitsVittori2012}, is the 
phase speed of the critical Tollmien-Schlichting wave.
For a given frequency $ \omega $, the wavenumber of 
the Tollmien-Schlichting wave can approximately be given
by $\Imag(a) $, where $ a $ is defined in equation (\ref{eq:PsiPSE}). 
In figure \ref{fig:phaseSpeed}, the phase speed
in the absolute frame of reference for the
critical Tollmien-Schlichting waves
is plotted as a function of $ \xi $.
The parameters for the plotted cases are given in 
table \ref{tab:amplification}. The Tollmien-Schlichting waves
seem to first propagate in the direction of the 
solitary wave and then reverse their direction of
propagation. This result has also been obtained by 
\cite{BlondeauxPralitsVittori2012} for their 
perturbations. They proposed that the flow reversal in 
the boundary layer is causing the Tollmien-Schlichting
waves to reverse their direction of propagation too. From
figure \ref{fig:phaseSpeed}, we observe that for increasing
amplitudes $\epsilon $, the Tollmien-Schlichting waves travel with an increasing phase speed. The reason may be that
the magnitude of the particle velocities in the base flow  become 
higher for increasing $\epsilon $. 

\begin{table}
\centering
\caption{Critical parameters for the case $ \delta = 8 \cdot 10^{-4} $,
for different values of $ \epsilon $}
\[
\begin{array}{l|l|l|l|l|l|l}
\epsilon & c &\xi_c & t_c = \xi_c/c&\omega_c & k_c & A(\xi=19.5)/A_{\min} \\
\hline
0.1 & 1.049 & 2.125 & 2.03 &0.218  & 0.212 & 10^{2.7}\\
0.2 & 1.094 & 1.249 & 1.14 &0.228  & 0.218 & 10^{5.0}\\
0.3 & 1.138 & 0.969 & 0.85 &0.230  & 0.216 & 10^{6.5} \\
0.4 & 1.179 & 0.820 & 0.70 &0.240  & 0.220 & 10^{7.5}
\end{array}
\]
\label{tab:amplification}
\end{table}

\clearpage
\section{Conclusions} \label{sec:conclusions}
In the present treatise, the nonlinear solution of the
boundary layer under a solitary wave is computed. Unlike 
 \cite{LiuParkCowen2007} we find that the
linear and nonlinear profiles differ significantly for
increasing amplitude $ \epsilon $ of the solitary wave. 
Consequently, our further analysis is based on a fully nonlinear 
boundary layer theory, unlike the preceding studies from
the literature.
The stability of the boundary layer flow is investigated by means of
two methods based on model equations,
the Orr-Sommerfeld equation and the parabolic stability
equation, and a direct numerical simulation
by a spectral Navier-Stokes solver. 
We confirmed the
result by \cite{BlondeauxPralitsVittori2012}
that in the sense of linear stability, the boundary
layer flow is always unstable for the parameter range considered, meaning
that Tollmien-Schlichting waves will start to grow once they reach the
critical position $ \xi_c $. This critical position can be obtained
by computing the stability domains for different amplitudes $ \epsilon $ 
of the solitary wave
and different viscosity parameters $ \delta $, which is inversely proportional
to the Reynolds number.
Increasing $ \epsilon $ or decreasing $ \delta $ will increase
the unstable region of the flow. 
By comparing the stability domain by the linear solution by
\cite{LiuParkCowen2007} to the one of the present nonlinear
solution, we found that perturbations start to grow earlier (smaller $\xi_c$)
in the 
linear case and that the critical frequency is
higher than in the nonlinear case. Using the Orr-Sommerfeld
equation and the parabolic stability equation, the nonparallel effects
of the flow were analyzed. For small $ \delta $ the nonparallel
effects become less significant. However, for larger values
of $ \delta $, they will retard the critical position. 
For low values of $ \epsilon $ and/or
high values of $ \delta $, the unstable region is in the deceleration
region of the wave, where the pressure gradient is favorable to instability. 
However, decreasing $ \delta $ sufficiently will lead to the growth of
a 'tongue' shaped region extending into the acceleration region of the 
flow, which is by Rayleigh's inflection point theorem a stable region
for inviscid flow. Therefore viscosity plays a major role for this
instability. Although only qualitatively, this supports the observation
of instabilities in the acceleration region
by \cite{SumerJensenSorensenFredsoeLiuCarstensen2010} in their experiments. 
We underline, however, that such instabilities may only be found for
water depths larger than those used in normal wave tank experiments. 
\cite{VittoriBlondeaux2008,VittoriBlondeaux2011} did not observe
instabilities before the crest, since the Reynolds numbers
they considered were too small. 
However, opposed to the present results,
\cite{SumerJensenSorensenFredsoeLiuCarstensen2010}
by means of experiments 
and \cite{VittoriBlondeaux2008,VittoriBlondeaux2011}
by means of simulations
found that the flow turns unstable only beyond a certain critical
Reynolds number $ \Rey_\summer $, which differed for the
experiments and the simulations. Similar
\cite{BlondeauxPralitsVittori2012} postulated
that a critical set of parameters $ (\epsilon,\delta) $
can be defined for this flow. By looking at the amplification
of the perturbation it was shown herein that not only differences in the
base flow field but also the initial amplitude of the perturbation
might be the reason why for the experiments 
by \cite{SumerJensenSorensenFredsoeLiuCarstensen2010}
the flow turned unstable for one Reynolds number
whereas it did not for the simulations
by  \cite{VittoriBlondeaux2008,VittoriBlondeaux2011}. As such we found that
a critical Reynolds number or a critical set of parameters $ (\epsilon,\delta) $
for this flow does not exist and 
that we at most might give a criterion dependent on the
initial amplitude of the perturbation for the 
appearance of transition in the flow field. 
A possible direction for future research is
investigation of the boundary layer flow under a solitary
wave by means of nonlinear stability analysis where Tollmien-Schlichting waves
with different frequencies interact with each other. 
In addition, the roll up of the Tollmien-Schlichting waves into vortices
in connection with the initial seeding
of the perturbation needs to be investigated more in detail.
This topic touches
also the appearance of turbulent spots observed by
\cite{SumerJensenSorensenFredsoeLiuCarstensen2010} leading
to the question of how the growth of Tollmien-Schlichting waves,
the roll up into vortices and the interaction with turbulent
spots produce transitions in the boundary layer under a solitary wave. \\[1ex]
Prof. Arnold Bertelsen is cordially acknowledged for interesting discussions. \\
The work was supported by the Norwegian Research Council 
under the project 205184/F20.\\
The computations were partly performed on the Abel Cluster,
owned by the University of Oslo
and the Norwegian
metacenter for High Performance Computing (NOTUR).

\begin{appendix} 
\section{Verification and validation of the boundary layer solver} \label{sec:appendixBoundaryLayer}

The verification of the boundary layer solver,
derived in subsection \ref{sec:boundaryLayerSolver},
was done by means of a manufactured solution, cf. subsection
\ref{sec:appendixBoundaryLayer1}. We validated the method by
applying it to the Blasius boundary layer problem,
for which an accurate reference solution exists, cf.
subsection \ref{sec:appendixBoundaryLayer2}. Finally
in subsection \ref{sec:appendixBoundaryLayer3},
we investigated the convergence of the method, when applied
to the problem of the boundary layer under a solitary wave.

\subsection{Verification by means of a toy problem}
\label{sec:appendixBoundaryLayer1}

\begin{figure}

  \centerline{\includegraphics[width=\linewidth]{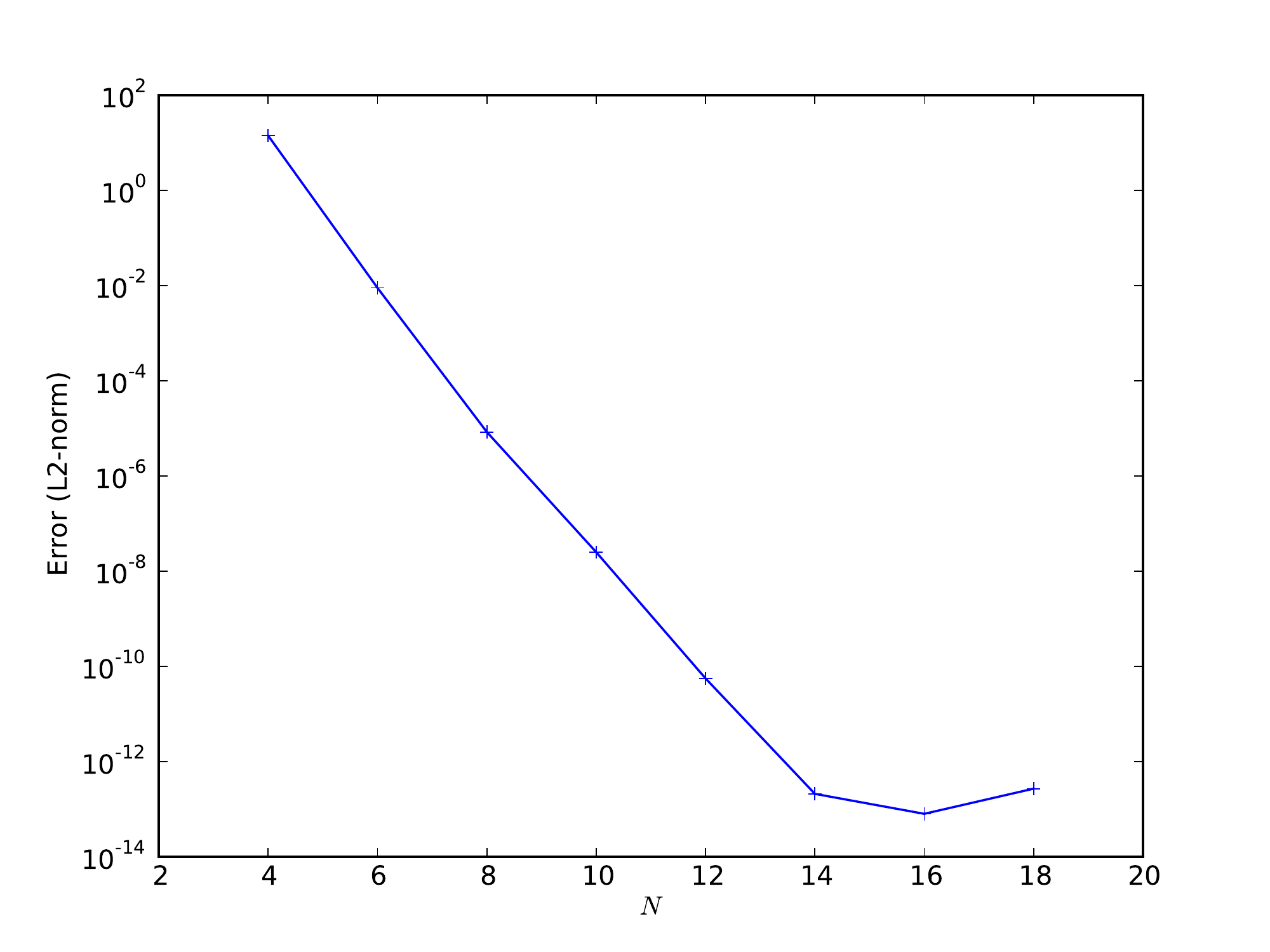}}
  \caption{Convergence of the boundary layer solver described
    in subsection \ref{sec:boundaryLayerSolver} for the toy problem
\ref{eq:toy}}
\label{fig:convergenceToy}
\end{figure}

In order to verify  the boundary layer equations solver
proposed in 
subsection \ref{sec:boundaryLayerSolver}, we used the artificial
field
\be
u^{\rf}= - y^{n-1} \cos x \quad v^{\rf} = \frac{1}{n} y^n \sin x,
\label{eq:toy}
\ee
which satisfies the continuity equation (\ref{eq:BL1}) with an external
pressure gradient given by
\be
-\frac{\partial p^{\rm ext}}{\partial y} = 
u^{\rf} \frac{\partial u^{\rf}}{\partial x} + v^\rf 
\frac{\partial u^{\rf}}{\partial y} - \frac{1}{2}
\frac{\partial^2 u^{\rf}}{\partial y^2}. \label{eq:toy2}
\ee
This pressure gradient 
should reproduce the velocity field (\ref{eq:toy}). We remark
that the external pressure gradient (\ref{eq:toy2}) does not
satisfy equation (\ref{eq:BL3}). This does, however, not represent
a flaw for verifying the boundary layer solver derived in subsection
\ref{sec:boundaryLayerSolver}. Figure \ref{fig:convergenceToy} shows
the convergence of the method for this problem, when choosing $ n = 4 $. 
For this simple problem we observe spectral convergence of the method.
The round off accuracy is already obtained with only $ N_\bl = 14 $. The
error was measured in the $ L2 $ norm. 

\subsection{Validation by means of the Blasius boundary layer flow}
\label{sec:appendixBoundaryLayer2}

\begin{figure}

  \centerline{\includegraphics[width=\linewidth]{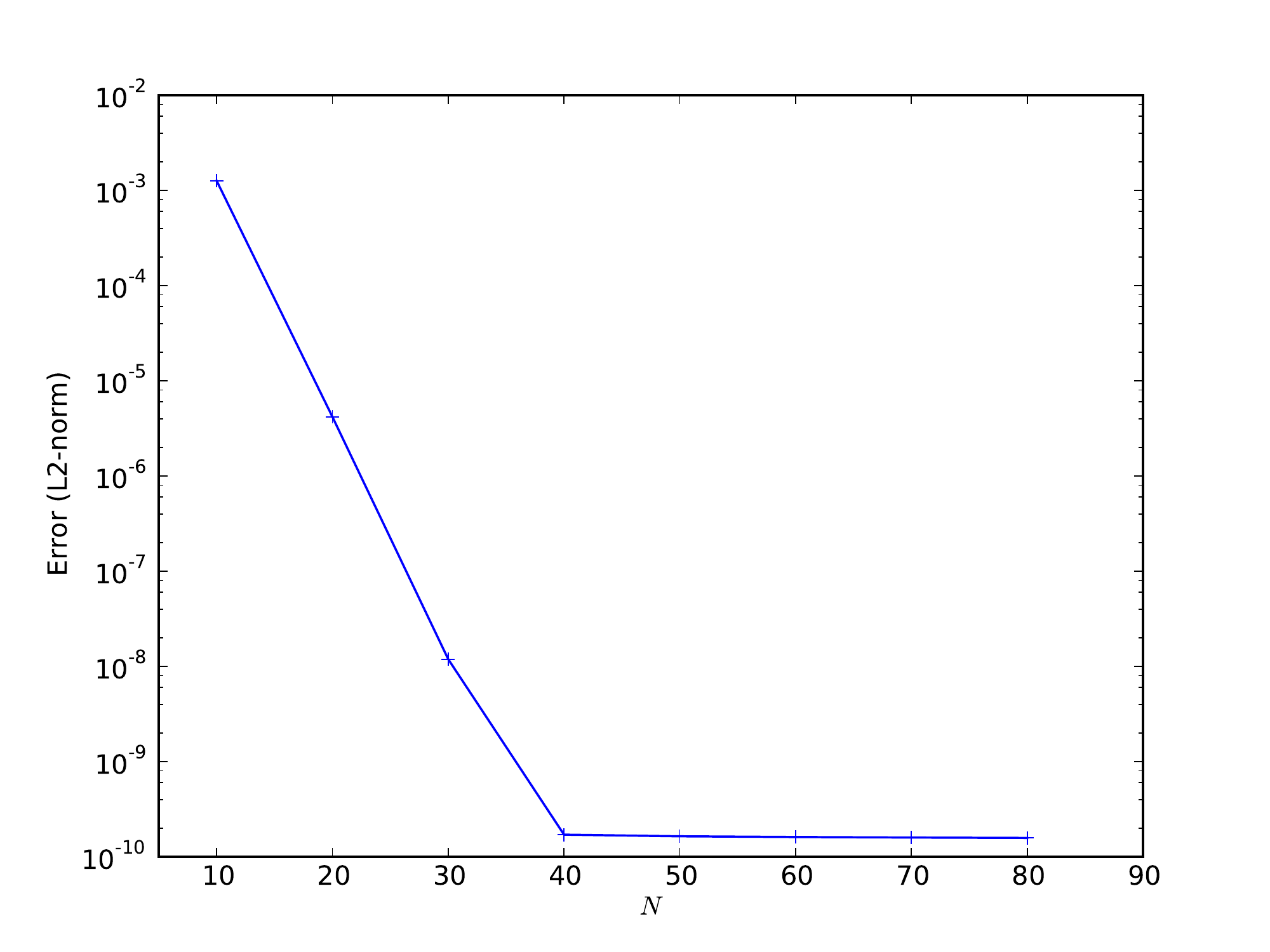}}
  \caption{Convergence of the boundary layer solver described
in subsection \ref{sec:boundaryLayerSolver} for the Blasius boundary layer.}
\label{fig:convergenceBlasius}
\end{figure}

For the Blasius boundary layer problem \cite{Ryhming2004} a well
known numerical solution exists. In order to validate 
the boundary layer solver in subsection \ref{sec:boundaryLayerSolver},
we applied the present scheme to solve the boundary layer equations
for a flat plate. The result was then compared to the
Blasius boundary layer solution by means of the $ L2 $ norm.
The domain of integration was from $ \Rey_x = 100 $ to $ \Rey_x = 200 $.
The convergence of the method for increasing $ N_\bl $ can be observed
in figure \ref{fig:convergenceBlasius}. The error decreases spectrally up 
to a value of $ 10^{-10} $ after which it stagnates. The reason for this
might be the accumulation of round off error or due to some coarse internal 
parameters in the inbuilt ordinary differential equation solver by MATLAB
used to generate the reference solution. 

\subsection{Validation by means of the boundary layer flow under
a solitary wave}
\label{sec:appendixBoundaryLayer3}

\begin{figure}

  \centerline{\includegraphics[width=\linewidth]{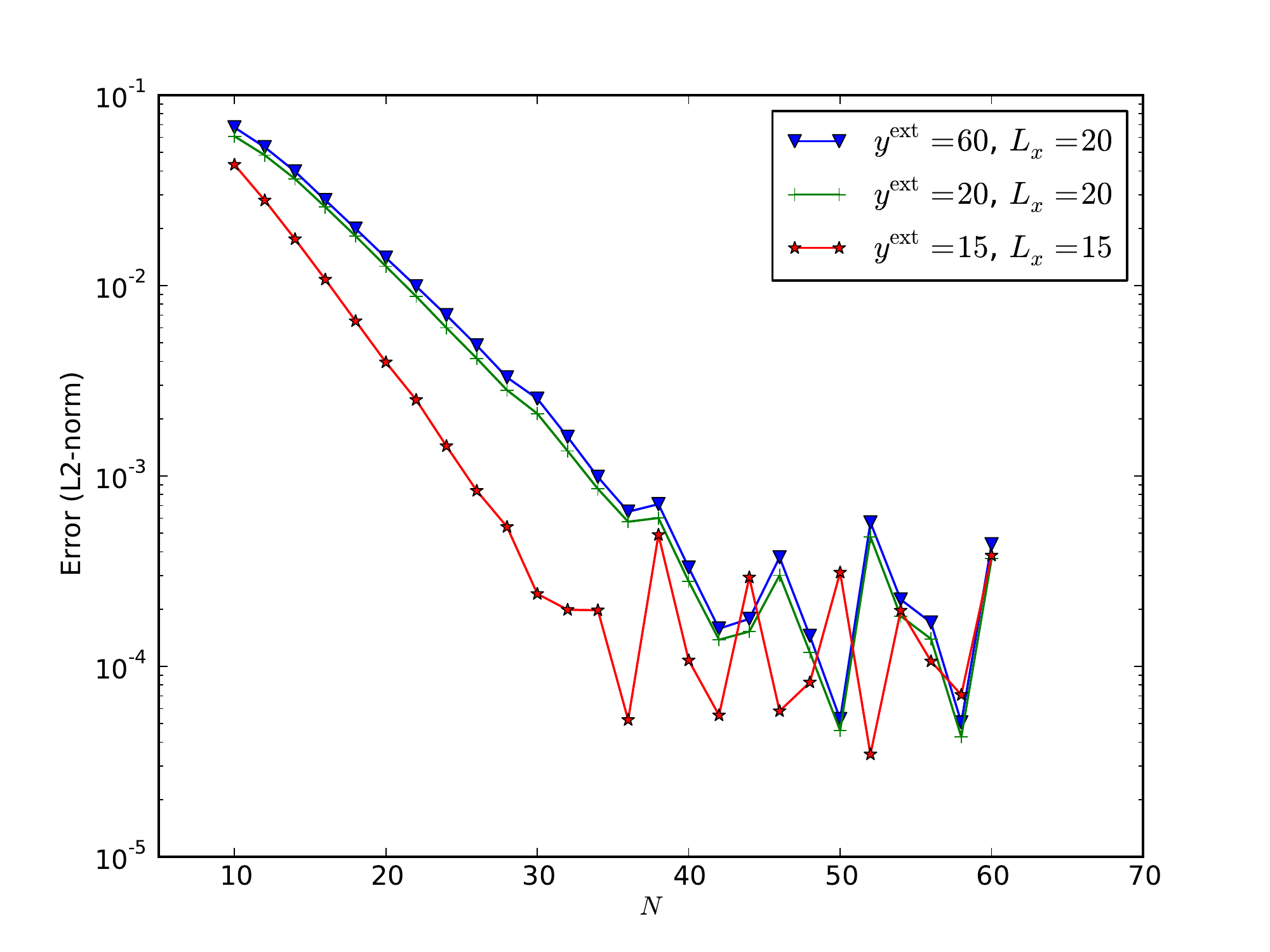}}
  \caption{Convergence of the boundary layer solver described
in subsection \ref{sec:boundaryLayerSolver} for the boundary layer under a solitary wave with $ \epsilon = 0.4 $.}
\label{fig:convergenceSolitary Wave1}
\end{figure}

As a last benchmark test, we tested the convergence of the boundary layer
solver, cf. subsection \ref{sec:boundaryLayerSolver}, for the
boundary layer flow under a solitary wave itself, in particular
for the parameter $ \epsilon = 0.4 $ and for different domain
sizes given by $ y^{\rm ext} $ and $ L_x $. 
The reference solutions have in this case
been produced by using the boundary layer solver on a fine grid
with $ N_\bl = 80 $. The results can be seen in figure 
\ref{fig:convergenceSolitary Wave1}. 
The convergence appears to be only spectral for a resolution up to 
$ N_\bl \approx 40 $ for the larger domain sizes,
after which the error oscillates around $ 10^{-4}$.
Increasing the resolution
of the reference solution did not lead to better results. The
reason for this behavior is that the inviscid flow solution 
is only accurate up to $ 10^{-4} $. 
This can also be seen by the fact that for smaller 
domains the convergence is better until we reach a level of the
error of approximately $ 10^{-4} $ behind which the convergence stagnates,
cf figure \ref{fig:convergenceSolitary Wave1}, since 
the inviscid flow is the same for all domain sizes. 
For a smaller value of $ \epsilon $, namely $ \epsilon = 0.2 $ ($ y^{\rm ext} = 20 $ and $ L_x = 20 $ ),
the accuracy is somewhat better. 
In order to be on the safe side concerning the accuracy of the
numerical solution for different values of $ \epsilon $,
we chose a value of $ N_\bl = 80 $ for all simulations in section
\ref{sec:results}. The domain size parameters are chosen $ y^{\rm ext} = 60 $
and $ L_x = 20 $ such that the domain is large enough
in order to accommodate the full width of the Tollmien-Schlichting waves. 
This way the numerical solution 
to the boundary layer equations (\ref{eq:BL1}-\ref{eq:BL2}) 
can  be assumed to be accurate up to the fourth digit. 

\begin{figure}

  \centerline{\includegraphics[width=\linewidth]{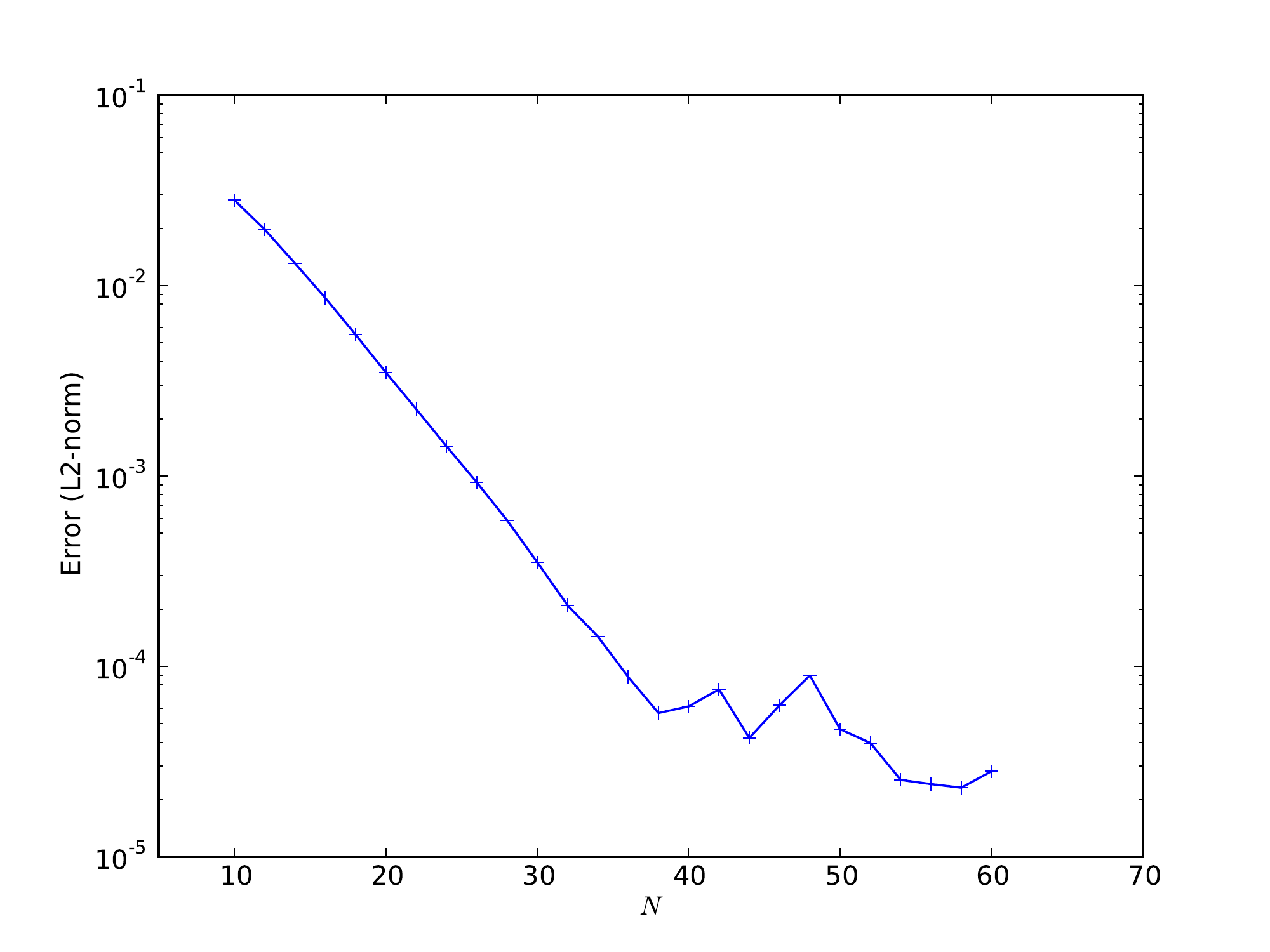}}
  \caption{Convergence of the boundary layer solver described
in subsection \ref{sec:boundaryLayerSolver} for the boundary layer under a solitary wave with $ \epsilon = 0.2 $.}
\label{fig:convergenceSolitary Wave3}
\end{figure}

\section{Verification and validation of the Orr-Sommerfeld solver} \label{sec:appendixOSE}

As for the boundary layer solver, we verified and validated the 
present Orr-Sommerfeld solver derived in subsection \ref{sec:OrrSommerfeldSolver} by means of the Blasius boundary layer and by means of
the boundary layer under solitary wave itself. 

\subsection{Verification and validation by means of the Blasius boundary layer}

\begin{figure}
  \centerline{\includegraphics[width=\linewidth]{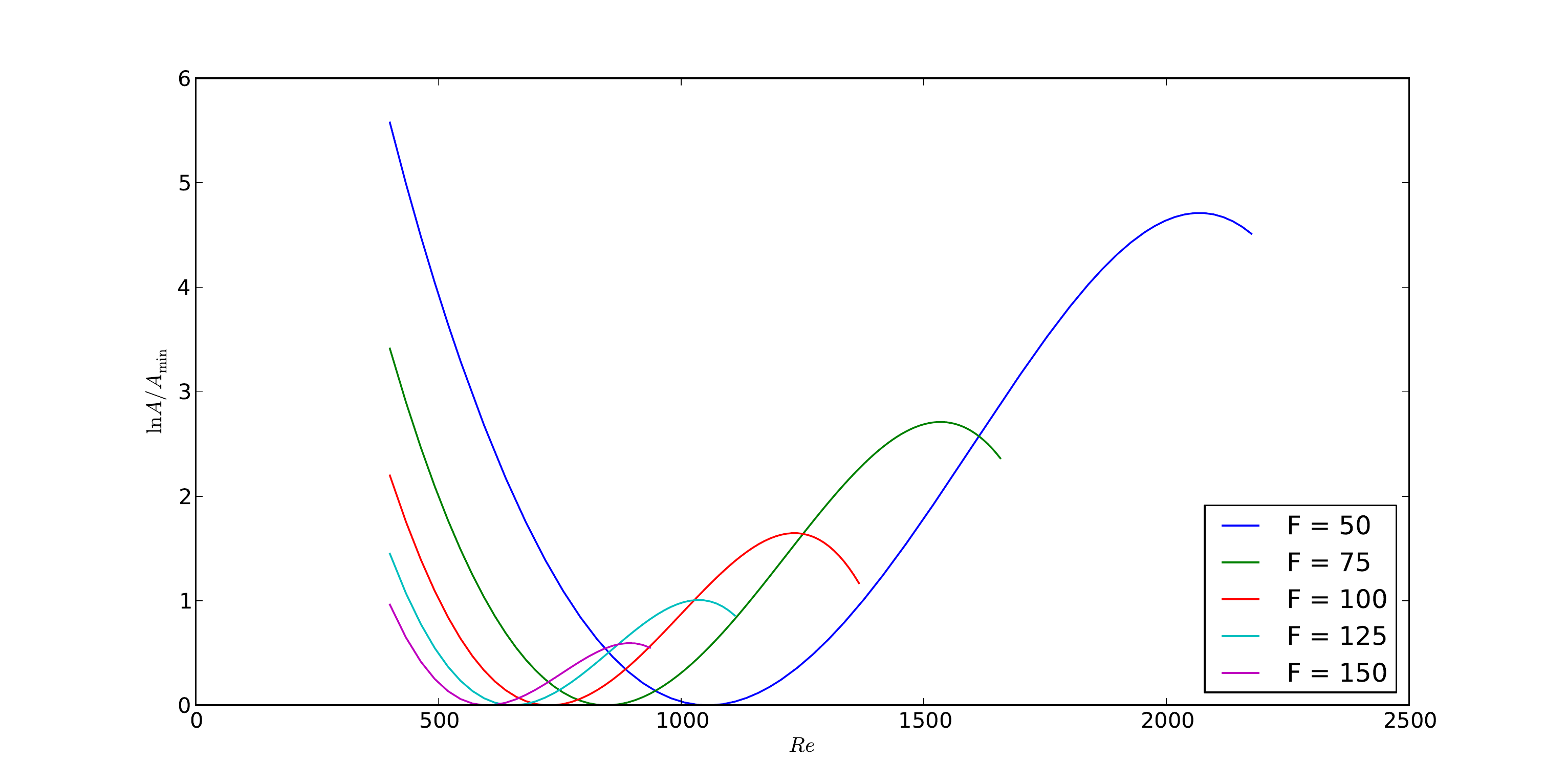}}
\caption{Amplification curves for the Blasius boundary layer
using the Orr-Sommerfeld solver, cf. subsection \ref{sec:OrrSommerfeldSolver},
for different frequencies $ F \times 10^{6} = 50,75,100,125,150 $.
The results can be compared to the results in
figure 4(a) in \cite{Jordinson1970}.}
\label{fig:AmplificationJordinson}
\end{figure}

\cite{Jordinson1970} was the first to use the Orr-Sommerfeld equation to
investigate spatial instabilities of the Blasius boundary layer. 
He presented the values for $ a $, equation (\ref{eq:PsiOSE}),

for a few selected cases. We computed the eigenvalues for these
cases, which are listed below. These results are obtained
by using $ N_\ose = 80 $ and $ N_\ose = 120 $. Only the 
coinciding digits of the solutions have been printed.
All results are given in the scaling used in 
\cite{Jordinson1970} and can be compared with the 
values given therein. 
\begin{itemize}
\item Case 1: $ \Rey = 336 $, $ \omega = 0.1297 $
\[
a = -0.007952136+ 0.308318511 {\rm i}
\]
\item Case 2: $ \Rey = 598 $, $ \omega = 0.1201 $
\[
a =  0.001893765+ 0.307831329 {\rm i}
\]
\item Case 3: $ \Rey = 998 $, $ \omega = 0.1122 $
\[
a =  0.005707382+ 0.308584442 {\rm i}
\]
\end{itemize}
By comparing the above values to the values given in \cite{Jordinson1970},
we see that some of the above growth rates and wave numbers display 
small differences on the fourth decimal. In addition, we computed the 
amplifications
for the frequencies $ F \times 10^{6} = 50,75,100,125,150 $, cf. figure \ref{fig:AmplificationJordinson}. For the precise definition of $ F $, we refer to
\cite{Jordinson1970}. When comparing the present results to the 
corresponding results in \cite{Jordinson1970}, 
figure 4(a) therein, we see that the amplification for lower frequencies seems
to coincide well. However, for higher frequencies the present amplifications
reach their maximum earlier than in \cite{Jordinson1970}. We accord this
difference to the different numerical methods used in \cite{Jordinson1970} and
in the present work. We remark that the graphs in figure \ref{fig:AmplificationJordinson} were obtained by using two resolutions $ N_\bl = 80 $ and $ N_\bl = 120 $ giving results identical to plotting accuracy. 

\subsection{Validation by means of the boundary layer under a solitary wave}

\begin{figure}
  \centerline{\includegraphics[width=\linewidth]{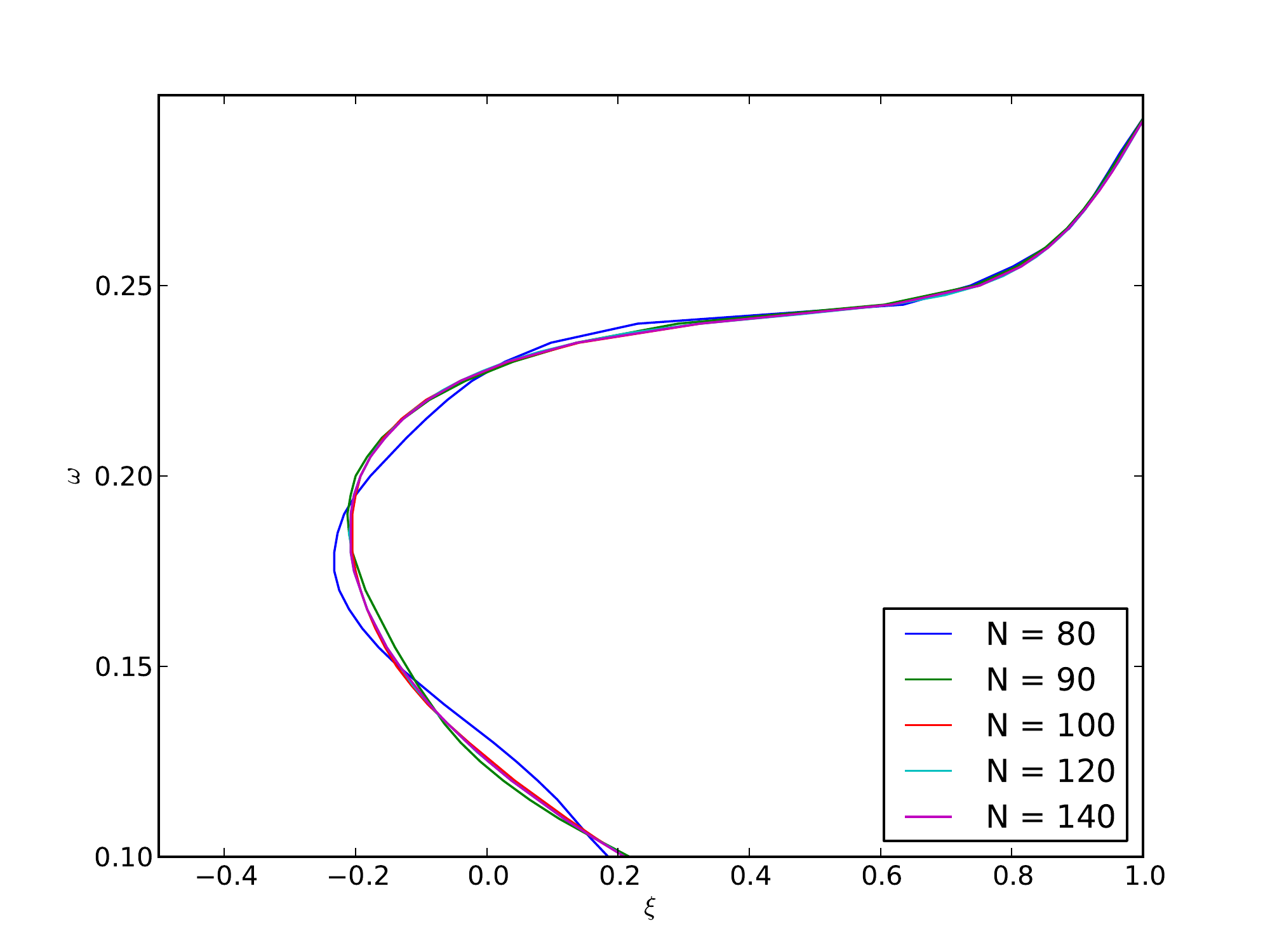}}
\caption{The stability domain for the case $ \epsilon = 0.4 $ and $\delta = 10^{-4} $ computed using the Orr-Sommerfeld solver derived in subsection
\ref{sec:OrrSommerfeldSolver} for different resolution $N_\ose $.}
\label{fig:domainOrrSommerfeld}
\end{figure}

In order to determine the resolution $ N_\ose $ necessary 
to obtain meaningful results, we used the 
present Orr-Sommerfeld solver to compute the stability domain for
the case $ \epsilon = 0.4 $ and $ \delta = 10^{-4}$ for different
resolutions. A zoom onto the neutral curve can be seen in figure
\ref{fig:domainOrrSommerfeld}. For resolutions $ N_\ose \ge 100 $ the
curves are almost identical. For finer resolutions, the neutral curve
oscillates around the curve of finer resolution. These oscillations become
smaller and smaller in amplitude and osculate to the limiting curve. 
The accuracy chosen for all simulations using the Orr-Sommerfeld 
solver in section \ref{sec:results} was $ N_\ose = 130 $. 
This was done in order to keep the error
contribution by the Orr-Sommerfeld solver subdominant comparing to
the error contribution by the potential solver.

\section{Verification and validation of the parabolic stability equation solver} \label{sec:appendixPSE}
The parabolic stability equation solver,
derived in subsection \ref{sec:ParabolicStabilityEquationSolver} 
was verified the same way as the Orr-Sommerfeld equation solver. First, we used
data in the literature to verify the correct implementation 
of the method by 
means of the Blasius boundary layer. After that we investigated the convergence of the
method when applied to the computation of the neutral curve for 
a stability domain for the boundary layer flow under a solitary wave
itself. 

\subsection{Validation by means of the Blasius boundary layer flow}

\begin{figure}
  \centerline{\includegraphics[width=\linewidth]{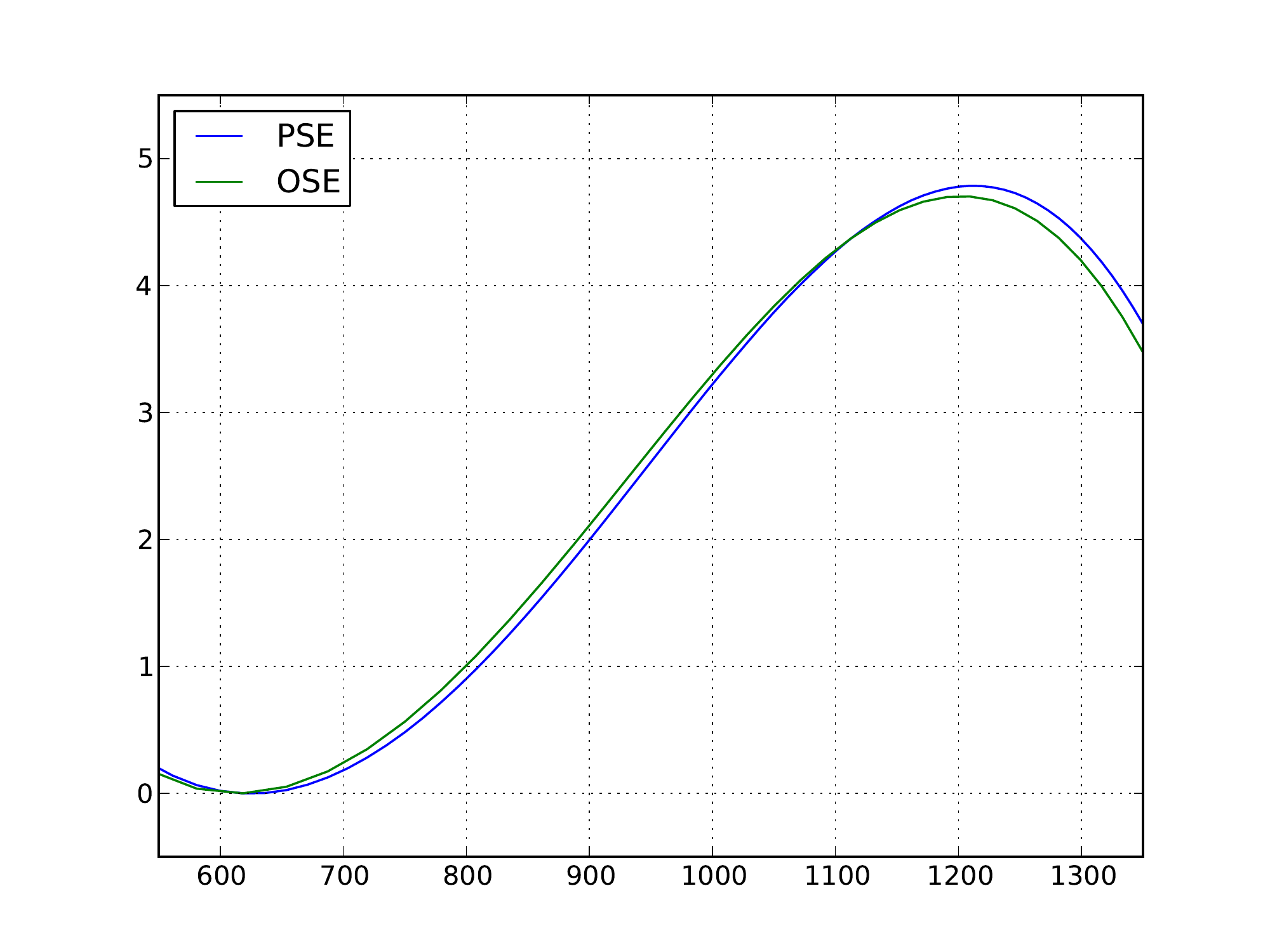}}
\caption{The amplification for a Tollmien-Schlichting wave with frequency $ F = 50 \cdot 10^{-6} $ for the Blasius boundary layer.
The graphs were computed by means of the present parabolic
stability equation solver (PSE) and the present Orr-Sommerfeld equation solver (OSE).
The graphs can be compared to figure 4(b) in \cite{BertolottiHerbertSpalart1992}.} 
\label{fig:bertolotti1}
\end{figure}

\begin{figure}
  \centerline{\includegraphics[width=\linewidth]{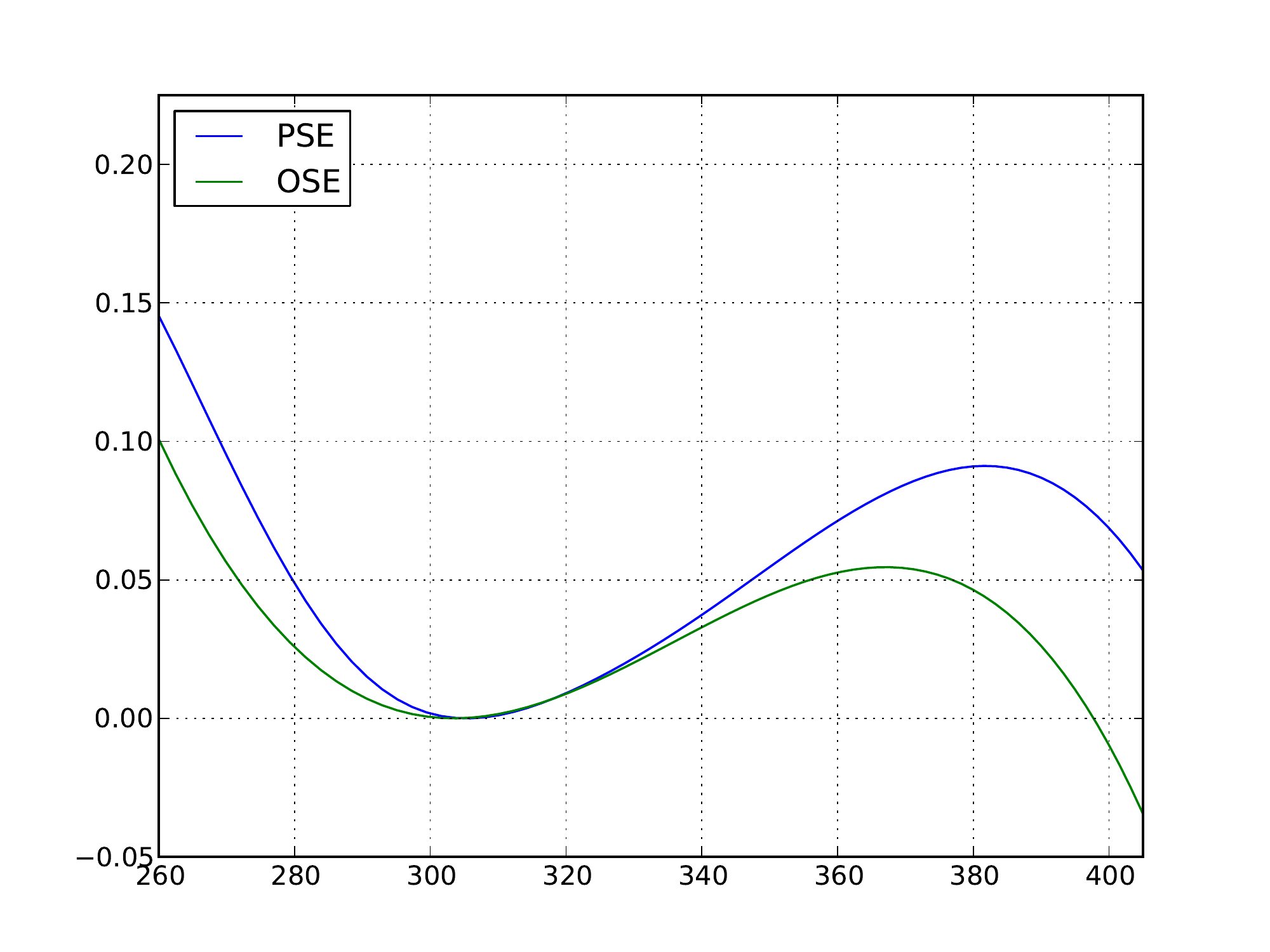}}
\caption{The amplification for a Tollmien-Schlichting wave with frequency $ F = 220 \cdot 10^{-6} $ for the Blasius boundary layer.
The graphs were computed by means of the present parabolic
stability equation solver (PSE) and the present Orr-Sommerfeld equation solver (OSE).
The graphs can be compared to figure 4(a) in \cite{BertolottiHerbertSpalart1992}.} 
\label{fig:bertolotti2}
\end{figure}

Opposed to the Orr-Sommerfeld solver, no explicit reference values
are given enabling to pin down the correct implementation
by means of a concrete number. Therefore, we recalculated two cases
given in \cite{BertolottiHerbertSpalart1992} to verify the
correctness of the present method. These two cases consist in
computing the amplification of a Tollmien-Schlichting wave for
the frequencies $ F= 50 \cdot 10^{-6} $ and $ F= 220 \cdot 10^{-6} $,
respectively. The results of the present computations can be 
seen in figures \ref{fig:bertolotti1} and \ref{fig:bertolotti2}. 
These can be compared directly to the graphs in 
figure 4(a) and 4(b) in \cite{BertolottiHerbertSpalart1992}. 
Digitizing the amplification curves from this reference we find that
the results agree to plotting scale. The amplification curves
computed by means of the Orr-Sommerfeld equation are slightly different. 
However, \cite{BertolottiHerbertSpalart1992} did not give any details
on the implementation and resolution used for their Orr-Sommerfeld equation solver. The resolution used for the computation 
of the amplification curves was $ N_\ose = 120 $ and $ N_\pse = 120 $ for
both cases.

\subsection{Validation by means of the boundary layer under a solitary wave}

\begin{figure}
  \centerline{\includegraphics[width=\linewidth]{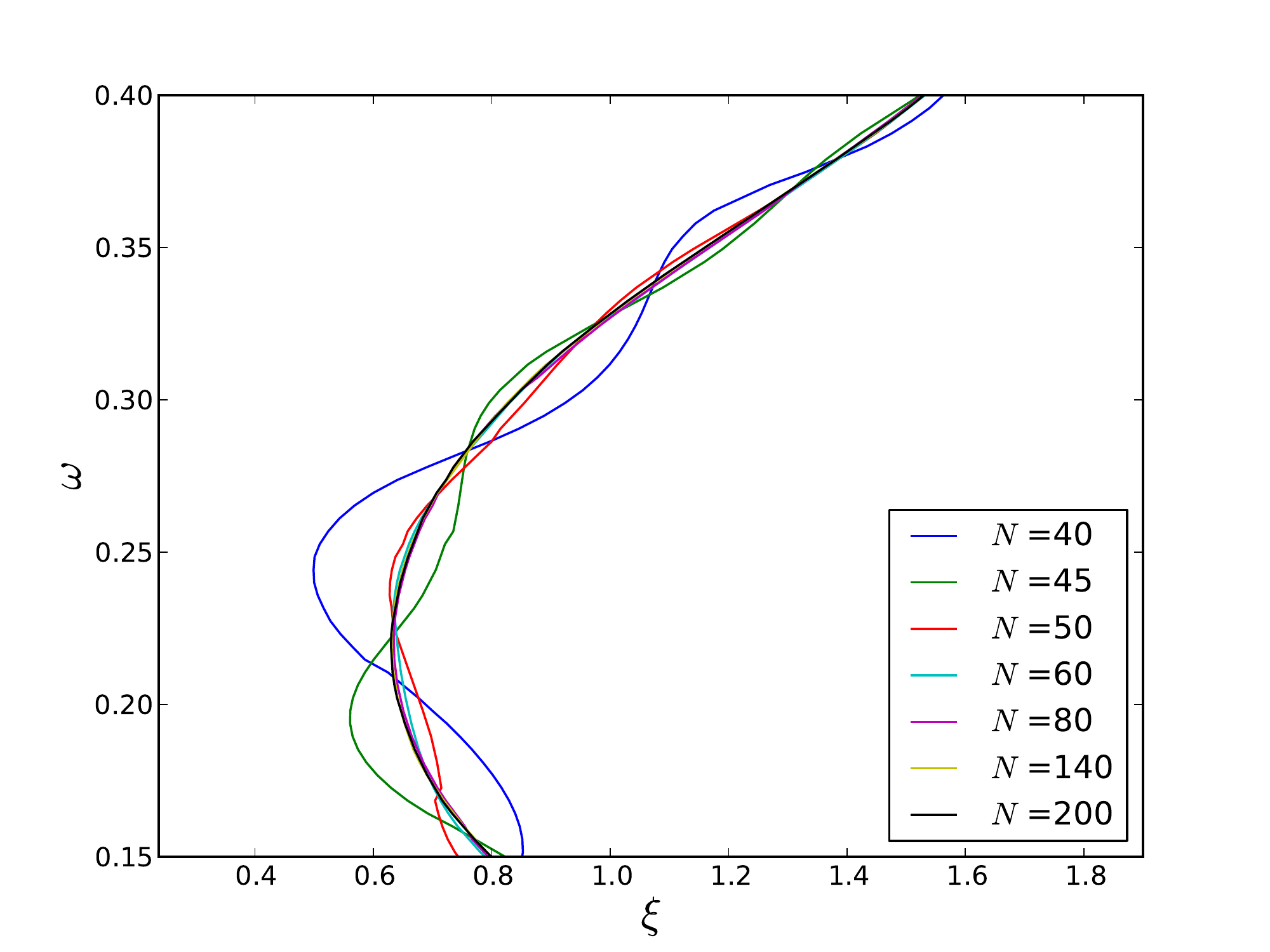}}
  \caption{Zoom on the stability curve of figure \ref{fig:domainDelta0.000475}
    for $ \delta = 4.75\times 10^{-4} $ and $ \epsilon = 0.3 $. The 
curves for different resolutions converge to one single curve for increasing
resolution $ N_\pse $.}
\label{fig:domainDelta0.000475Convergence}
\end{figure}

Similar to the verification of the Orr-Sommerfeld solver, we computed the neutral curve for the boundary layer flow under a solitary wave
using the present parabolic stability equation solver derived in 
subsection \ref{sec:ParabolicStabilityEquationSolver} for different resolutions
$ N_\pse $.
The case chosen was $ \epsilon = 0.3 $ and $ \delta =  4.75\times 10^{-4} $. 
The results can be seen in figure \ref{fig:domainDelta0.000475Convergence}. 
As can be observed the curves for $ N_\pse \ge 140 $ are coinciding on
a plotting accuracy. The higher the resolution, the better the curves follow
this ultimate curve. For all the simulations in section \ref{sec:results}, we
used a resolution of $ N_\pse = 180 $. 

\section{Validation of the Navier-Stokes solver} \label{sec:appendixNS}

The spectral Legendre-Galerkin solver NEK5000 has been verified and
validated over several years. A careful validation of 
the present case is nevertheless indispensable. As a first test, we validated the present solver by
performing a convergence test for the case of no perturbation. 
A second test is then given by introducing a Tollmien-Schlichting wave
and analyzing its amplification for different resolutions. 

\subsection{Validation by means of the boundary layer under a solitary
wave}

\begin{figure}
  \centerline{\includegraphics[width=\linewidth]{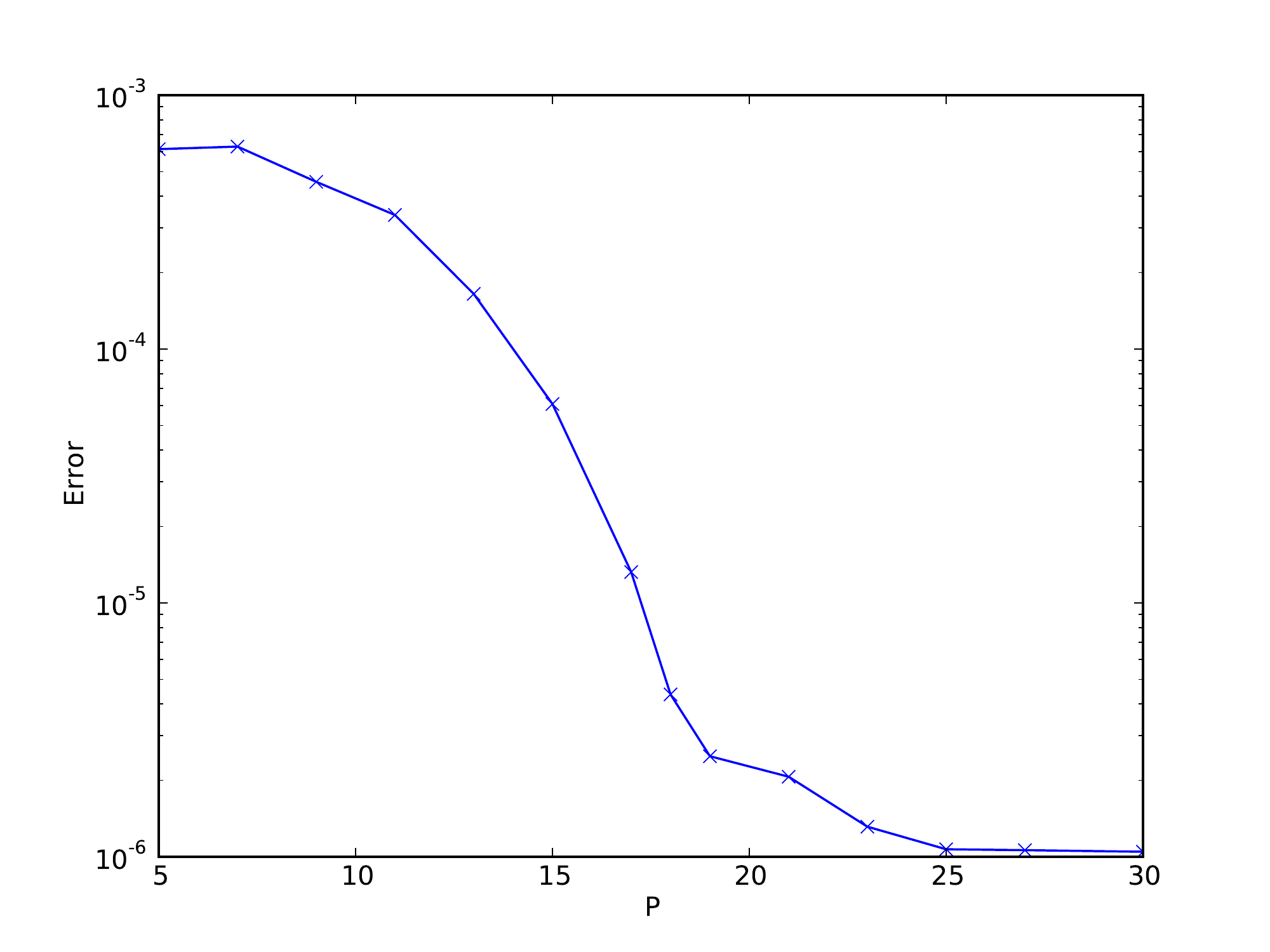}}
  \caption{Convergence of the Navier-Stokes solver when increasing
the degree $ P $ of the polynomials. The boundary layer was simulated
in a domain given by the extensions $ \xi_0 = -0.4 $, $ \xi_1=0.4$ and $ L_y = 41.7 $
for the case $ \epsilon = 0.4 $, $ \delta = 8 \cdot 10^{-4} $. The number
of elements in $ x $ and $ y $ was $ N_x = 20 $ and $ N_y = 1 $, respectively. 
The error was measured in the $ L2 $ norm.}
\label{fig:convergenceNavierStokes}
\end{figure}

\begin{figure}
  \centerline{\includegraphics[width=\linewidth]{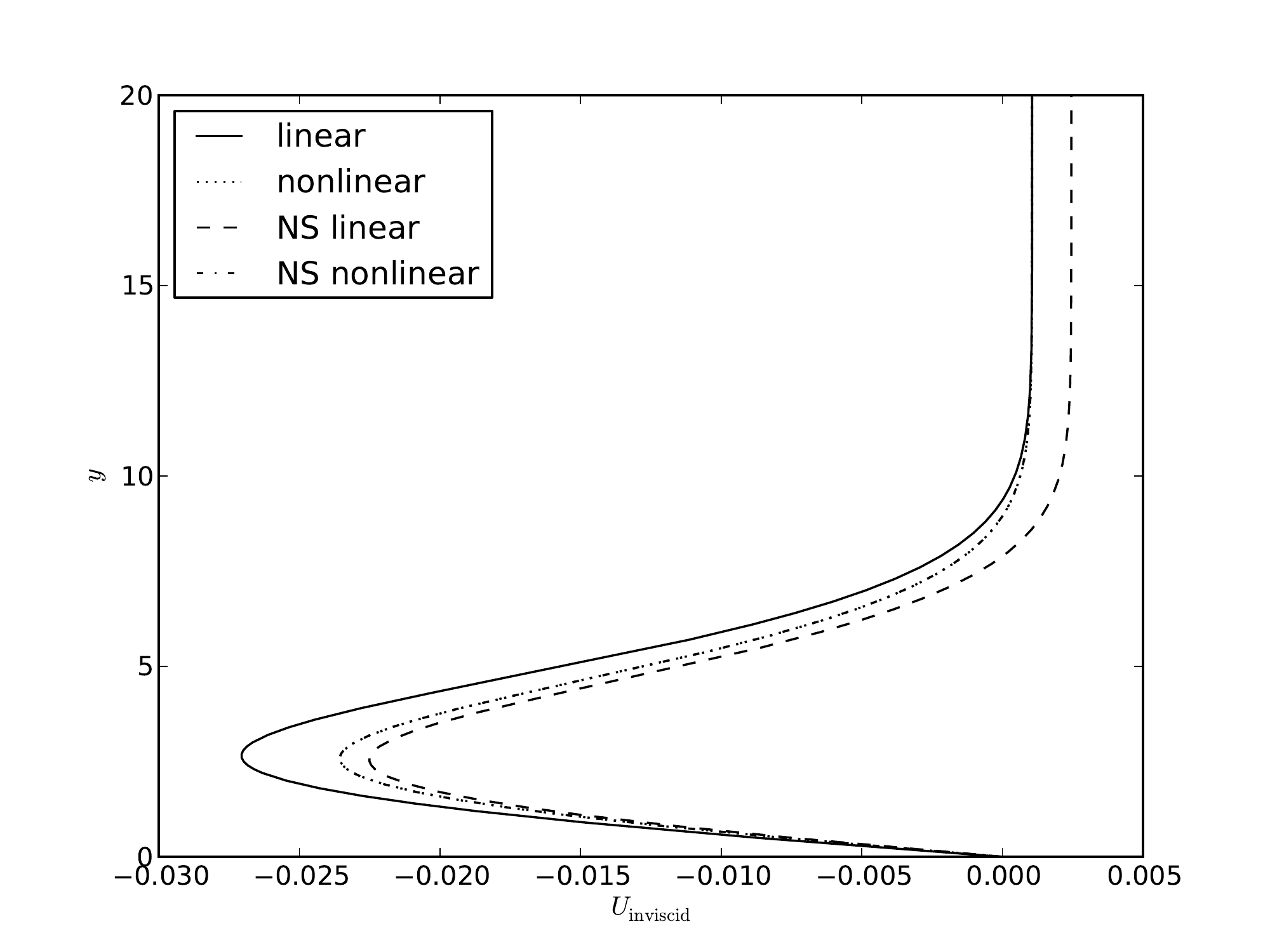}}
  \caption{Profiles of the horizontal velocity component
for the case $ \epsilon = 0.2 $ and $ \delta = 4.4 \cdot 10^{-3} $.
The profiles are computed by means of the linear boundary layer solution (linear),
equation (\ref{eq:linearLiu}), the nonlinear boundary layer
solution (nonlinear) or by means of the 
Navier-Stokes solver (NS), using either the linear (NS linear) or nonlinear
(NS nonlinear)
boundary layer solution at the boundaries of the domain.}
\label{fig:profilesLinearVSNonlinearNSSolver}
\end{figure}

By setting $ u'=0$ , $ v'=0 $ in equation (\ref{eq:inflowNS}), the present
Navier-Stokes solver should reproduce the solution given by the boundary layer
solver. This allowed us to test the convergence properties of the
present Navier-Stokes solver. We used the case $ \epsilon = 0.4 $, $ \delta
= 8 \cdot 10^{-4} $, on a grid given by $ \xi_0 = -0.4 $, $ \xi_1=0.4$ and $ L_y = 41.7 $. The number of elements was chosen relatively low
in order to observe some convergence, $ N_x = 20 $, $ N_y = 1 $. In figure
\ref{fig:convergenceNavierStokes}, we observe that after a relatively 
slow convergence up to $ P = 13 $, the solver displays a 
spectral convergence up to the point, where the error reaches a limit
of approximately $ 10^{-6} $, which happens at $ P = 19 $. This
saturation is due to the error for the inviscid flow computation
being larger than the error by the boundary layer solver and the
Navier-Stokes solver. Since it is more efficient to increase
the number of elements which can be distributed onto more
cores, we chose in general $ P = 11 $, $ N_x = 300 $ and $ N_y = 12 $ for
the simulations in section \ref{sec:results}. As a matter of fact
the choice of the velocity $ (U_\base, V_\base) $ at the
boundaries of the domain has a huge impact on the 
accuracy of the boundary layer solution. In figure
\ref{fig:profilesLinearVSNonlinearNSSolver}, we plotted
different profiles of the horizontal velocity component
for the case $ \epsilon =0.2 $ and $ \delta = 4.4\cdot 10^{-3} $
at the location $ \xi = 9.57 $, which was also investigated
in section \ref{sec:results} and by \cite{LiuParkCowen2007}. 
For all the profiles, we used Grimshaw's solution, equation 
(\ref{eq:grimshaw1}) for the inviscid free stream velocity. 
As can be seen from figure \ref{fig:profilesLinearVSNonlinearNSSolver},
the linear boundary layer profile computed by means of
\ref{eq:linearLiu} and the nonlinear boundary layer profile
computed by means of the boundary layer solver derived in 
subsection \ref{sec:boundaryLayerSolver}
have different courses in the boundary layer, as discussed in section
\ref{sec:results}. However, their value for $ y \rightarrow \infty $
is identical and corresponds to the value given by Grimshaw's solution
at the bottom of the solitary wave. If applying the values for $ (U_\base,V_\base) $
computed using the nonlinear boundary layer solver on the boundary
of the domain for the Navier-Stokes solver, we observe that the 
profiles computed by means of the nonlinear boundary layer solver
and the Navier-Stokes solver are identical up to plotting accuracy,
cf. figure \ref{fig:profilesLinearVSNonlinearNSSolver}. 
This is, however, not the case when applying the linear solution,
equation (\ref{eq:linearLiu}) together with the correct
normal velocity component, equation (\ref{eq:normalComponentLiu})
at the boundaries of the domain for the Navier-Stokes solver. The profile
displays in this case a different free stream velocity.
Due to the nonlinearity of the Navier-Stokes solver, the
linear boundary layer solution on the top boundary could not be satisfied without
violating continuity. Therefore, in order to satisfy continuity, 
additional fluid is pushed in horizontal direction, leading
to a different value of the free stream velocity and a
second boundary layer at the top of the computational domain. 

\subsection{Validation by means of the amplification of a Tollmien-Schlichting wave}

\begin{figure}
  \centerline{\includegraphics[width=\linewidth]{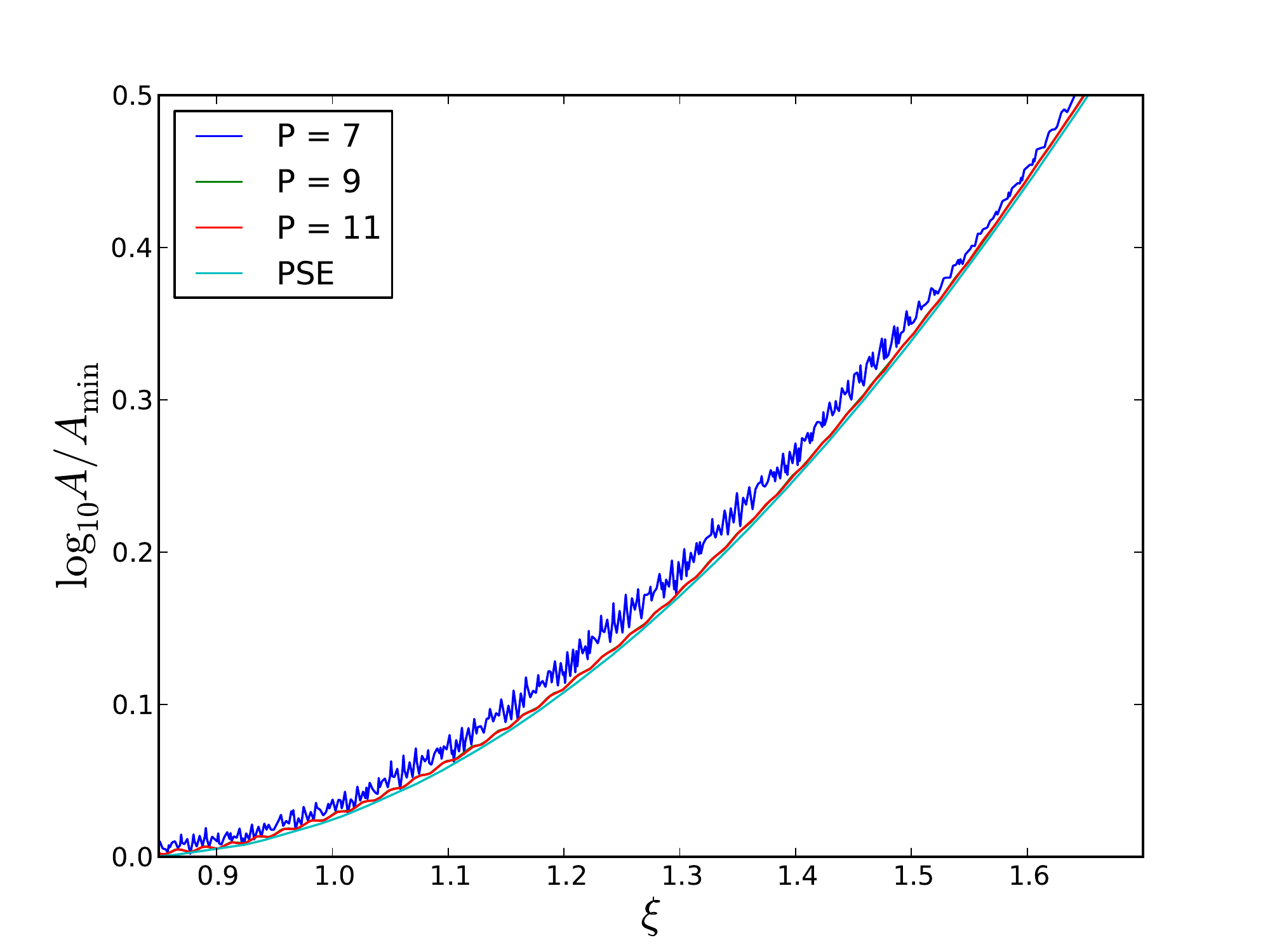}}
  \caption{Amplification curves for the case $  \epsilon = 0.4 $, $ \delta = 8 \cdot 10^{-4} $ and $ \omega = 0.24 $ 
computed by the Navier-Stokes solver for different polynomial degrees $ P $. }
\label{fig:convergenceAmplification}
\end{figure}

The good agreement between amplifications computed by the parabolic
stability equation solver and the Navier-Stokes solver in
section \ref{sec:results} is by itself a validation for both methods. However, a few words on the numerical issues when computing
such amplification curves must be said. In figure 
\ref{fig:convergenceAmplification}, we plotted a zoom on the curve
in figure \ref{fig:amplificationDelta0.0008} using different 
polynomial degrees $P$. As can be observed the curve for $ P = 7 $
displays some wiggles. The curves for $ P = 9 $ and $ P = 11 $ lie on top of
each other. However, a closer look reveals that also they display a
slight undulation which is not present in the amplification
curve by the parabolic stability equation method. 
The wiggles and the undulation might have
their origin in the fact that the Navier-Stokes solver solves
the nonlinear Navier-Stokes equations whereas the parabolic stability equation
method is
based on linearized equations. Therefore we might expect a
signal with an amplitude of the order of $ 10^{-7} $ (corresponding to the omitted nonlinear term in the parabolic stability equation $\approx (5 \cdot 10^{-4})^2$), which is
also the amplitude of the undulation in figure \ref{fig:convergenceAmplification}. 

\end{appendix}


\end{document}